\documentclass[reprint,superscriptaddress]{revtex4-2}

\usepackage[T1]{fontenc}
\usepackage[latin9]{inputenc}
\usepackage{babel}
\usepackage{enumitem}
\usepackage{pifont}
\usepackage{verbatim}
\usepackage{float}
\usepackage{amsmath}
\usepackage{physics}
\usepackage{mathtools}
\usepackage{amsthm}
\usepackage{amssymb}
\usepackage{graphicx}
\usepackage{appendix}
\usepackage[dvipsnames]{xcolor}
\usepackage{array}
\usepackage{multirow}
\newcolumntype{P}[1]{>{\centering\arraybackslash}p{#1}}

\usepackage{tikz}
\usepackage{pgfplots}
\pgfplotsset{compat=newest} 
\pgfplotsset{plot coordinates/math parser=false}
\usepackage{rotating}

\usetikzlibrary{decorations.markings,positioning,calc}

\usepackage[unicode=true,pdfusetitle,
bookmarks=true,bookmarksnumbered=false,bookmarksopen=false,
breaklinks=false,pdfborder={0 0 1},backref=false,colorlinks=false]
{hyperref}

\usepackage{subfigure,bbm}

\usepackage{mathtools}

\DeclarePairedDelimiter\floor{\lfloor}{\rfloor}

\newcommand{\san}[1]{\mathsf{#1}}

\newif\ifnotes
\notestrue

\makeatother

\newcommand{\be}{\begin{equation}}
\newcommand{\ee}{\end{equation}}
\newcommand{\ba}{\begin{align}}
\newcommand{\ea}{\end{align}}

\newcommand{\id}{\mathbb{I}}
\newcommand{\q}{\hat{q}}
\newcommand{\p}{\hat{p}}

\makeatletter

\begin{document}

\title{All-photonic GKP-qubit repeater using analog-information-assisted multiplexed entanglement ranking}

\author{Filip Rozp\k{e}dek}
\email{frozpedek@umass.edu}
\affiliation{Pritzker School of Molecular Engineering, University of Chicago, IL 60637, USA}
\author{Kaushik P. Seshadreesan}
\affiliation{College of Optical Sciences, University of Arizona, Tucson, AZ 85721, USA}
\affiliation{Department of Informatics and Networked Systems, University of Pittsburgh, Pittsburgh, PA 15260, USA}
\author{Paul Polakos}
\affiliation{Cisco Systems, New York, NY 10119, USA}
\author{Liang Jiang}
\affiliation{Pritzker School of Molecular Engineering, University of Chicago, IL 60637, USA}
\author{Saikat Guha}
\email{saikat@arizona.edu}
\affiliation{College of Optical Sciences, University of Arizona, Tucson, AZ 85721, USA}

\begin{abstract}
Long distance quantum communication will require the use of quantum repeaters to overcome the exponential attenuation of signal with distance. One class of such repeaters utilizes quantum error correction to overcome losses in the communication channel. Here we propose a novel strategy of using the bosonic Gottesman-Kitaev-Preskill (GKP) code in a two-way repeater architecture with multiplexing. The crucial feature of the GKP code that we make use of is the fact that GKP qubits easily admit deterministic two-qubit gates, hence allowing for multiplexing without the need for generating large cluster states as required in previous all-photonic architectures based on discrete-variable codes. Moreover, alleviating the need for such clique-clusters entails that we are no longer limited to extraction of at most one end-to-end entangled pair from a single protocol run. In fact, thanks to the availability of the analog information generated during the measurements of the GKP qubits, we can design better entanglement swapping procedures in which we connect links based on their estimated quality. This enables us to use all the multiplexed links so that large number of links from a single protocol run can contribute to the generation of the end-to-end entanglement. We find that our architecture allows for high-rate end-to-end entanglement generation and is resilient to imperfections arising from finite squeezing in the GKP state preparation and homodyne detection inefficiency. In particular we show that long-distance quantum communication over more than 1000 km is possible even with less than 13 dB of GKP squeezing. We also quantify the number of GKP qubits needed for the implementation of our scheme and find that for good hardware parameters our scheme requires around $10^3-10^4$ GKP qubits per repeater per protocol run.
\end{abstract}

\maketitle

\section{Introduction}

Quantum repeaters~\cite{briegel1998quantum, Munro_15} are central to the realization of a vision for the global-scale quantum internet. They can overcome losses in the communication channel by working around the limitations imposed by the no cloning theorem~\cite{wootters1982single}. Specifically, this theorem tells us that direct noiseless signal amplification cannot be realised in the quantum domain, while it is also known that the type of quantum amplifiers allowed by quantum mechanics introduce too much noise and that a quantum amplifier on its own cannot compensate for the channel losses~\cite{namiki2014gaussian}.
A myriad different architectures and designs for quantum repeaters have been proposed over the years.

A widely-used strategy in quantum repeaters to significantly increase the rate of generating long-distance entangled links is multiplexing. 
Generally, there exist two types of multiplexing strategies depending on the experimental capabilities of the quantum memories at the repeaters. Physical platforms with a single communication qubit with an optical interface and large number of ancilla memories offer the possibility of time multiplexing~\cite{jones2016design}. For this type of multiplexing, one needs to attempt optical interfacing between adjacent nodes sequentially due to the limitation of the single communication qubit. However, one can increase the repetition rate to be limited only by the local gate time between the communication qubit and the memories rather than the duration of the single entanglement generation attempt limited by the classical communication time between the nodes. This strategy is particularly applicable to defects in diamond such as NV or SiV centres~\cite{van2017multiplexed} as well as to dual-species trapped ion architectures~\cite{inlek2017multispecies,dhara2022multiplexed}. Unfortunately, in many of these platforms, such as e.g.~NV centres in diamond, a too long two-qubit gate time is still a strong limiting factor for such multiplexing schemes~\cite{van2017multiplexed}.

On the other hand, physical platforms that offer the possibility of a parallel transfer of large number of optical qubits into the quantum memory enable parallel entanglement generation attempts where an end-to-end link can be established provided that at least one of the multiplexed links over every elementary segment is successfully generated~\cite{guha2015rate}. Atomic ensemble platforms are promising candidates for such multiplexed quantum memories as it has already been experimentally demonstrated that they can absorb and store multiple frequency, spatial and temporal modes in parallel~\cite{sinclair2014spectral, yang2018multiplexed}. 
However, entanglement swapping in such platforms still needs to be performed optically, which unfortunately cannot be performed deterministically.

As an alternative strategy, it has been observed that quantum memories can be emulated using photonic states encoded in quantum error correcting codes protecting the state against photon loss so that the photonic qubits can simply be reliably stored in the optical fiber~\cite{Azuma_15,pant2017rate}. One such promising code is the tree code~\cite{varnava2006loss} which forms a basis for multiple repeater proposals~\cite{Azuma_15,pant2017rate,borregaard2019one}. 
It has also been shown that by the so called ``time-reversed'' entanglement swapping, where each repeater pre-generates a clique-cluster, a deterministic entanglement swap can be effectively engineered even with a purely optical system based on dual-rail encoded qubits~\cite{Azuma_15,pant2017rate}. Combining the simulated photonic quantum memories with the ``time reversed'' entanglement swapping allows for efficient multiplexing without the need for any physical quantum memories at all~\cite{Azuma_15,pant2017rate}.
However, the cost of this architecture is the necessity of pre-generating a large highly-entangled cluster with qubits encoded in large tree codes. Moreover, even for a large number of multiplexing levels, at most one end-to-end entangled link can be extracted through this procedure per run of the repeater protocol.

\begin{figure*}
    \centering
    \includegraphics[width=1\textwidth]{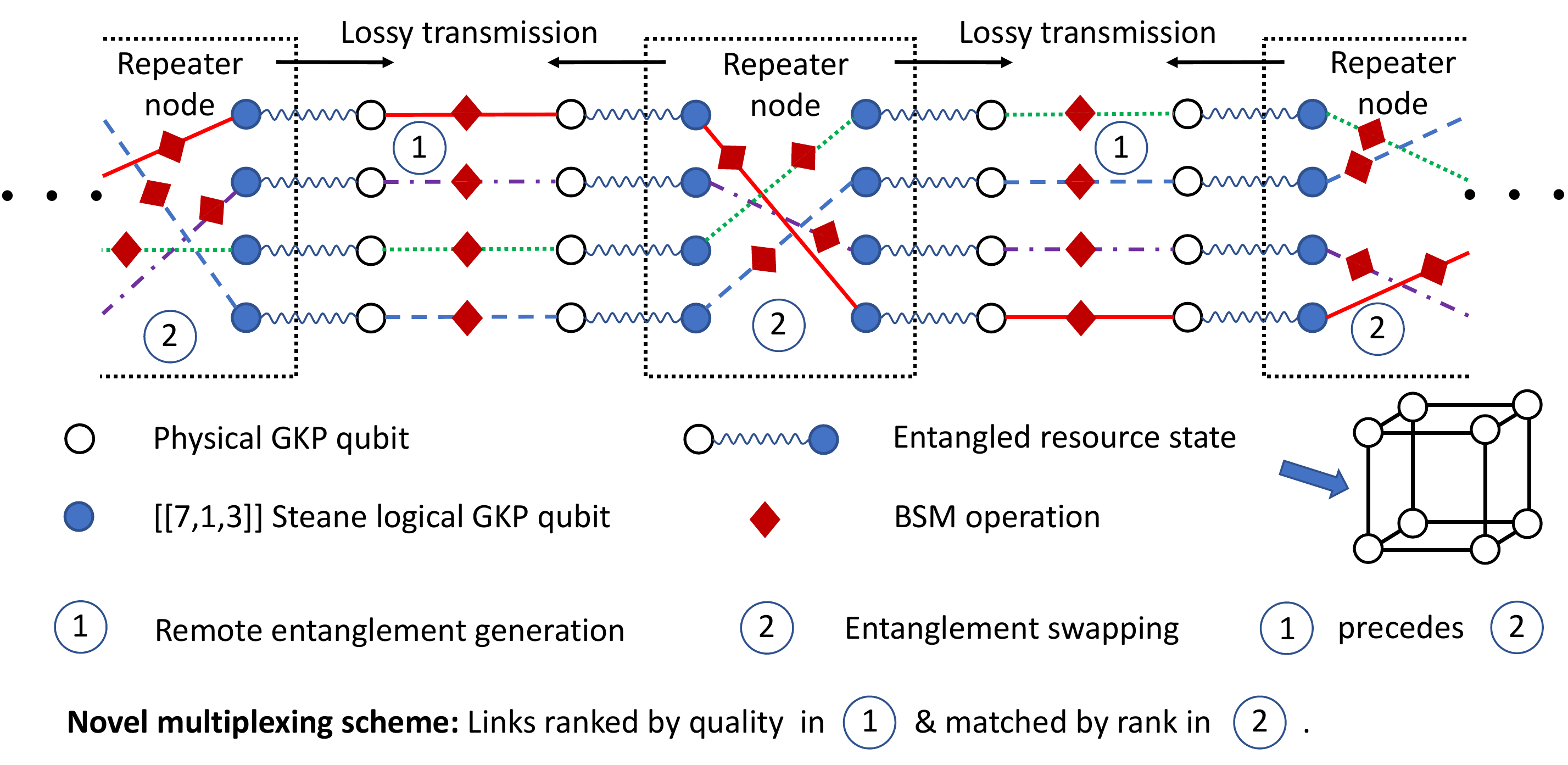}
    \caption{A schematic of the proposed multiplexed two-way all-photonic quantum repeater scheme based on GKP-encoded qubits and the [[7,1,3]] Steane code. At the beginning repeaters need to prepare the entangled resource states that correspond on the logical level to a Bell pair between a concatenated-coded qubit and a bare GKP qubit and on the physical level to a cube graph state (up to Hadamards) of eight GKP qubits. Then remote entanglement generation is performed by sending the bare physical GKP qubits (white/empty circle) towards each other for Bell state measurement (BSM). The BSMs are ranked according to their reliability estimated from the GKP syndromes obtainable from the continuous BSM outcomes themselves. The ranking is illustrated by the lines of different color joining outer leafs. This ranking information as well as logical BSM outcomes are sent to the repeaters which perform entanglement swapping on the concatenated-coded qubits (blue/filled circle) based on that ranking information (lines of different color joining inner leafs).}
    \label{fig:schematic}
\end{figure*}

In this work we propose a new two-way all-photonic repeater architecture with multiplexing which does not require the pre-generation of a large clique-cluster. 
It requires much less optical modes and allows for extraction of entanglement or secret-key for the cryptographic task of quantum key distribution (QKD)~\cite{Bennett_84,ekert1991quantum} from all the multiplexed links rather than only a single link. 
The crucial component of our scheme which enables these features is the use of the bosonic Gottesman-Kitaev-Preskill (GKP) code~\cite{gottesman2001encoding} as our encoded qubit. 
This encoding has been proven to be very efficient against photon loss~\cite{albert2018performance,noh2018quantum} and has already been introduced as a promising error-correcting code for quantum repeaters~\cite{fukui2021all,rozpkedek2021quantum}. 
Moreover, the GKP code has certain useful properties that are particularly relevant in the context of all-optical two-way repeaters. 
Firstly, GKP qubits admit deterministic two-qubit gates~\cite{gottesman2001encoding} which alleviates the need of ``time-reversed'' entanglement swapping through the generation of the clique-cluster. 
Secondly, measurement of the GKP qubits generates additional analog information~\cite{fukui2017analog}. 
This analog information created during the generation of elementary multiplexed links (specifically during the heralded Bell state measurements) enables us to rank all the multiplexed links according to observed reliability of each individual Bell State Measurement (BSM). 
Then, by a suitable strategy of connecting the ranked elementary links inside the repeaters, one can efficiently extract entanglement or secret key from all the links, hence significantly improving the performance relative to the previous all-optical repeater architectures. 

Moreover, as GKP qubits already offer significant protection against photon loss and since it has already been shown that using the GKP analog information can significantly boost the error-correction capabilities of the concatenated-coded schemes, here we emulate quantum memories by concatenating the GKP code with the [[7,1,3]] Steane code~\cite{steane1996error}. 
This concatenation strategy has already been proposed in the context of one-way repeaters based on GKP qubits~\cite{rozpkedek2021quantum} where it was shown that together with the help of the analog information, such encoding can probabilistically correct most of the single- and two-qubit errors on the higher level. On the other hand in our new two-way scheme we implement the [[7,1,3]]-code correction through entanglement swapping on a logical level which is more resilient to hardware imperfections than the previously considered method of stabilizer measurements using additional ancilla GKP qubits. 

We find that thanks to the features described above, our scheme achieves the entanglement/secret-key generation rate per mode which for good but at the same time reasonable hardware parameters can stay above 0.7 for 750 kilometers. To our knowledge there has been only one other repeater scheme proposed so far which can achieve such performance~\cite{fukui2021all}. That scheme also utilises GKP qubits, yet we show here that our scheme can achieve this performance with a significantly relaxed requirement on the amount of GKP squeezing relative to the previous scheme. Moreover, our scheme can reach much longer distances than comparable previously proposed GKP based repeater strategies and can do so with larger inter-repeater spacing.

Finally, we note that thanks to the [[7,1,3]]-code syndrome information obtained at each repeater we can further significantly boost the achievable distances~\cite{namiki2016role}. While over these longer distances we might not be able to generate the amount of entanglement or secret key needed for high-speed QKD or distributed quantum computing, we show that using the [[7,1,3]]-code syndrome information we can significantly extend the distance regime over which the performance of our scheme stays above the PLOB bound~\cite{pirandola2017fundamental}, which is the ultimate limit of repeater-less quantum communication and corresponds to the two-way assisted capacity of the direct transmission pure loss channel. In the limit of high losses this bound scales linearly with the channel transmissivity~\cite{takeoka2014fundamental}.

Additionally we also analyse the number of GKP qubits needed for the realisation of our scheme. We find that for reasonable hardware parameters one needs around $10^3-10^4$ GKP qubits per repeater per single protocol run for 20 multiplexed links. This is around four orders of magnitude lower than the number of single photons needed in the discrete-variable all-photonic scheme of~\cite{pant2017rate}. Clearly GKP qubits and single photons correspond to very different type of resources. Therefore additionally we also provide a high level discussion of how many Gaussian Boson Sampling (GBS) circuits and single mode squeezers would be needed to generate this required number of GKP qubits, if that strategy for GKP qubit generation was used.

Our results have been obtained using numerical Monte-Carlo simulations done in MATLAB. The code used to obtain these results is freely available~\cite{Note2}. However, we also provide analytical analysis of some of the features of our scheme.

For the convenience of the reader, we sum up here the main results of the paper:
\begin{enumerate}
\item We propose a novel form of multiplexing based on GKP analog information, where parallel entangled links can be ranked according to their expected quality. This strategy is summarised in Section~\ref{sec:rankingOuter} and its performance evaluated in Section~\ref{sec:RankingPerformance}. This broadly applicable technique could also be useful in other contexts, such as entanglement routing, where different user pairs in a network might require different quality entanglement or in measurement based quantum computation where the resource states of different quality could be constructed for different computational tasks.
\item We propose a novel all-photonic GKP-based repeater scheme and develop a broad framework for analyzing its performance under various hardware imperfections. This framework is based on numerical simulations but it also includes an analytical model which captures specific features of our scheme (see Section~\ref{sec:AnalyticalModel}). Our repeater protocol achieves high performance and has significantly reduced experimental requirements relative to the previous concatenated-coded GKP-based repeater architectures. The relaxation of these hardware requirements in our scheme include:
\begin{itemize}
    \item Reduced GKP squeezing requirement (see Fig.~\ref{fig:AchievableDistancesL50Main}).
    \item Robustness to finite homodyne detection efficiency (see Fig.~\ref{fig:AchievableDistancesL50Main}).
    \item Assumption of perfect GKP-qubit storage only during the generation of the eight-qubit resource states.
\end{itemize}
\item We perform a detailed resource analysis for our novel repeater scheme (see Section~\ref{sec:OptResourcesResults}). Our analysis establishes the total number of GKP qubits needed to implement our scheme, i.e., it includes all the GKP qubits that would be discarded in the post-selected fusions during resource state creation. We also provide simple estimates of the total number of single-mode squeezers if the GKP qubits were to be prepared using the Gaussian Boson Sampling technique (see Section~\ref{sec:GBS}).
\item We analyse the trade-off between the scheme performance and the required resources (see Section~\ref{sec:OptResourcesResults}). We find that for a specific choices of hardware parameters, our scheme is able to generate between 2-8 ebits/secret bits using in total $10^7-10^8$ GKP qubits per protocol run across the distance of 5000 km. This can be compared against the discrete-variable all-photonic scheme of~\cite{pant2017rate} which over the same distance achieves the rate of 0.14 secret bits per protocol run using in total $10^{12}$ single photons.
\end{enumerate}

The paper is structured as follows. In Section~\ref{sec:GKPintro} we introduce the GKP code and the GKP stabilizers. In Section~\ref{sec:RepeaterScheme} we describe our repeater scheme. In Section~\ref{sec:ResourceState} we describe the procedure to generate the resource state for our repeater scheme. In Section~\ref{sec:rankingOuter} we provide more details on our novel multiplexing procedure based on the ranking of the outer leaves. In Section~\ref{sec:innerleaves} we describe the error correction procedure for the concatenated-coded inner leaves and explain how the [[7,1,3]] code syndrome information can boost the achievable distances. In Section~\ref{sec:performanceMetrics} we provide a high-level description of how we quantify the performance of our scheme. In Section~\ref{sec:Simulation} we provide a high-level overview of our numerical simulation for the repeater performance. In Section~\ref{sec:results} we describe our findings and results. In Section~\ref{sec:discussion} we provide further discussion of our scheme. Specifically we describe a simple analytical model that we develop in order to attempt to analytically capture and approximate the complex behaviour of our scheme, we discuss how the GKP qubits can be prepared in the optical regime using Gaussian boson sampling circuits, we compare our scheme's performance to that of certain other proposed repeater schemes and finally we discuss the requirements of the end-nodes of Alice and Bob depending on the specific communication task in mind. We conclude in Section~\ref{sec:conclusion} with the future outlook.

\section{GKP qubits and GKP error correction}
\label{sec:GKPintro}

The single-mode square-lattice based GKP code~\cite{gottesman2001encoding} is defined as the simultaneous plus one eigenspace of the two operators:

\begin{equation}
\hat{S}_q = \exp(i2\sqrt{\pi}\hat{q}), \quad \quad \hat{S}_p = \exp(-i2\sqrt{\pi}\hat{p})\, .
\label{eq:GKPstabilizers}
\end{equation}

The standard basis states of the GKP code are then defined as:

\begin{equation}
\begin{aligned}
\ket{0_{\text{GKP}}} &= \sum_{n \in \mathbb{Z}} \ket{q = 2n\sqrt{\pi}}, \\
\ket{1_{\text{GKP}}} &= \sum_{n \in \mathbb{Z}} \ket{q = (2n+1)\sqrt{\pi}}.
\end{aligned}
\label{eq:GKPStandardBasis}
\end{equation}

while the $X$-basis states as:

\begin{equation}
\begin{aligned}
\ket{+_{\text{GKP}}} &= \sum_{n \in \mathbb{Z}} \ket{p = 2n\sqrt{\pi}}, \\
\ket{-_{\text{GKP}}} &= \sum_{n \in \mathbb{Z}} \ket{p = (2n+1)\sqrt{\pi}}.
\end{aligned}
\label{eq:GKPXBasis}
\end{equation}

We note that such ideal GKP qubits require infinite amount of squeezing and therefore are unphysical. We therefore consider finitely squeezed GKP qubits which we describe mathematically in Section~\ref{sec:ResourceState}.

The two stabilizers in Eq.~\eqref{eq:GKPstabilizers} allow us to measure displacements in $q$ and $p$ quadratures modulo $\sqrt{\pi}$. Hence the GKP code can correct small displacements whose component along each of the two quadratures has a magnitude smaller than $\sqrt{\pi}/2$, i.e. half of the resolution of the stabilizers.

Let us now briefly describe on a high level how a GKP stabilizer measurement can be performed. We note that there are different methods for measuring the GKP stabilizers and performing GKP error correction such as Steane Error Correction (Steane EC) or Teleportation-based Error Correction (TEC). However, all of them require ancilla GKP qubits with GKP peak spacing of $\sqrt{\pi}$ in the measured quadrature. A GKP stabilizer measurement should only reveal the stabilizer value without revealing the encoded information. This can be realised by making the data qubit interact with the ancilla in such a way that in the measured quadrature the data and ancilla information become combined. For example, for the measurement of $\hat{S}_p$ we apply an interaction that provides us with $\hat{p}_{\text{data}} + \hat{p}_{\text{ancilla}}$ on the ancilla qubit to be measured. Then we measure the ancilla GKP qubit in $p$ quadrature and obtain outcome $p_0$. We note that since the ancilla had a spacing of GKP peaks of $\sqrt{\pi}$, it has masked the encoded information so that the outcome $p_0$ fundamentally contains no information whether $\hat{p}_{\text{data}}$ was centered at odd or even multiples of $\sqrt{\pi}$. The outcome $p_0$ can only tell us about any shifts on $\hat{p}_{\text{data}}$ away from the GKP code space consisting of multiples of $\sqrt{\pi}$. Hence let us define a function
\begin{equation}
R_{s}(x) = x - s \floor*{\frac{x}{s} + \frac{1}{2}},
\end{equation}
so that $R_{\sqrt{\pi}}(.)$ corresponds to the classical processing that maps the input value modulo $\sqrt{\pi}$ to the interval $[-\sqrt{\pi}/2, \sqrt{\pi}/2)$. Therefore it is clear that $R_{\sqrt{\pi}}(p_0)$ provides us exactly with the measured stabilizer value of $\hat{S}_p$ on $\hat{p}_{\text{data}}$. If $\hat{p}_{\text{data}}$ had an error shift smaller than $\sqrt{\pi}/2$ then $R_{\sqrt{\pi}}(p_0)$ would correspond exactly to that shift and hence the error could be corrected (the specific strategy of how the GKP syndrome information should be used for correcting the error depends on the applied GKP error correction technique). If the error shift had a magnitude from the interval $[\sqrt{\pi}/2, 3\sqrt{\pi}/2)$, there would be a $\sqrt{\pi}$ discrepancy between the error shift and $R_{\sqrt{\pi}}(p_0)$ leading to the logical error. We note that a net shift by an even multiple of $\sqrt{\pi}$ along the $p$ quadrature corresponds exactly to the action of the $\hat{S}_p$ stabilizer under which our state is invariant. Hence the logical error is caused by any shift belonging to any of the intervals $[(4m+1)\sqrt{\pi}/2, (4m+3)\sqrt{\pi}/2)$ for $m \in \mathbb{Z}$, i.e. any interval centered at an odd multiple of $\sqrt{\pi}$.

Let us now also briefly describe the TEC, which is the error correction strategy used in our repeater scheme. This strategy effectively allows us to perform a noise-resilient BSM between the data GKP qubit and an ancillary GKP Bell pair. Specifically, the two classical outcomes of the BSM are obtained by measuring two GKP qubits, one in each quadrature. Then these continuous outcomes are discretized into bits by thresholding them as described in Appendix~\ref{sec:GKPBSMs}. If the error shift on the input data qubit does not cross this threshold boundary the BSM outcome is reliable and effectively the continuous error becomes corrected. This is because it has no effect on the logical BSM outcome and hence no effect on the GKP qubit onto which the data become teleported. We note that, by construction, a logical BSM does not reveal any information about the logical state of the data qubit. Additionally to TEC, in our scheme we also employ EC through a BSM between two data GKP qubits rather than a single data qubit and an ancilla Bell pair. More details about these schemes as well as the analysis of noise affecting these processes is provided in Appendix~\ref{sec:NoisyGKPMeasurements}.

\section{Repeater scheme}
\label{sec:RepeaterScheme}

The proposed repeater scheme is illustrated in FIG.~\ref{fig:schematic}. 
The repeater nodes are assumed to be equipped with GKP qubit factories that generate approximate GKP qubits at high repetition rates. 
These ``physical'' GKP qubits are then suitably combined to form $[[7,1,3]]$ Steane logical qubits comprised of 7 physical GKP qubits each. 
These Steane-GKP qubits are subsequently combined further with physical GKP qubits to form physical GKP - Steane-GKP Bell states that serve as the resource states for the repeater scheme. 
For a multiplexed repeater scheme of multiplexing level $k$, each repeater node generates $2k$ such resource Bell states per entanglement generation cycle i.e. $k$ for the channel segment to the left of the node and $k$ for the channel segment to the right. 
The ``half'' of each resource Bell state comprised of the bare physical GKP qubit, which we refer to as the ``outer leaf'' qubit, is then transmitted across the channel segment to optically interface with the adjacent repeater node. 
Outer leaf qubits from a pair of adjacent repeater nodes are interfered with each other at the middle of the channel segment connecting the nodes using a GKP BSM, establishing an ``elementary link''. The physical outcomes of the BSM measurements provide not only the GKP logical BSM outcomes but also analog information $R_{\sqrt{\pi}}(p_0)$ and $R_{\sqrt{\pi}}(q_0)$ about the reliability of the corresponding logical outcomes. This reliability information from all the elementary multiplexed links over that segment are compared, and the links are ranked according to that reliability. 

While the outer leaf qubits are still in transit towards their neighboring repeater nodes in the process of elementary link generation, the ``other half'' of each resource Bell state, namely, the [[7,1,3]]-GKP qubits that we refer to as the ``inner leaf'' qubits, are locally retained at the repeater nodes all optically using fiber spools. 
Until the information about the outcomes of the outer leaf qubit BSMs and the link ranking arrive at the respective repeater nodes, the corresponding inner leaves at the repeater nodes are periodically GKP-error corrected in the fiber spools using TEC. When the outer leaf BSM outcomes and the ranking information arrive, the outer links to the left and to the right of the repeater are matched based on their respective ranking. 
This is followed by inner leaf logical BSMs between each pair of matched links across the repeater node. 

We note that our scheme inherits the benefit of all all-photonic schemes that the protocol repetition rate is only limited by the rate of generating the resources. The fact that no quantum memories are required at the repeater nodes effectively means that we can store as many states in our simulated fiber-based memories as needed. In our specific case the protocol repetition rate will be to a large extent dictated by the rate at which we can prepare GKP qubits.  

While GKP qubits have been demonstrated to perform well against photon loss~\cite{albert2018performance, noh2018quantum}, in our scheme two different strategies of using GKP qubits against photon loss have been considered. The first one is based on pre-amplifying the GKP state with the gain given by the inverse of the transmissivity of the lossy channel, before sending it through the transmission channel. In this way the effective channel of phase-insensitive pre-amplification followed by the lossy channel with transmissivity $\eta$ is equivalent to a Gaussian random displacement channel with standard deviation $\sigma = \sqrt{1-\eta}$~\cite{kim1996quantum,sabapathy2011robustness,ivan2011operator,noh2018quantum}. The Gaussian random displacement channel is given by:
\begin{equation}
\mathcal{N}_{\text{disp}}[\sigma](\rho) = \frac{1}{\pi \sigma^2}\int d^2\alpha \exp\left[-\frac{\abs{\alpha}^2}{\sigma^2}\right] \hat{D}(\alpha)\rho \hat{D}^\dag(\alpha),
\label{eq:randomgaussiandisplChannel}
\end{equation}
where $\hat{D}(\alpha)$ is a displacement operator by $\alpha$. We use this strategy on all the physical inner leaf GKP qubits while storing them in the local fiber, i.e., we pre-amplify them before every fiber segment after which we perform GKP TEC on them.

\begin{figure*}
\includegraphics[width=1\textwidth]{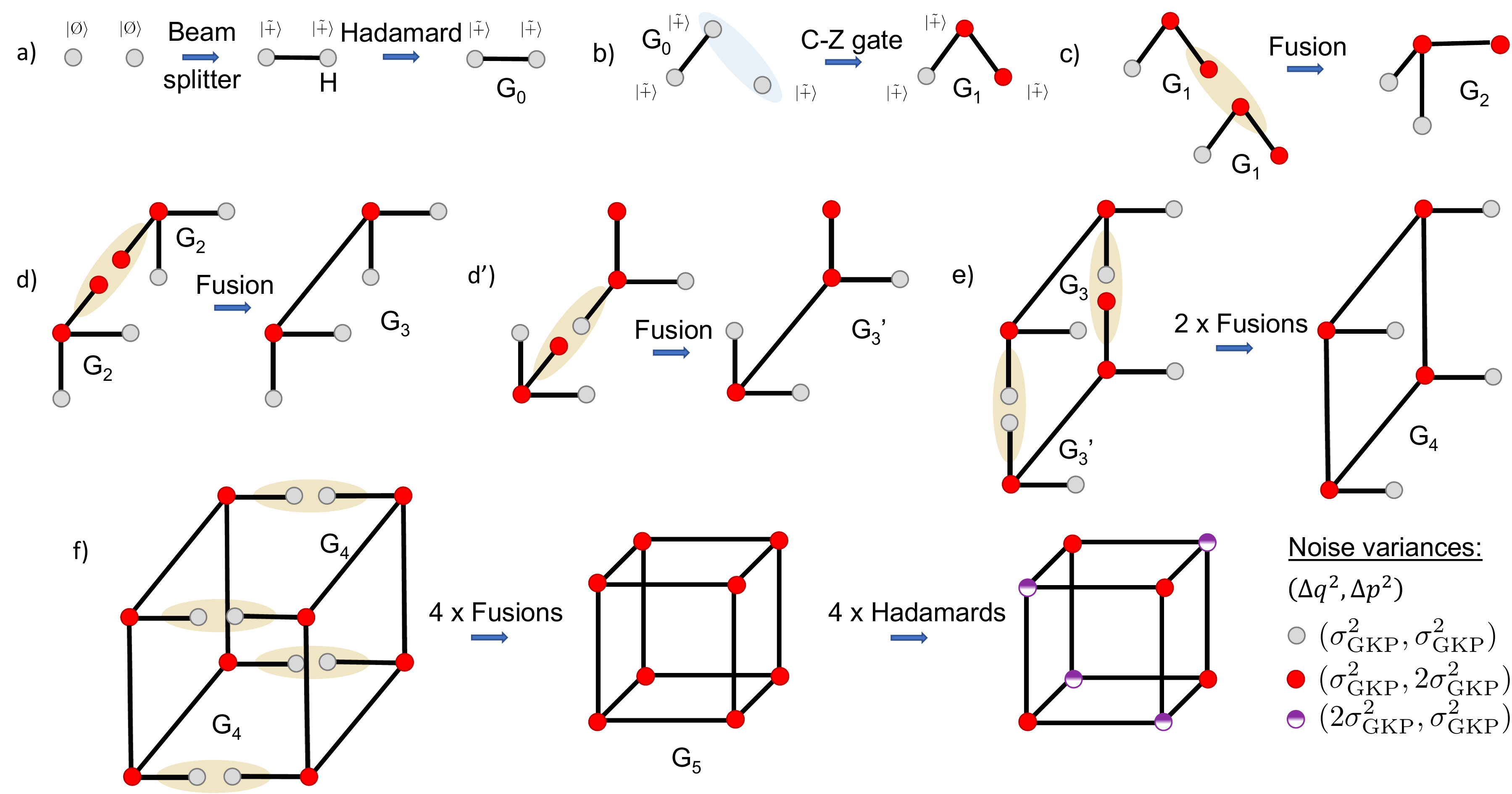}
\caption{A modular procedure for resource state generation at the repeaters. The procedure acts on finitely squeezed GKP qubits prepared in the qunaught $|\oslash\rangle$ and $|+\rangle$ states. In step a), two $|\oslash\rangle$ states are combined on a 50:50 beam splitter followed by a Hadamard gate operation on one of the two qubits via local oscillator phase delay of $\pi/2$, which results in a two-qubit GKP graph state $G_0$. In step b), a $C_Z$ gate is applied between one of the two qubits in the graph state $G_0$ and a fresh GKP qubit in the $|+\rangle$ to produce a 3-tree graph state $G_1$. In step c) a pair of $G_1$ states are stitched together via a fusion operation to generate a 4-tree graph state $G_2$. In steps d-f), $G_2$ graph states are further stitched together by successive fusion operations to form increasingly larger graph states culminating in the generation of the cube graph state $G_5$. After that applying Hadamard gates on 4 out of the 8 constituent qubits results in our desired resource state.}
\label{fig:resource_prep}
\end{figure*}

The second strategy can be applied in the specific scenario when two GKP modes that have both passed through the same amount of loss are interfered together for error correction and entanglement swapping. In this case no amplification is necessary as the outcomes of the homodyne measurements performed after the interference can simply be rescaled by the inverse of the square root of the lossy channel transmissivity. This method has been proposed in~\cite{fukui2021all} and referred to as CC-amplification, where CC stands for ``classical computer'', signifying that no actual physical amplification needs to be performed on the quantum states themselves. In this case if the total channel separating the two GKP qubits has transmissivity $\eta$, i.e. the two qubits are sent towards each other each over a channel with transmissivity $\sqrt{\eta}$, then the effective channel of photon loss followed by interference and measurements with outcome rescaling can be described by a Gaussian random displacement channel with standard deviation $\sigma = \sqrt{\frac{1-\sqrt{\eta}}{\sqrt{\eta}}}$~\cite{fukui2021all}. We use this strategy on outer leaf GKP qubits when generating the elementary links as well as on the last fiber segment for the inner leaf GKP qubits when there is no subsequent TEC. This is because the final GKP correction for the inner leaf qubits is performed together with the [[7,1,3]] correction through the concatenated-coded BSM between the matched links.

We note that for simplicity of our model we assume that the classical information about the outer leaf BSMs is transmitted back to the repeaters through the same type of fiber as the one for transmission of outer leaves. This means that the length of the local fiber acting as quantum memory for the inner leaf GKP qubits is the same as inter-repeater spacing $L$ ($L/2$ is the ``memory distance'' until the two outer leaves have met for BSM and then another $L/2$ ``memory distance'' is needed while waiting for the BSM outcomes and ranking information to come back). For direct transmission of quantum states through fiber of length $l$ we model the corresponding channel transmissivity as $\eta=e^{-l/L_0}$, where the channel attenuation length is $L_0 = 22$ km. Moreover, for simplicity we assume that fiber loss during all the other local processes such as the resource state creation at the repeaters is negligible. We also assume that the GKP qubits are generated directly in the fiber and all the required linear optics operations are performed fiber-based.

\section{Resource state generation}
\label{sec:ResourceState}
The resource state, which is a Bell state between a bare physical GKP qubit and a [[7,1,3]] logical qubit comprised of 7 physical GKP qubits, is equivalent to a graph state of cube topology up to Hadamard gates on 4 out of the 8 qubits. 
Here, we briefly describe the generation of this resource state (see Appendix \ref{sec:resource_generation} for full details).

\subsection{Finite GKP squeezing}
First of all, we consider finitely squeezed GKP qubits. 
We model the imperfections due to finite squeezing as ideal GKP states undergoing additional Gaussian random displacement channel with standard deviation $\sigma_{\text{GKP}}$ where the amount of squeezing in dB $s$ can be linked to $\sigma_{\text{GKP}}$ as:
\begin{equation}
    s = -10 \log_{10}(2\sigma_{\text{GKP}}^2).
\end{equation}
This way of modelling the effect of finite GKP squeezing is widely used, and can be justified as follows~\cite{noh2020fault}. 
A finitely squeezed GKP state is given by $\ket{\psi^{\Delta}_{\text{GKP}}} \propto \exp(-\Delta^2 \hat{n})\ket{\psi_{\text{GKP}}}$, where $\ket{\psi_{\text{GKP}}}$ is the corresponding ideal GKP state and $\Delta$ describes the width of each peak in the approximate GKP grid. 
Randomly applying displacements along $q$ and $p$ to $\ket{\psi^{\Delta}_{\text{GKP}}}$ by any even integer multiple of $\sqrt{\pi}$ according to a uniform distribution, known as twirling, produces a state:
\begin{equation}
\rho_{\text{GKP}}[\sigma_{\text{GKP}}] = \mathcal{N}_{\text{disp}}[\sigma_{\text{GKP}}](\dyad{\psi_{\text{GKP}}})\, ,
\end{equation}
where $\sigma_{\text{GKP}}^2 = (1-e^{-\Delta})/(1+e^{-\Delta})$~\footnote{Describing finite squeezed GKP qubits and graph states thereof without the twirling operation is possible, but cumbersome; c.f.,~\cite{SDPJG22}}. In other words, adding more noise to it produces a state that is akin to an ideal GKP state being passed through a Gaussian random displacement channel with standard deviation $\sigma_{\text{GKP}}$. 

\subsection{Procedure}
\label{sec:ResourceStateGeneration}
We consider a modular procedure for the resource state generation, which is similar to the one used in~\cite{fukui2018high, fukui2021all}. 
The procedure is as illustrated in FIG.~\ref{fig:resource_prep}. Firstly, two-qubit GKP Bell states are deterministically generated by mixing pairs of finitely squeezed GKP qubits of displacement noise variance (also referred to as GKP squeezing variance) $\sigma^2_{\text{GKP}}$ prepared in the qunaught state on 50:50 beam splitters, where the ideal qunaught state is defined as
\begin{equation}
\begin{aligned}
|\oslash\rangle &\propto\sum_{n=-\infty}^{\infty}\exp(-i\sqrt{2\pi}n\hat{p})|0\rangle_q\nonumber\\
&=\sum_{n=-\infty}^{\infty}\exp(i\sqrt{2\pi}n\hat{q})|0\rangle_p.
\end{aligned}
\label{eq:qunaughtstate}
\end{equation}
Note that $|\oslash\rangle$ has a periodicity of $\sqrt{2\pi}$ in both $q$ and $p$ quadratures. 
The two-qubit GKP Bell states produced have the same GKP squeezing variance as the input qubits, which are denoted as gray (lighter shade fill) qubits in FIG.~\ref{fig:resource_prep}. 
Hadamard gates that transform the quadratures as $q\rightarrow p$ and $p\rightarrow -q$ acting on one of the two qubits of the GKP Bell states transform them into two-qubit graph states denoted as $G_0$. 
A Hadamard gate can be implemented by introducing a phase delay of $\pi/2$ to the local oscillator. 

This is followed by the generation of tree graph states of size 3 denoted as $G_1$. 
A $G_1$ state is obtained by the action of a $C_Z$ gate between one of the two qubits in a $G_0$ state and a third GKP qubit prepared in the $|+\rangle$ state, also of GKP squeezing variance $\sigma^2_{\text{GKP}}$, where a $C_Z$ gate can be deterministically implemented using either inline or offline squeezers and linear optics. 
The $C_Z$ gate that transforms the quadratures of the two participating qubits as $q_1\rightarrow q_1$, $p_1\rightarrow p_1-q_2$, $q_2\rightarrow q_2$, and $p_2\rightarrow p_2-q_1$, results in their GKP squeezing variances to change from $(\sigma_{\text{GKP}}^2, \sigma_{\text{GKP}}^2)$ along $q$ and $p$ quadratures, respectively, to $(\sigma_{\text{GKP}}^2, 2\sigma_{\text{GKP}}^2)$, which are denoted as red (darker shade fill) qubits in FIG.~\ref{fig:resource_prep}. Subsequently, pairs of $G_1$ states are fused through the node qubit of one and the noisier (red/darker shade fill) of the two leaf qubits of the other to generate 4-tree graph states denoted as $G_2$, where a fusion operation involves a Hadamard gate, a beam splitter and homodyne detection and is deterministic as described in Appendix~\ref{sec:GKPBSMs}. 
Note that we model the homodyne detectors used in the fusion operations as realistic, lossy detectors with finite detection efficiencies $0<\eta_d<1$. 

Pairs of $G_2$ states are then inter-fused to generate 6-qubit graph states, where $G_3$ and $G_3'$ denote two different configurations with regard to the squeezing variances of the GKP qubits participating in the fusion operation, namely, one involving the noisier (red/dark shade fill) of the three leaf qubits on both $G_2$ states being fused ($G_3$), and other involving the noisier qubit only on one of the $G_2$ states being fused ($G_3'$). 
Likewise, pairs of 6-qubit graph states consisting of a $G_3$ and a $G_3'$ state are in turn inter-fused (via 2 fusion operations) to generate 8-qubit graph states $G_4$. 
Finally, pairs of $G_4$ states are inter-fused (via 4 fusion operations) to generate the cube graph states $G_5$. 
Further, 4 of the 8 GKP qubits (denoted as purple/half filled) in the cube graph state $G_5$ are acted on by Hadamard gates that flip their $q$ and $p$ quadratures, resulting in the desired resource state.

It should be noted that the cube graph state could in principle be generated all the way by repeated applications of $C_Z$ gates starting from individual GKP qubits prepared in the $|+\rangle$ state. 
However, in the face of finite GKP squeezing, such an approach would evidently result in a highly noisy resource state in terms of the GKP squeezing variances of the constituent qubits, which would in turn result in an increased likelihood of logical errors when the qubits are measured. 
The modular approach presented above avoids this build up of noise on the resource state qubits by making minimal use of $C_Z$ gates, and instead relying on fusion operations. One might also ask why we need to generate different states $G_3$ and $G'_3$ rather than only generating states $G_3$ and fusing two of them later into states $G_4$. This is because of the certain correlated two-qubit logical errors arising during the fusions as discussed below and in Appendix~\ref{sec:resource_generation}. The proposed way of performing fusions enables us to avoid such correlated errors between the outer leaf qubit and any of the inner leaf qubits.  

Unfortunately, the fusion operations also suffer due to finite GKP squeezing, albeit differently. 
The problem is the possibility of logical errors emanating from the homodyne measurements carried out as part of the fusion operations that can propagate onto the resource state qubits. The rule for this error propagation during such fusions is derived in Appendix~\ref{sec:ErrorPropagationFusion}.
For the different configurations of the participating qubits, namely, two red/darker shade fill qubits (fusion 1), or one gray/lighter shade fill qubit and one red/darker shade fill qubit (fusion 2), or two gray/lighter shade fill qubits (fusion 3), the two homodyne measurements involved in the fusion operation deal with noise variance $3\sigma_{\textrm{GKP}}^2 + (1-\eta_d)/\eta_d$ each, or one each of variances $2\sigma_{\textrm{GKP}}^2 + (1-\eta_d)/\eta_d$ and $3\sigma_{\textrm{GKP}}^2 + (1-\eta_d)/\eta_d$, or $2\sigma_{\textrm{GKP}}^2 + (1-\eta_d)/\eta_d$ each, respectively. The greater the noise variance of the homodyne measurements involved, the greater the probability of logical errors. To control these error probabilities, we use post-selection at the expense of the determinism of the fusion operation. 
The fusion operations under post-selection thus only succeed probabilistically, but the probability of success can be boosted by multiplexing the fusion attempts. 
The use of post-selection and multiplexing in the resource state generation are discussed next. A detailed analysis of error propagation after the discussed fusion operations is provided in Appendix~\ref{sec:resource_generation}.

\subsection{Post-selected fusions}

As mentioned above, during the fusions leading to the generation of the resource state we apply post-selection to increase the quality of the generated resource conditioned on passing the post-selection.

Here we describe in more detail how such post-selection works. Let us define a parameter $v$ describing the size of the discard window. This parameter defines a window such that if the GKP syndrome $R_{\sqrt{\pi}}(q_0)$ (or $R_{\sqrt{\pi}}(p_0)$) is in the interval $-\frac{\sqrt{\pi}}{2} \le R_{\sqrt{\pi}}(q_0) \le -\frac{\sqrt{\pi}}{2} + v$ or $\frac{\sqrt{\pi}}{2} - v \le R_{\sqrt{\pi}}(q_0) \le \frac{\sqrt{\pi}}{2}$, the measurement is deemed unreliable and the state is discarded. The state is only kept if the syndrome satisfies $-\frac{\sqrt{\pi}}{2} + v< R_{\sqrt{\pi}}(q_0) < \frac{\sqrt{\pi}}{2} - v$. This means that the probability of passing the post-selection and having a logical error is then given by
    \begin{equation}
        E_1(v;\sigma^2)=\sum_{m \in \mathbb{Z}} \int_{\frac{4m+1}{2}\sqrt{\pi} + v}^{\frac{4m+3}{2}\sqrt{\pi}-v}f_{\sigma}(x)dx \, ,
    \end{equation}
    where
    \begin{equation}
        f_{\sigma}(x) = \frac{e^{-\frac{x^2}{2 \sigma^2}}}{\sqrt{2\pi \sigma^2}}
        \label{eq:GaussianDist}
    \end{equation}
    is the Gaussian noise distribution. The probability of passing the post-selection without a logical error is then:
    \begin{equation}
        E_0(v;\sigma^2)=\sum_{m \in \mathbb{Z}} \int_{\frac{4m-1}{2}\sqrt{\pi} + v}^{\frac{4m+1}{2}\sqrt{\pi}-v}f_{\sigma}(x)dx \, .
    \end{equation}
    We note that:
    \begin{equation}
    \begin{aligned}
       &\int_{-\frac{3}{2}\sqrt{\pi} + v}^{-\frac{1}{2}\sqrt{\pi}-v}f_{\sigma}(x)dx + \int_{\frac{1}{2}\sqrt{\pi} + v}^{\frac{3}{2}\sqrt{\pi}-v}f_{\sigma}(x)dx \\ &=  2\int_{\frac{1}{2}\sqrt{\pi} + v}^{\frac{3}{2}\sqrt{\pi}-v}f_{\sigma}(x)dx
       \end{aligned}
    \end{equation}
    acts as a lower bound on $E_1(v;\sigma^2)$ and
    \begin{equation}
    \begin{aligned}
       &\int_{-\infty}^{-\frac{1}{2}\sqrt{\pi}-v}f_{\sigma}(x)dx + \int_{\frac{1}{2}\sqrt{\pi} + v}^{\infty}f_{\sigma}(x)dx \\& =  2\int_{\frac{1}{2}\sqrt{\pi} + v}^{\infty}f_{\sigma}(x)dx
    \end{aligned}
    \end{equation}
    as an upper bound i.e.
    \begin{equation}
        2\int_{\frac{1}{2}\sqrt{\pi} + v}^{\frac{3}{2}\sqrt{\pi}-v}f_{\sigma}(x)dx \leq E_1(v;\sigma^2) \leq 2\int_{\frac{1}{2}\sqrt{\pi} + v}^{\infty}f_{\sigma}(x)dx \, .
    \end{equation}
    Similarly, we also obtain the following bounds on $E_0(v;\sigma^2)$:
    \begin{equation}
    \begin{aligned}
        &\int_{-\frac{1}{2}\sqrt{\pi} + v}^{\frac{1}{2}\sqrt{\pi}-v}f_{\sigma}(x)dx \leq E_0(v;\sigma^2) \\ &\leq \int_{-\frac{1}{2}\sqrt{\pi} + v}^{\frac{1}{2}\sqrt{\pi}-v}f_{\sigma}(x) dx +  2\int_{\frac{3}{2}\sqrt{\pi} + v}^{\infty}f_{\sigma}(x)dx \, .
    \end{aligned}
    \end{equation}
    We find that for the values of $\sigma$ and $v$ used in this paper these bounds are numerically tight for our purposes. Specifically their relative difference given by the upper bound minus lower bound divided by the upper bound is at most of the order of $10^{-4}$ for $E_1(v;\sigma^2)$ and at most of the order of $10^{-14}$ for $E_0(v;\sigma^2)$. Hence in our simulation and resource analysis we use the lower bounds, i.e. we define:
    \begin{equation}
        \begin{aligned}
        \tilde{E}_1(v;\sigma^2) &= 2\int_{\frac{1}{2}\sqrt{\pi} + v}^{\frac{3}{2}\sqrt{\pi}-v}f_{\sigma}(x)dx \, , \\
        \tilde{E}_0(v;\sigma^2) &= \int_{-\frac{1}{2}\sqrt{\pi} + v}^{\frac{1}{2}\sqrt{\pi}-v}f_{\sigma}(x)dx \, .
        \end{aligned}
    \end{equation}
    Then the probability of passing post-selection and the probability of having a logical error given that the post-selection has been passed are given by
    \begin{equation}
    \begin{aligned}
        P_\textrm{PS}(v;\sigma^2)&=\tilde{E}_1(v;\sigma^2) + \tilde{E}_0(v;\sigma^2),\\
        E_\textrm{PS}(v;\sigma^2)&=\tilde{E}_1(v;\sigma^2)/(\tilde{E}_1(v;\sigma^2) + \tilde{E}_0(v;\sigma^2)).
    \end{aligned}
    \label{eq:PostselectionProbAndPostselectedError}
    \end{equation}

We note that as discussed in Section~\ref{sec:ResourceStateGeneration}, our post-selected fusions involve measurements of GKP qubits with two different variances. Some of them have variance $\sigma^2 = 3\sigma_{\text{GKP}}^2 + (1-\eta_d)/\eta_d$ and some have variance $\sigma^2 = 2\sigma_{\text{GKP}}^2 + (1-\eta_d)/\eta_d$. For practical purposes it suffices to suppress errors at different measurements to roughly the same order of magnitude. Clearly if we used the same value of $v$ for measurements of the qubits with both of these variances, the errors for the ones with the latter variance will be suppressed to much smaller order of magnitude than the errors after the measurements of the first variance. This will not provide any observable benefit but would cost us large number of additional fusion attempts and hence large number of additional resources. Therefore based on our numerical investigation for the relevant hardware parameter regime we have established a high level heuristic that whenever we use a discard window $v$ for the measurement of a GKP qubit with the first noise variance, for the corresponding measurement of a GKP qubit with the latter noise variance we use $0.7v$. Hence from now on whenever we refer to discard window $v$ we always refer to the value of the discard window we use for the measurements of the GKP qubits with the first variance. For the corresponding latter one we always use 0.7 of that value. 
For a GKP squeezing variance $\sigma_{\textrm{GKP}}^2$ and a discard window size $v$, the relevant probabilities are tabulated in TABLE~\ref{tab:fusion_psucc}.
\begin{table}
    \centering
    \begin{tabular}{c|c}
        Fusion Type & Success Probability $p$ \\
        \hline
        Fusion 1 & $\left(P_{\textrm{PS}}(v;3\sigma_{\text{GKP}}^2 + (1-\eta_d)/\eta_d)\right)^2$\\
        \hline
        \multirow{2}{3.75em}{Fusion 2} & $P_{\textrm{PS}}(v;3\sigma_{\textrm{GKP}}^2 + (1-\eta_d)/\eta_d)$\\
        &$\times P_{\textrm{PS}}(0.7v;2\sigma_{\text{GKP}}^2 + (1-\eta_d)/\eta_d)$\\
        \hline
        Fusion 3 & $\left(P_{\textrm{PS}}(0.7v;2\sigma_{\text{GKP}}^2 + (1-\eta_d)/\eta_d)\right)^2$
    \end{tabular}
    \caption{Success probabilities of fusion operations for different GKP squeezing variances of the participating qubits. Fusions 1, 2 and 3 correspond to the cases of the input qubits being two red/darker shade fill qubits, one gray/lighter shade fill qubit and one red/darker shade fill qubit, and two gray/lighter shade fill qubits in reference to FIG.~\ref{fig:resource_prep}, respectively. In our model, the numerical values of these success probabilities will range from around 0.3 to more than 0.999 depending on the chosen values of $\sigma_{\text{GKP}}$, $\eta_d$, and $v$.}
    \label{tab:fusion_psucc}
\end{table}

\begin{figure*}
\includegraphics[width=1\textwidth]{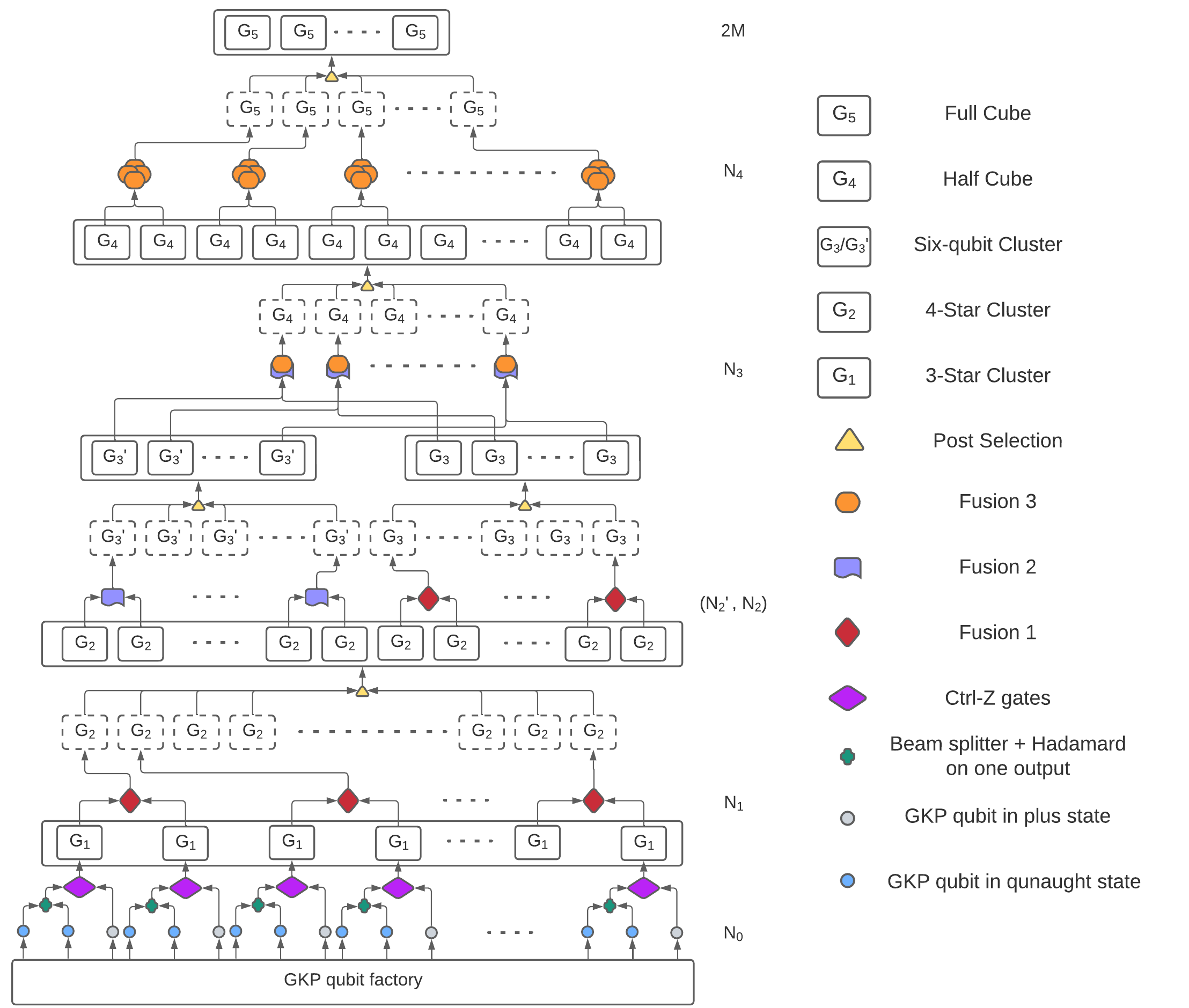}
\caption{A flowchart showing multiplexed fusions for near-deterministic resource state generation as per the procedure described in FIG.~\ref{fig:resource_prep}.}
\label{fig:flowchart}
\end{figure*}

\subsection{Multiplexed fusions}
In spite of the probabilistic, post-selected fusion operations involved in the modular approach, here we show that the repeaters can still generate the resource states near deterministically with the help of multiplexed fusion attempts at each step along the procedure outlined in Section~\ref{sec:ResourceStateGeneration}. 
The idea is to start with a sufficiently large number of GKP qubits at each repeater so as to be able to support sufficiently large numbers of fusion attempts at each successive layer of the flowchart in FIG.~\ref{fig:flowchart} from bottom to top. 
The end goal is to make the probability of simultaneously succeeding in generating the required $2k$ number of resource states at each repeater ($k$ being the repeater multiplexing level) along a chain of repeaters between two end nodes in a repeater-enhanced quantum network approach 1 during each time step of the repetition cycle of the repeater protocol.

For a given number of repeaters $n_{\text{rep}}$ connecting two end nodes and for $2n_1$ number of $G_1$ three-trees per repeater, the probability of successfully generating the necessary resource states at all repeaters during each time step $p^{\textrm{res-gen}}_\textrm{succ}$ can be expressed in terms of different number of attempts at the various kinds of fusions along the layers of the flowchart $N_i \, \, \forall i\in\{1,2,3,4\}$ and $N'_2$ as
\begin{equation}
\begin{aligned}
&\left(p^{\textrm{res-gen}}_\textrm{succ}\right)^{1/n_{\text{rep}}}\\
&=\sum_{n_2,n_3,n'_3,n_4}P(S_5\geq 2k|N_4=n_4) \\
& \times P(S_4= 2n_4|N_3=\min(n_3,n'_3))\\
&\times P\left(S'_3= n'_3|N'_2=\frac{p_{G_3}}{p_{G_3} + p_{G'_3}}n_2\right) \\
&\times P\left(S_3= n_3|N_2=\frac{p_{G'_3}}{p_{G_3} + p_{G'_3}}n_2\right)\\
&\times P(S_2= 2n_2|N_1=n_1) \, , 
\end{aligned}
\label{multi-res-gen-psucc}
\end{equation}
where, $S_i \, \, \forall i\in\{2,3,4,5\}$ denote the number of successes in the different fusions. Note that here we do not count successes in individual fusions but rather successes in all the fusions needed to simultaneously succeed in order to successfully fuse two states $G_i$ into a state $G_{i+1}$. Let us refer to all such fusions as fusion sets. Then the number of successful fusion sets $S_{i+1}$ corresponds to the number of successfully generated states $G_{i+1}$. Since at the next level we are fusing pairs of states $G_{i+1}$ into states $G_{i+2}$, the maximum number of attempts $N_{i+1}$ that we can perform is equal to half of the number of successes $S_{i+1}$. The exception to this are the pairwise fusions of the states $G_3$ with states $G_{3'}$ and so $N_3$ is given by the minimum of the number of available states $G_3$ and states $G_{3'}$.
The constituent probabilities $P(S|N \textrm{\ fusion \ attempts})$ can be evaluated as
\begin{align}
P(S_i|N_{i-1})=\binom{N_{i-1}}{S_i}p_{G_{i}}^{S_i}(1-p_{G_{i}})^{N_{i-1}-S_i},
\end{align}
where $p_{G_i}$ is the success probability of a single fusion set attempt at generating a state $G_i$. 
We note that when having in total $n_2$ attempts of generating the states $G_3$ and $G'_3$ we use a fraction $\frac{p_{G'_3}}{p_{G_3} + p_{G'_3}}$ of that $n_2$ for generating $G_3$ and fraction $\frac{p_{G_3}}{p_{G_3} + p_{G'_3}}$ for generating $G'_3$. This choice enables for generating on average an equal number of both states. The quantity $p^{\textrm{res-gen}}_\textrm{succ}$ can be made arbitrarily close to 1 by increasing the number of GKP qubits $N_0 = 6n_1$.

We note that the choice of fusion 1 instances early on in the procedure of FIG.~\ref{fig:resource_prep} is not accidental, but to encounter the more noisy fusion operations earlier in the workflow rather than later. This is because post-selection failures at higher layers amount to discarding larger number of GKP qubit resources, which we want to minimize. 
For the same reason, likewise, fusions 2 and 3 follow fusion 1s higher up in the resource generation procedure in that order.

Since analytical evaluation of $p^{\textrm{res-gen}}_{\textrm{succ}}$ is challenging, we perform a Monte-Carlo simulation to evaluate $(p^{\textrm{res-gen}}_\textrm{succ})^{1/n_{\text{rep}}}$ for different numbers of input GKP qubits $N_0$.

\section{Ranking of multiplexed outer leaves}
\label{sec:rankingOuter}

A key feature of our scheme which enables us to achieve such high performance is the new type of multiplexing that offers the possibility of ranking all the links based on their estimated quality. The quality estimate of each of the links can be obtained from the GKP analog information generated during the BSM performed on the two outer leaf qubits sent from the left and right. In other words the encoded GKP BSM is error-protected and the analog value of the outcome provides the likelihood that we wrongly decode the logical outcome. Based on these error-likelihoods, we can rank the links according to their estimated quality.

The specific strategy to rank the links is then as follows. Firstly for each of the $k$ multiplexed links and $k$ outer leaf fused pairs, we evaluate the likelihood of error in each quadrature separately. Specifically, we evaluate $p[\sigma](R_{\sqrt{\pi}}(p_0))$ (which we denote as $P_p$), the likelihood of error after the communication channel given the GKP stabilizer $\hat{S}_p$ outcome $R_{\sqrt{\pi}}(p_0)$. Here $\sigma$ is the standard deviation of the Gaussian random displacement channel acting on our measured GKP qubit, see Appendix~\ref{sec:errorsOuterLeaves} for more details on evaluating the error likelihood based on the GKP syndrome. 

Similarly we evaluate $p[\sigma](R_{\sqrt{\pi}}(q_0))$ (which we denote as $P_q$) for the $q$ quadrature. For a given link, the likelihood that there was no logical error on the outer leaf qubit during and after the BSM is:  

\begin{equation}
P_{\text{no-error}} = (1-P_p)(1-P_q) \, .
\label{eq:PnoErrorOuterLeaf}
\end{equation}

We then rank the links according to $P_{\text{no-error}}$. Then inside the repeaters we connect the links based on the ranking information. In Appendix~\ref{sec:strategiesConnectingLinks} we prove that in order to maximise the hashing rate (or equivalently a secret-key rate for the one-way six-state protocol) the optimal entanglement swapping strategy in the regime of small error probabilities is to connect the best link with the best, second best with the second best etc. in all the repeaters. We also refer the reader to Appendix~\ref{sec:errorsOuterLeaves} for a detailed analysis of all the error processes affecting the outer leaves.

\section{Inner leaves}
\label{sec:innerleaves}

While the outer leaves are being sent through the fiber for establishing the elementary links, the inner leaf logical qubits, encoded in a concatenation of the GKP and the [[7,1,3]] code are stored locally in the fiber. We note that each of those logical qubits needs to be stored in the fiber for a distance twice longer than the fiber length covered by each of the outer leaf qubits, because the inner leaf qubits need to also be stored for the time during which the classical information, in particular the analog ranking information, from the outer leaves comes back to the repeaters. That is also why we want to offer more protection to the inner leaf qubits by encoding them in a concatenated code. However, thanks to the boosting capability of the GKP analog information for the higher level code, using a small [[7,1,3]] code is sufficient as observed and used for one-way repeaters in~\cite{rozpkedek2021quantum}. Similarly to the outer leaves, the BSM/entanglement swapping on the inner leaves is performed on the encoded level and hence is error protected but this time on two levels. That means that effectively this BSM implements both GKP and [[7,1,3]] code correction but as it is destructive, it does not require any ancillas that could introduce extra noise. Hence this measurement of the GKP and [[7,1,3]] code stabilizers is only limited by the homodyne detection inefficiency. We note that as the [[7,1,3]] code admits transversal CNOT gates~\cite{steane2003overhead}, the encoded BSM can actually be done just using pairwise interference of individual GKP qubits from two [[7,1,3]]-code logical qubit blocks. This is because the BSM can be implemented using a CNOT gate followed by the X and Z measurements. The [[7,1,3]] code CNOT gate corresponds to 7 pairwise physical CNOT gates between the two blocks and the logical X and Z measurements can be implemented by measuring individual physical qubits as the logical X operator for this code is given by $X_{L} = XXXXXXX$ and the logical Z operator is given by $Z_{L} = ZZZZZZZ$.  Hence effectively we just need to perform 7 pairwise physical BSMs which for GKP qubits can be implemented using a beamplitter and two homodyne detections of the opposite quadratures as done on the outer leaves. 

\begin{figure*}
\includegraphics[width =\textwidth]{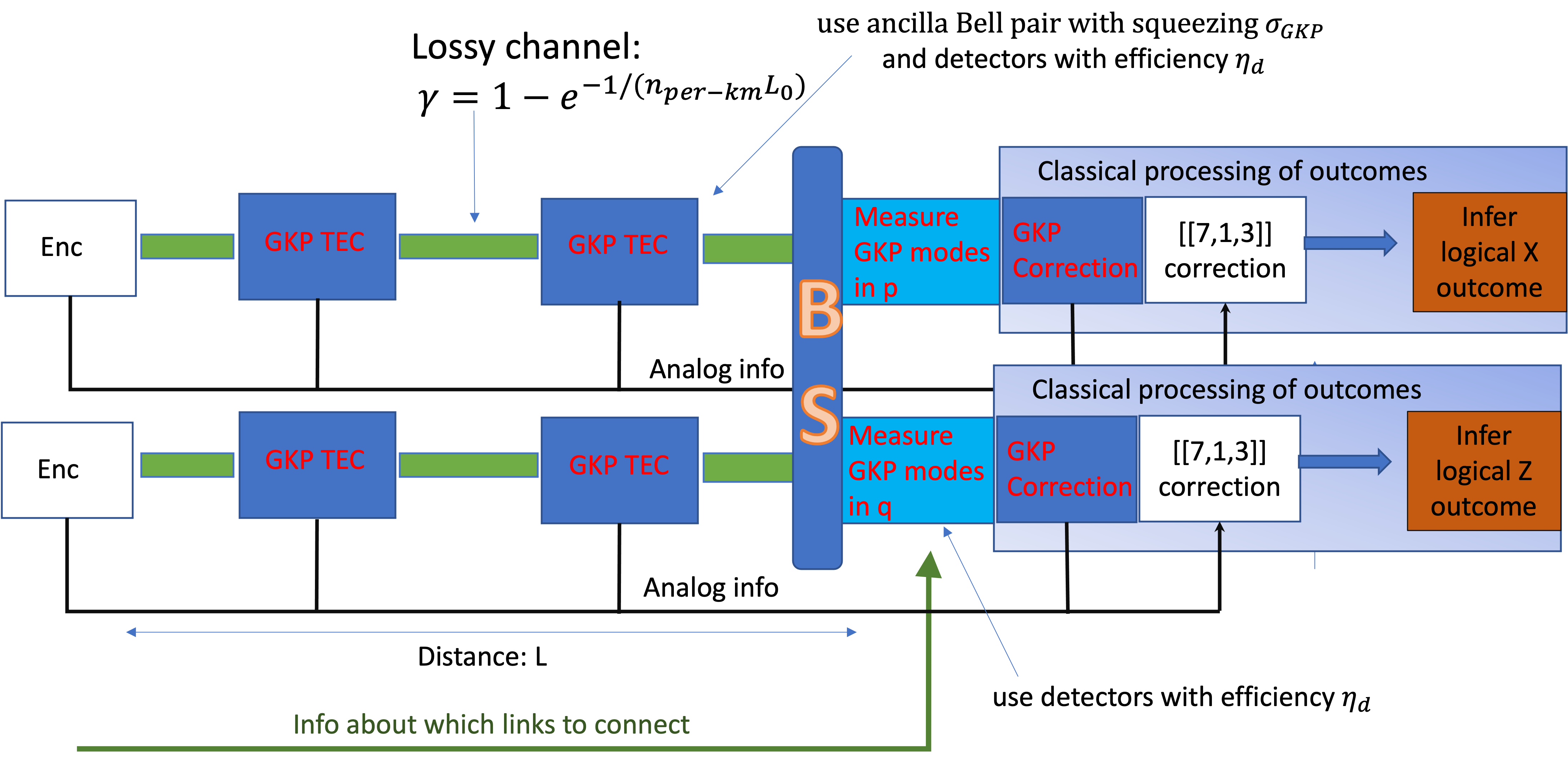}
\caption{Storage and error correction of the inner leaf qubits. The concatenated-coded qubits are stored in the fiber with periodic lower-level TEC GKP corrections $n_{\text{per-km}}$ times per km of fiber. Each of those GKP corrections generates analog information that is stored. When the ranking information arrives, a logical error-protected BSM is performed between the relevant inner leaf qubits. In the figure $\gamma$ denotes the photon loss probability in each segment between two consecutive GKP corrections.}
\label{fig:InnerLeavesSchematic}
\end{figure*}

Now, in order to better protect the inner leaf qubits during storage we additionally periodically perform TEC-based GKP correction on each of the 7 physical GKP qubits from each block. Hence effectively we perform multiple rounds of noisy GKP correction using noisy ancilla Bell pairs followed at the end by a close to perfect GKP correction and close to perfect [[7,1,3]] code stabilizer measurement. Such a simulated quantum memory is itself very similar to the one-way repeater of~\cite{rozpkedek2021quantum}. The main difference here is that all but the last one GKP correction are performed in the TEC-based fashion and the final GKP correction and the [[7,1,3]] code stabilizer measurement is close to perfect, i.e. without using noisy ancillas. Hence we can use the same strategy as in~\cite{rozpkedek2021quantum} of collecting the analog information from all the GKP corrections and then use it to reliably read off and correct the final logical BSM outcome. Specifically, after measuring the 14 GKP qubits, 7 in $\hat{p}$ quadrature, 7 in $\hat{q}$ quadrature we firstly perform GKP correction on the measured outcomes by bringing the measured values to the nearest multiple of $\sqrt{\pi}$. Then we decode them from the GKP code by reading them as 0 (1) for the even (odd) multiple of $\sqrt{\pi}$. Then we apply the parity check matrix of the [[7,1,3]] code to the resulting string of 7 bits. Each possible non-zero syndrome obtained from the 3 checks in $X$ and 3 checks in $Z$ corresponds to either a single-qubit error or one of three two-qubit errors (which are equivalent to each other up to the stabilizers). We can then decide which of these four cases is most likely based on the GKP analog information from all the GKP corrections during storage of the inner leaf qubits. After estimating the most-likely error the corresponding bits are flipped and the logical value of the outcome can be obtained by xor-ing all the 7 bits. This entire procedure is depicted in FIG.~\ref{fig:InnerLeavesSchematic}.

We note that as this last GKP and the [[7,1,3]] correction is done on the classical level, we are guaranteed that the final outcome must be in the code space i.e. after the correction the measured 7 bits will satisfy all the three parity checks. Similarly, while the GKP TEC uses noisy ancilla GKP modes and lossy homodyne detections, it can actually be modelled as perfect GKP correction with additional noisy channels before and after the correction, see~\cite{fukui2021all} and Appendix~\ref{sec:NoisyGKPMeasurements} for details. We also refer the reader to Appendix~\ref{sec:ErrChforInnerLeaves} for a detailed analysis of all the error processes affecting the inner leaf GKP qubits.

It is important to contrast the way we use inner leaves here with the way they are used in the clique-cluster-based GKP scheme of~\cite{fukui2021all}. That scheme utilises the inner leaves in the same way as the discrete-variable all-photonic scheme of~\cite{Azuma_15} where the inner leaves are also transmitted through the communication channel. In that case they can be measured immediately after the outer leaves and so they are effectively transmitted through a twice shorter lossy channel than in our case. However, transmitting the inner leaves requires transmission of many more modes per single raw bit or EPR pair. In our scheme we retain these inner leaves at the repeaters which is similar to the scheme of~\cite{pant2017rate} for the discrete-variable case. As mentioned, the total channel experienced by each qubit is much longer now but additionally to reducing the number of transmitted optical modes, retaining these qubits has two additional benefits that are specific to our scheme. Firstly it enables us to do GKP correction as frequently as we want on these qubits. This can compensate for the extra length of the lossy channel. Secondly, since GKP qubits admit deterministic two-qubit gates, in this case no clique-cluster is needed at all and the entanglement swapping can be done at the end rather than in the time-reversed way. Alleviating the need for the clique cluster significantly reduces the size of the required resource states.

An additional component that we use that can significantly increase the distance over which a non-zero performance is achieved is based on binning the end-to-end links based on the discrete inner leaf syndrome as proposed in~\cite{namiki2016role,jing2020quantum}. Specifically, as mentioned, when measuring the $X$ and $Z$ stabilizers of the [[7,1,3]] code for each of the two syndromes we can either obtain a zero-syndrome or a non-zero syndrome. If a zero-syndrome is measured, it is very unlikely that we make an error during the encoded entanglement swapping, because this undetected error would then need to be a logical error i.e. an error of weight three. On the other hand if we obtain a non-zero error syndrome it is much more likely that even using the GKP analog information we incorrectly identify between the single- and one of three two-qubit errors. Hence the end-to-end links for which more repeaters measured a zero-error syndrome during entanglement swapping will be on average of higher quality than the links for which more repeaters measured a non-zero error syndrome. In our scheme the end nodes extract secret key or entanglement separately from each bin corresponding to different number of non-zero error syndromes across the whole repeater chain. That is, the best links will be those for which all repeaters had a zero-error syndrome for both $X$ and $Z$ stabilizers. Then we will separately have a bin for which all repeaters had a zero-error syndrome in $X$ but in $Z$ there was one repeater that had a non-zero error syndrome. Then there will be a bin with that single non-zero error syndrome in $X$, then a bin with a single non-zero error syndrome in both and so on. This binning gets then combined with the ranking of the outer links. Hence for every bin there are actually $k$ separate links of different quality where $k$ is the number of multiplexing levels. Hence the total number of different quality links from which we extract secret key (entanglement) separately is $(n+1)^2 k$, where $n$ is the total number of repeaters.

An important requirement of our scheme which we additionally need to mention here is that the end nodes of Alice and Bob will have different requirements depending on whether the aim is to extract secret key or entanglement. For secret key, the end nodes do not need to have any inner-leaf qubits at all. In every protocol run, they just need to be able to prepare $k$ randomly chosen six-state protocol basis states within their outer leaves, see the discussion Section~\ref{sec:RequirementsEndNodes}. Hence for QKD the requirements on the resources of Alice and Bob are very limited and certainly much less than those on the required resources of the repeaters.

On the other hand if the goal is to establish long distance entanglement, then Alice and Bob require much more resources in terms of quantum memories. We will discuss this aspect more in Section~\ref{sec:RequirementsEndNodes} but here we just want to emphasize that in order for our scheme to work also for remote entanglement generation, we need to assume that Alice and Bob are in possession of quantum memories that can near-perfectly store the state for the duration of classical communication between the end-nodes.

\section{End-to-end error and performance metrics}
\label{sec:performanceMetrics}

Let us now put together all the pieces described above and provide the formulas for calculating the end-to-end performance.

We have already pointed out that the convexity of secret key and entanglement measures with respect to the shared quantum state means that we can increase the performance by connecting the outer links according to their ranking. Then we can extract the key (entanglement) from links of different rank separately. We have also explained how to use the additional information from the inner-leaf discrete syndromes to further exploit that convexity and increase the performance even further. Let us now make these ideas more precise mathematically.

When performing entanglement swapping on the inner leaves we use the [[7,1,3]] code syndrome to group the links. When measuring each of the $X$ or $Z$ stabilizers, there are two possible syndromes: the syndrome corresponding to no error which we will denote as $s=0$ and the one corresponding to an error which we will denote as $s=1$. We use that binary information as follows.

Firstly, the logical error probability over the single link consists of two pieces, the probability of the logical error on the inner leaves $Q_{X/Z, \text{inner}}(s)$ (where we separate between the $s=0$ and $s=1$ case) and the probability of the logical error on the outer leaves for the outer links ranked as $j \in \{1,...,k\}$ (in the order of their estimated quality), for $k$ multiplexed links $Q_{X/Z, \text{outer}}(j)$. Then the total error probability over the single link depending on $s$ and ranking $j$ is:

\begin{equation}
\begin{aligned}
 Q_{X/Z}(s,j) &=  Q_{X/Z, \text{inner}}(s)\left(1-Q_{X/Z, \text{outer}}(j)\right) \\
&+ \left(1- Q_{X/Z, \text{inner}}(s)\right)Q_{X/Z, \text{outer}}(j) \, . 
 \end{aligned}
\end{equation}

Let $n$ be the total number of inner leaf repeaters between Alice and Bob. Let $\vec{s} \in \mathbb{F}_2^n$ be the binary vector of dimension $2^n$ (bit string of length $n$) describing which of the $n$ repeaters had the inner leaf error syndrome during swapping. Then the end-to-end error probability for the end-to-end link with $j$'th ranking and when $m \in \{0,...,n\}$ of the inner leaves had a non-zero error syndrome is:
\begin{equation}
\begin{aligned}
 &Q_{X/Z,\text{end}}(\abs{\vec{s}_{X/Z}} = m_{X/Z},j) = \\  &\frac{1-(1-2Q_{X/Z}(s=1,j))^{m_{X/Z}} (1-2Q_{X/Z}(s=0,j))^{n-m_{X/Z}}}{2} \, .
 \end{aligned}
 \label{eq:elementaryLinkFlip}
\end{equation}
Here $\abs{\vec{s}_{X/Z}}$ denotes the Hamming weight of the bit string $\vec{s}_{X/Z}$. Let $p_{X/Z}(\abs{\vec{s}_{X/Z}} = m_{X/Z})$ be the probability that $m$ of the $n$ repeaters measured the non-zero error syndrome on the inner leaves. Then the total secret key (entanglement) rate per mode when distilled separately for different $j$'s and different $m$'s is:
\begin{equation}
\begin{aligned}
    R &= \frac{1}{k}\sum_{j, m_X,m_Z} p_{X}(\abs{\vec{s}_{X}} = m_{X})p_{Z}(\abs{\vec{s}_{Z}} = m_{Z})\\ &r\left(Q_{X,\text{end}}(\abs{\vec{s}_X} = m_X,j), Q_{Z,\text{end}}(\abs{\vec{s}_Z} = m_Z,j)\right)
    \end{aligned}
\end{equation}
where $r$ is the secret-key fraction or a lower bound on distillable entanglement.

Note that:
\begin{equation}
  p_{X/Z}(\abs{\vec{s}_{X/Z}} = m_{X/Z}) = \binom{n}{m_{X/Z}} t_{X/Z}^{m_{X/Z}}(1 - t_{X/Z})^{n-m_{X/Z}} \, ,
\end{equation}
where $t_{X/Z}$ is the probability of an error syndrome ($s=1$) on the inner leaves and together with $Q_{X/Z, \text{inner}}(s=\{0,1\})$ and $Q_{X/Z, \text{outer}}(j=\{1,...,k\})$ it is obtained through the simulation.

We note that $r$ describes the repeater performance by providing a lower bound both on the achievable distillable entanglement and secret key extractable from the state generated in our setup.
In Appendix~\ref{sec:PerformanceMetricsFormulas} we provide the details of how the secret key or entanglement is extracted from those binned links parameterised by $\{m_X, m_Z, j\}$ and provide the formula for $r$. We also note that in all the above equations in this section we have assumed the number of repeaters to be equal to the number of elementary links and labeled both by $n$. Clearly the number of elementary links is actually larger by one than the number of repeaters. We clarify this point in Section~\ref{sec:RequirementsEndNodes}.

\section{Simulation}
\label{sec:Simulation}

In this section we briefly summarise how we numerically simulate the performance of our repeater setup. On a high-level our simulation is very similar to that of~\cite{noh2020fault} or~\cite{rozpkedek2021quantum} in the sense that we do not simulate evolution of any quantum state but only the error propagation through the noisy channels and noisy operations as well as the error correction. The inputs to our simulation are as follows.

Hardware parameters:
\begin{itemize}
    \item GKP squeezing $\sigma_{\text{GKP}}$ of all the GKP qubits used i.e. GKP qubits used to construct each cube and the ancilla GKP qubits used for TEC on inner leaf qubits.
    \item Detector efficiencies during homodyne detection $\eta_d$. These appear in numerous BSMs.
    \item Number of multiplexing levels $k$. 
    \item The discard window $v$ used during the resource state preparation. 
\end{itemize}

Parameters over which we later want to optimise:

\begin{itemize}
    \item Repeater spacing $L$. We consider five possible values of $L$ over which we optimise $L \in \{0.5, 1, 2, 2.5, 5\}$ km.
    \item How often we do GKP TEC on the inner leaf GKP qubits $n_{\text{per-km}}$. This parameter describes the number of GKP TECs we do per 1 km of fiber.
\end{itemize}

The output of the simulation is:

\begin{itemize}
    \item The independent logical $X$ and $Z$ error probability of each of the outer multiplexed links $k$, $Q_{X/Z, \text{outer}}(j=\{1,...,k\})$ for a single elementary segment.
    \item The independent logical $X$ and $Z$ error probability for the inner leaves $Q_{X/Z, \text{inner}}(s=0,1)$ for the two scenarios where the logical BSM on the inner leaves produced no-error syndrome $s=0$ or an error syndrome $s=1$ for a single elementary segment.
    \item The probability $t_{X/Z}$ of inner leaf logical BSM leading to the $s=1$ error syndrome for both $X$ and $Z$ errors.
\end{itemize}

The accuracy of our simulations is quantified as described in Appedix~\ref{sec:SimAcc}.

\section{results}
\label{sec:results}

In this section we present the results of our numerical Monte Carlo simulations. We also present the analytical results of our resource count estimates.

\subsection{Performance of the outer-leaf based ranking}
\label{sec:RankingPerformance}

\begin{figure}
\includegraphics[width =\columnwidth]{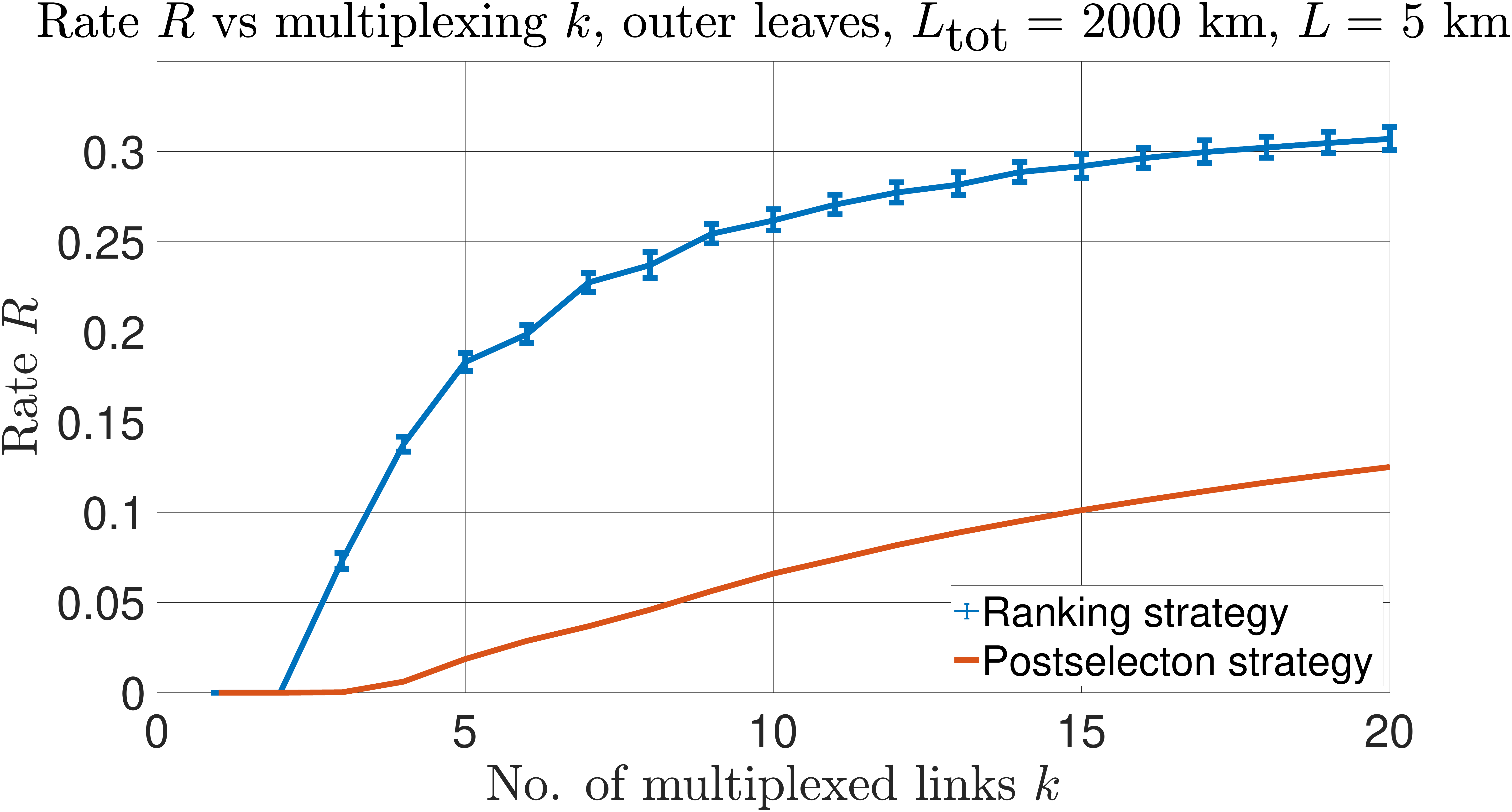}
\caption{Comparison between the proposed ranking strategy vs. post-selection strategy based on optimal discard window $v$ that maximises the rate. We plot the rate vs number of multiplexed links for those two strategies. As the purpose of this plot is only to compare the two strategies, we consider here ideal devices with infinite GKP squeezing and perfect homodyne detection. Moreover, we assume that the inner leaves are stored in perfect quantum memories, so that the only source of noise is the photon loss in the communication channel. We see that the ranking strategy significantly outperforms the optimised post-selection-based strategy. Moreover, we observe a strong saturation for the ranking strategy around $k=20$ multiplexed links. The error bars on the numerical data corresponding to the ranking strategy reflect the simulation accuracy described in Appendix~\ref{sec:SimAcc}.}
\label{fig:RateVsMultiplexing}
\end{figure}

To demonstrate the benefit of the ranking strategy, we can compare its performance against a simpler strategy based on post-selection. In this simpler strategy instead of ranking the links, a link is kept or discarded after the outer leaf BSM depending on whether the measured outcome falls in the post-selection or discard window. Clearly that strategy actually corresponds to a family of strategies parameterised by the size of the post-selection/discard window $v$. Let us for a moment assume that the only source of noise is the communication channel, while the resource state preparation is perfect (i.e. infinite squeezing and perfect homodyne detection). Moreover, let us assume that the storage of the inner leaf qubits is also perfect. Under that simplified paradigm the performance of the post-selection-based strategy can be evaluated analytically for a given window $v$, number of multiplexed links $k$ and repeater separation $L$ as described in Appendix~\ref{sec:postselectionstrategy}. The size of the window $v$ determines the trade-off between the rate of generating the raw pairs and their quality/fidelity. For a specific figure of merit such as secret-key rate or ebit rate we can optimise that performance metric over the window $v$. We note that both the ranking and the post-selection strategies allow to extract multiple end-to-end links from a single protocol run, in contrast to the previous all-photonic architectures based on the clique-cluster~\cite{Azuma_15,pant2017rate,fukui2021all}. For the post-selection strategy, the number of end-to-end extracted links in a single protocol run is determined by the number of elementary multiplexed links that pass post-selection on the elementary segment with the smallest number of successfully extracted multiplexed links. Below we compare the performance of the ranking strategy which we evaluate using our numerical simulation against the post-selection-based strategy which we evaluate analytically for each $v$ and then numerically optimise over that parameter. The performance is evaluated under the simplified model with the only source of noise being the communication channel and with perfect memories for the inner leaves. We see from the plot in FIG.~\ref{fig:RateVsMultiplexing} that the proposed more complex ranking strategy significantly outperforms the post-selection-based strategy. Moreover, we also see that the performance per mode saturates at around $k=20$ multiplexed links. In fact, we observe a similar saturation behaviour when including the finite GKP squeezing, finite efficiency of the homodyne detectors as well as lossy inner leaf qubit storage. Therefore from now on we fix the number of multiplexed levels to $k=20$.

\subsection{Discard window for post-selection during resource state preparation}
\label{sec:discardWindow}

A natural question to ask is what the reasonable value of the discard window $v$ for the resource state preparation should be. Clearly the larger the discard window, the more we can suppress the errors at that stage of the protocol at the expense of increasing the amount of GKP qubits we need to generate. However, we find that even from the perspective of the repeater performance alone, there are diminishing returns when increasing $v$ beyond a certain threshold which actually depends on the repeater spacing $L$. Here $L$ is the spacing between the major repeater nodes, i.e. halfway between them we still need to place the outer leaf BSM stations. This dependence arises because it is only helpful to decrease the errors from the resource state generation to just below the order of magnitude of the errors from the communication channel and inner leaf storage channel. At that point the communication channel or inner leaf storage channel errors completely dominate the overall error contribution and so there is no benefit in suppressing the resource errors further which would actually cost much more in terms of the required number of GKP qubits. Since the magnitude of the errors arising from the communication channel over each elementary link depends on the repeater separation $L$, the value of this threshold $v$ will also depend on $L$. In particular, the smaller $L$ is, the smaller the communication channel errors for the single elementary link and so the larger the threshold discard window will be. Here we will consider five possible inter-repeater spacings $L \in \{0.5, 1, 2, 2.5, 5\}$ km. By numerical investigation, we find that a reasonable choice for the corresponding discard window $v$ that minimizes the resource state generation errors to close to maximum level where it has a visible impact on the performance is $v = \{7,6,5,4,3\} \times \sqrt{\pi}/20$ for the five values of $L$ respectively. Hence, since we use post-selection at all stages of our resource state preparation, denser placement of repeaters should lead to the best performance. However the use of larger discard window for these smaller values of $L$ will lead to much larger resource requirements for the configurations with dense repeater spacing. We remind the reader that the values of $v$ discussed here represent the size of the discard window for the measurements during resource state preparation where the variance is given by $\sigma^2 = 3\sigma_{\text{GKP}}^2 + (1-\eta_d)/\eta_d$. For the measurements with variance $\sigma^2 = 2\sigma_{\text{GKP}}^2 + (1-\eta_d)/\eta_d$ we use $0.7v$.

Clearly one could also consider denser repeater spacing with smaller discard windows in order to reduce the required resources. However, we believe that this would not be very efficient because in that case the repeaters would actually contribute substantial amount of extra noise that would affect the performance. As mentioned, if one wants to reduce the amount of required GKP qubits, we believe that it is then more efficient to just consider larger repeater spacing with the maximal discard window suitable for that larger repeater spacing.

\subsection{Inner-leaf based discrete syndrome information}

\begin{figure}
\includegraphics[width =\columnwidth]{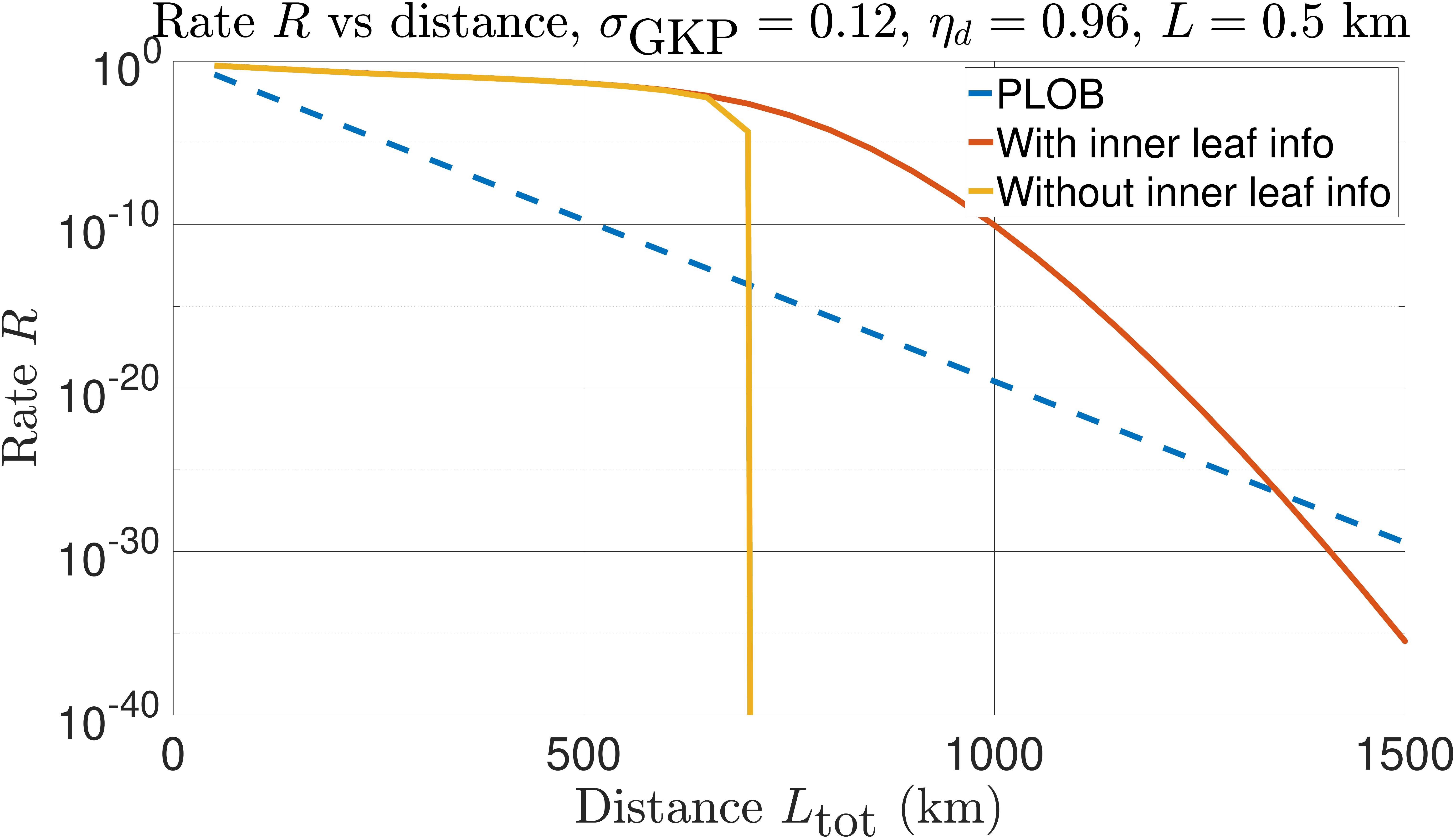}
\caption{Comparison between the scenario when we use and do not use the inner leaf information. We see that the inner leaf information enables us to achieve positive rate for much larger distances. The yellow line corresponding to the scenario without the inner leaf information drops rapidly to zero already around 700 km. The parameters $L$, $\sigma_{\text{GKP}}$ and $\eta_d$ refer to inter-repeater spacing, GKP squeezing and homodyne detection efficiency respectively.}
\label{fig:InnerLeafInfo}
\end{figure}

Similarly as in~\cite{namiki2016role} we observe that while the inner-leaf discrete syndrome information does not boost the performance for short distances, it can significantly improve the achievable distances, in particular distances over which our repeater scheme beats the PLOB bound. Specifically, as illustrated in FIG~\ref{fig:InnerLeafInfo} for a specific parameter choice, without the inner leaf information at certain distance the rate rapidly drops to zero, while with the information it continues its decay much more slowly hence maintaining reasonable performance over much longer distances.

Since the rate formulas are convex in the whole QBER regime one might wonder why the benefit of using the inner-leaf information manifests itself vividly only close to the distance where without this information the rate already rapidly drops to zero. Our investigation suggests that this is because it is not the continuous convexity of the secret-key fraction/achievable entanglement $r$ that matters but rather its convexity at a kink where $r$ becomes zero, since $r$ actually takes a form $r = \max\{\tilde{r},0\}$. Here $\tilde{r}$ just denotes a continuous convex function which actually can become negative for large errors, see Appendix~\ref{sec:PerformanceMetricsFormulas}. Hence the benefit of using the inner leaf information becomes visible in the regime where for some of the bins occurring with high-probability the rate is already zero. That only happens in the regime, where without using this discrete-syndrome information the total error is already so high that the rate starts rapidly dropping to zero. We also investigate the contribution of all the different inner leaf information bins on the final rate using the tools of typical sequences. Here the sequences of interest are the sequences of bits $\vec{s}_{X/Z}$ describing the syndromes in the inner leaves. We find that when using the inner leaf information the gradual decay of the rate, which stays positive yet reaches very small values, is dictated by how the number of the sequences corresponding to the bins with positive rate change with distance, see Appendix~\ref{sec:typicalsequences} for more details.

\subsection{Achievable distances}

Having understood the role of the outer-leaf multiplexing and inner leaf discrete syndrome, we can now proceed to investigate the achievable distances of our scheme. The different parameters that we vary are different homodyne detection efficiencies $\eta_d$ and different values for GKP squeezing during the GKP state preparation $\sigma_{\text{GKP}}$. We note that the dependence on GKP squeezing manifests itself not only through the squeezing of the GKP qubits being part of the resource state preparation but also through the squeezing of the GKP qubits used to create an ancillary Bell pair for GKP TEC. Similarly homodyne detection is used throughout the fusion gates during resource state creation, during GKP TEC and during outer leaf and inner leaf BSMs. We optimise the achievable distance over five possible inter-repeater spacings $L$, which, as mentioned before, correspond to $L \in \{0.5, 1, 2, 2.5, 5\}$ km. For each of them we consider a specific discard window, as explained in Section~\ref{sec:discardWindow}. The frequency of the GKP TEC on the inner leaves is described by the parameter $n_{\text{per-km}}$, the number of GKP corrections during inner leaf storage per km of fiber (this includes the final perfect destructive GKP correction during entanglement swapping). Here we fix $n_{\text{per-km}} = 4$. However, we have also done some exploratory simulation runs with $n_{\text{per-km}} = 2$ and found that $n_{\text{per-km}} = 4$ leads to much better performance for almost all the parameters. Only for bad squeezing and bad homodyne detection efficiency for which the achievable distance is already very low, of the order of at most 100 or 200 km, it might be better to use $n_{\text{per-km}} =2$ as then adding additional GKP TECs can actually ruin the performance by adding more very noisy operations. Therefore all the results presented here are based on the data obtained for $n_{\text{per-km}} = 4$.

\begin{figure*}
    \centering
    \includegraphics[width=1\textwidth]{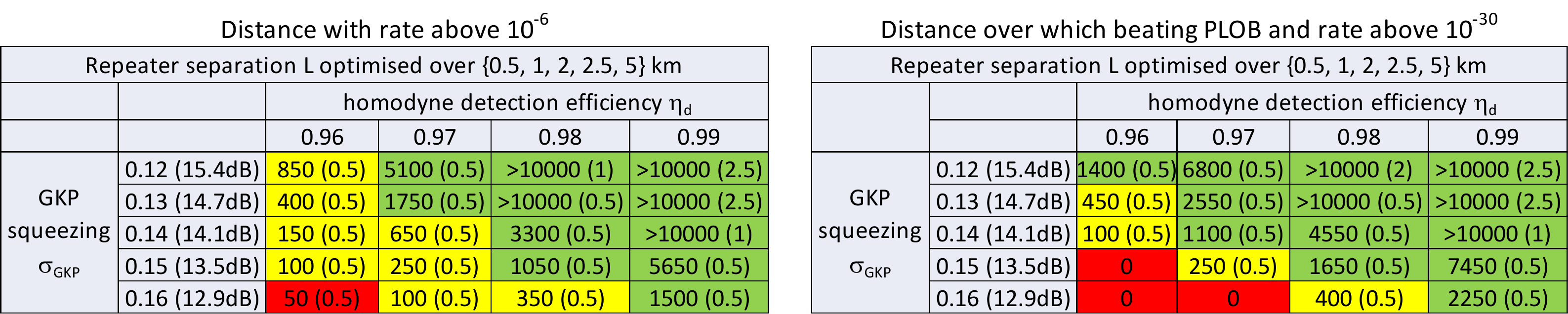}
    \caption{Achievable distances in km for different amount of GKP squeezing and homodyne detection efficiency. The left table shows the largest distances for which the rate stays above $10^{-6}$, while the right table shows the largest distances for which the rate stays above $10^{-30}$ and beats the PLOB bound. Red entries correspond to distances below 100 km, yellow to distances 100 km to 950 km and green to distances of 1000 km and above. Each entry depicts the largest distance optimised over five possible repeater spacings of $L=0.5$ km, $L=1$ km, $L=2$ km, $L=2.5$ km, $L=5$ km. The optimal repeater spacing for the given configuration is stated in brackets after the achievable distance. If for any of the repeater spacings the achievable distance is above 10000 km, we list in brackets for that configuration the largest spacing that still achieves 10000 km distance.}
\label{fig:AchievableDistances}
\end{figure*}

\begin{figure}
    \centering    \includegraphics[width=\columnwidth]{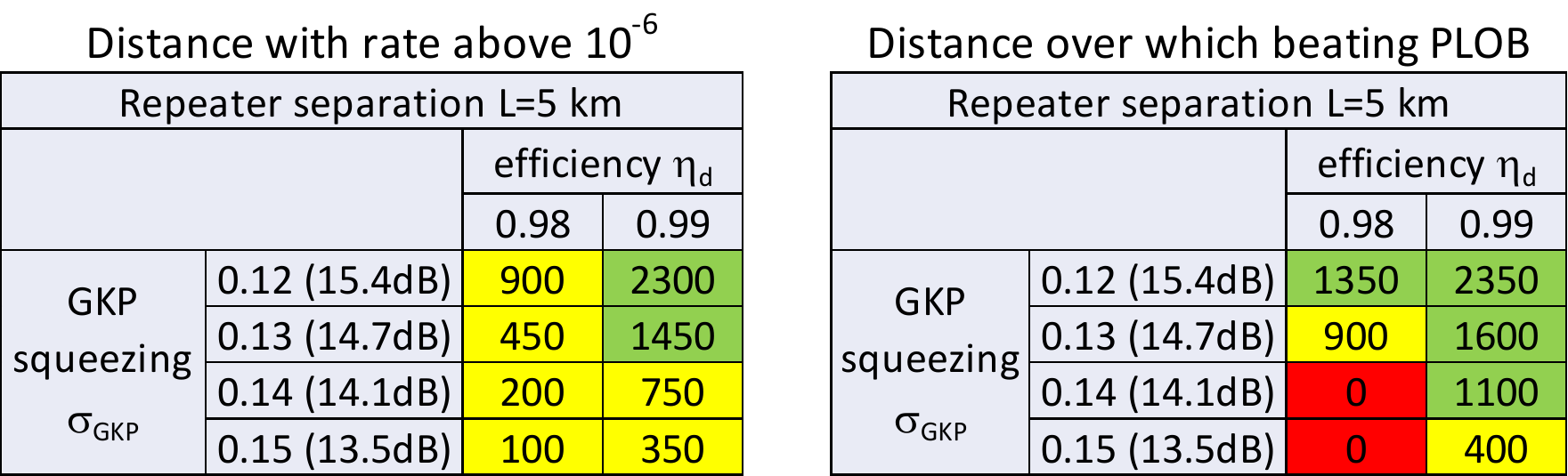}
    \caption{Achievable distances in km for different amount of GKP squeezing and homodyne detection efficiency for repeater separation of $L=5$ km.}
\label{fig:AchievableDistancesL50Main}
\end{figure}

Our final results in terms of achievable distances are shown in FIG.~\ref{fig:AchievableDistances}. We provide two tables, one showing the largest distance in km for which the repeater rate stays above $10^{-6}$ and one for describing the largest distance for which our repeater outperforms the PLOB bound, the ultimate limit of repeater-less communication. For the first table the choice of $10^{-6}$ is motivated by the fact that if the repetition rate of our protocol could be in the regime of at least 1 MHz, then our scheme with $k=20$ multiplexing levels could generate at least 20 Bell pairs/bits of secret key per second for these distances (generation of Bell pairs is counted based on asymptotic entanglement distillation). For the second table we also introduce a cut-off in terms of maximum value, i.e. we only consider distances for which the performance stays above $10^{-30}$ even if going below $10^{-30}$ still maintains the rate above the PLOB bound.

As mentioned for all these parameter settings we perform optimisation over the inter-repeater spacing $L \in \{0.5, 1, 2, 2.5, 5\}$ km. For each parameter set the value of $L$ that maximises the performance is quoted in brackets. Note that the largest possible achievable distance we consider is 10000 km. If this distance can be achieved for different values of $L$ we quote the largest one, based on the assumption that it is always cheaper to consider larger inter-repeater spacing. Note that this does not take into account the actual value of the rate in the sense that it is possible that a smaller inter-repeater spacing actually achieves higher rate.

Finally we use color-coding to help the reader see the pattern of the data. We color the parameter sets for which the achievable distance is below 100 km as red, those for which the achievable distance is between 100 and 1000 km as yellow and those with achievable distance above 1000 km as green.

We see that small improvements in the parameters can significantly increase the achievable distance. In particular, we see that the achievable distances are very sensitive to the efficiency of the homodyne detection for the parameter variation steps chosen here. This can be explained by the fact that we expect the dominant contribution to the rate to be coming from the first few best ranked links, for which the outer leaf error is suppressed below the level of the inner leaf error. The dominating inner leaf error channels are affected by the GKP squeezing through terms $2\sigma_{\text{GKP}}^2$ or $3\sigma_{\text{GKP}}^2$, see Table~\ref{tab:tableInnerLeaves} in Appendix~\ref{sec:ErrChforInnerLeaves}.  A decrease in $\sigma_{\text{GKP}}^2$ by 0.01 leads to a smaller change in those terms than the change of the terms containing the contribution of the detection noise $(1-\eta_d)/\eta_d$ and $(1-\eta\eta_d)/(\eta\eta_d)$, when $\eta_d$ is increased by 0.01. Moreover, as expected, we see that for all the parameter configurations for which the achievable distance is below 10000 km, the optimal repeater spacing is $L = 0.5$ km which stays in agreement with the fact that with our optimised discard windows $v$ the resource generation always contributes local noise below the level of the channel noise and so it is always best to put repeaters as densely as possible if performance is the considered figure of merit.

We note that even for the best considered parameters, the configurations with the repeater separation of $L = 5$ km can never achieve 10000 km. However, they might still be useful for communication over shorter distances, since placing repeaters further away is economically cheaper, and, as we will see, these configurations require significantly less resources. Therefore we additionally include the achievable distances for $L=5$ km in FIG.~\ref{fig:AchievableDistancesL50Main} for the configurations of the better parameters. We see that if we had access to hardware with such good parameters and if our goal was only to communicate over shorter to medium distances, then we could still achieve this goal with repeaters spaced so much more apart which we assume would be more appealing from the implementation perspective.

We see that for this large repeater separation and for the very good parameters, the achievable distance with respect to the rate being above $10^{-6}$ and being above the PLOB bound is not that much different. This is due to the fact that for this scenario the probabilities of the best sequences $\vec{s}$ (describing the inner leaf syndromes) which contribute to the non-zero rate at the critical distances are generally much larger than for configurations with smaller values of $L$. This leads to a fast drop of the rate to 0 from $10^{-6}$ for slightly larger distance when the given sequences do not contribute any positive rate anymore. This contrasts with the case of more densely place repeaters where the sequences $\vec{s}$ will be in general longer, leading to much smaller probabilities of individual strings and hence more fine-graining to the level significantly smaller than $10^{-6}$ allowing for a more gradual decay of the rate. For more detailed discussion of this phenomenon, see Appendix.~\ref{sec:typicalsequences}.

\subsection{Rate vs distance}

Having already established the general dependence of the achievable distances when we vary the amount of GKP squeezing and homodyne detection efficiency, we can now pick some of these parameter sets to plot the performance rate vs distance.

\begin{figure*}
\centering
\includegraphics[width = \textwidth]{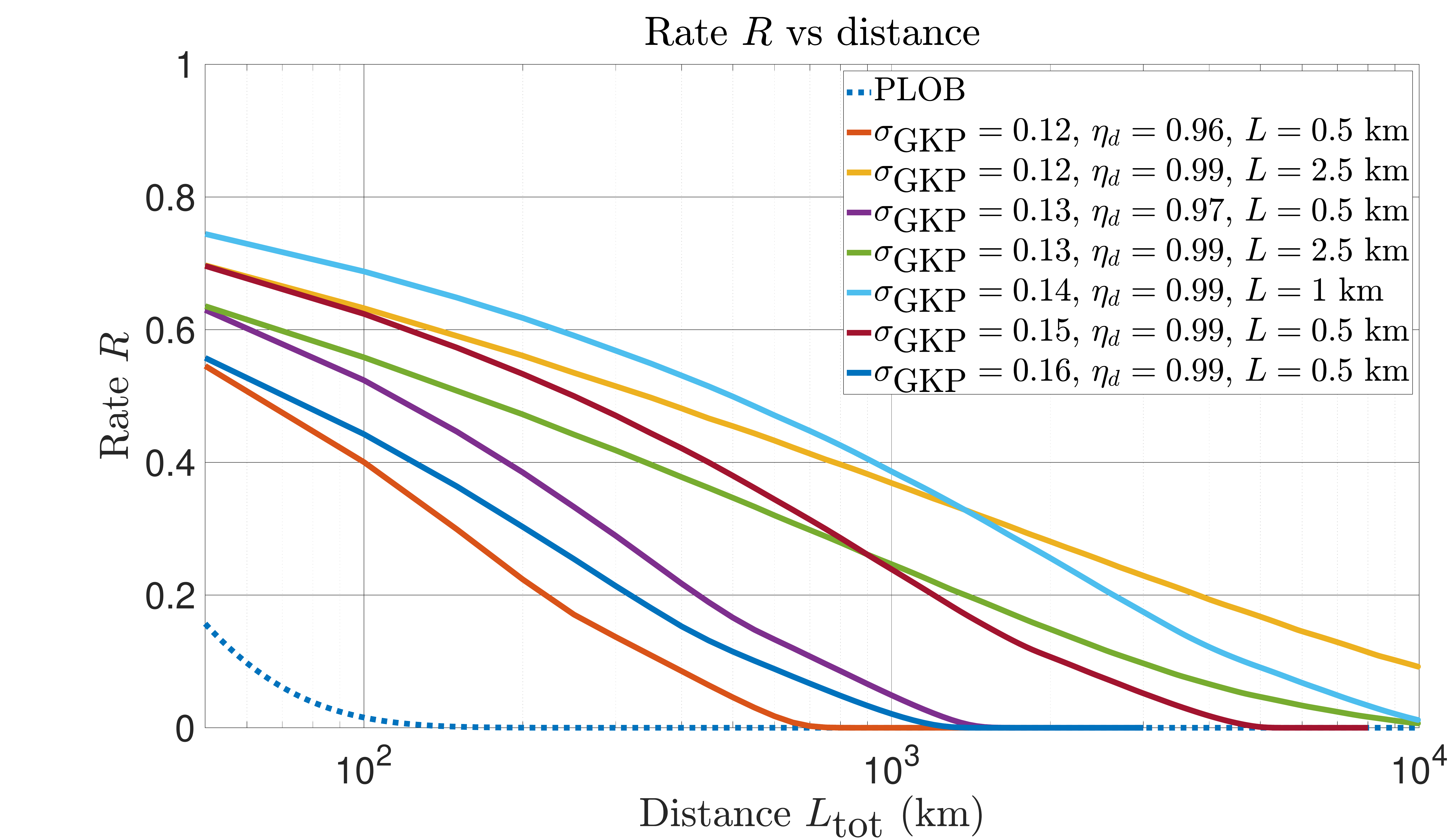}
\caption{Rate vs. distance for different hardware parameter configurations. The corresponding repeater separation for each configuration is the one listed in the table in FIG.~\ref{fig:AchievableDistances}. }
\label{fig:ratevsdist}
\end{figure*}

In FIG.~\ref{fig:ratevsdist} we plot the rate versus distance for different squeezing and homodyne detection efficiencies where the repeater spacing for each curve is the one quoted in the tables in FIG.~\ref{fig:AchievableDistances}. We see that, as expected, better hardware parameters allow to achieve larger rates for larger distances. However, for shorter distances we see that it is not only the hardware parameters that matter but also the repeater spacing $L$. In particular, we see that for certain configurations actually a higher rate can be obtained with worse parameters and smaller repeater spacing than for better parameters with larger repeater spacing. This leads to the fact that we see multiple curves crossing each other on the plot.

\begin{figure}
\centering
\includegraphics[width = \columnwidth]{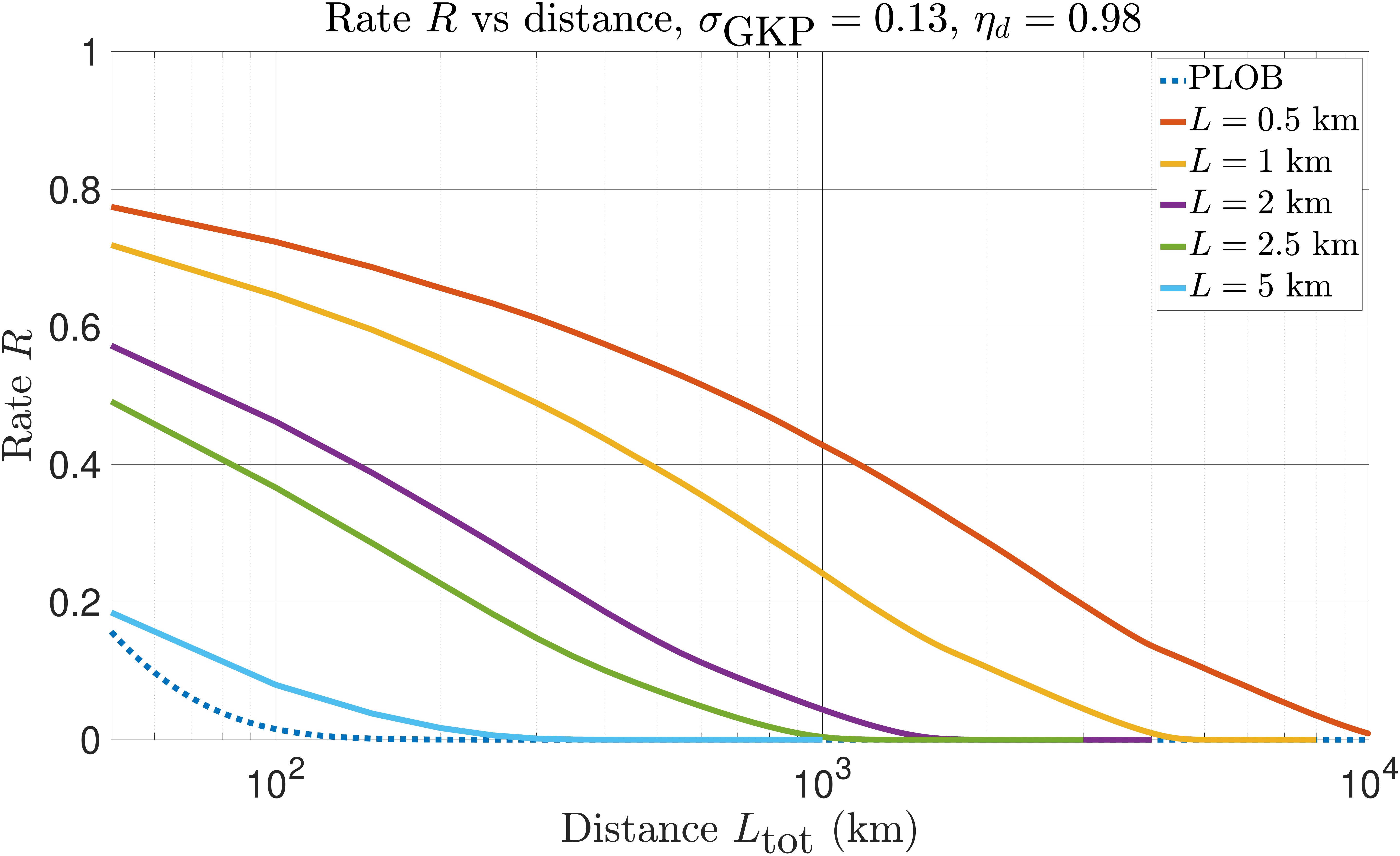}
\includegraphics[width = \columnwidth]{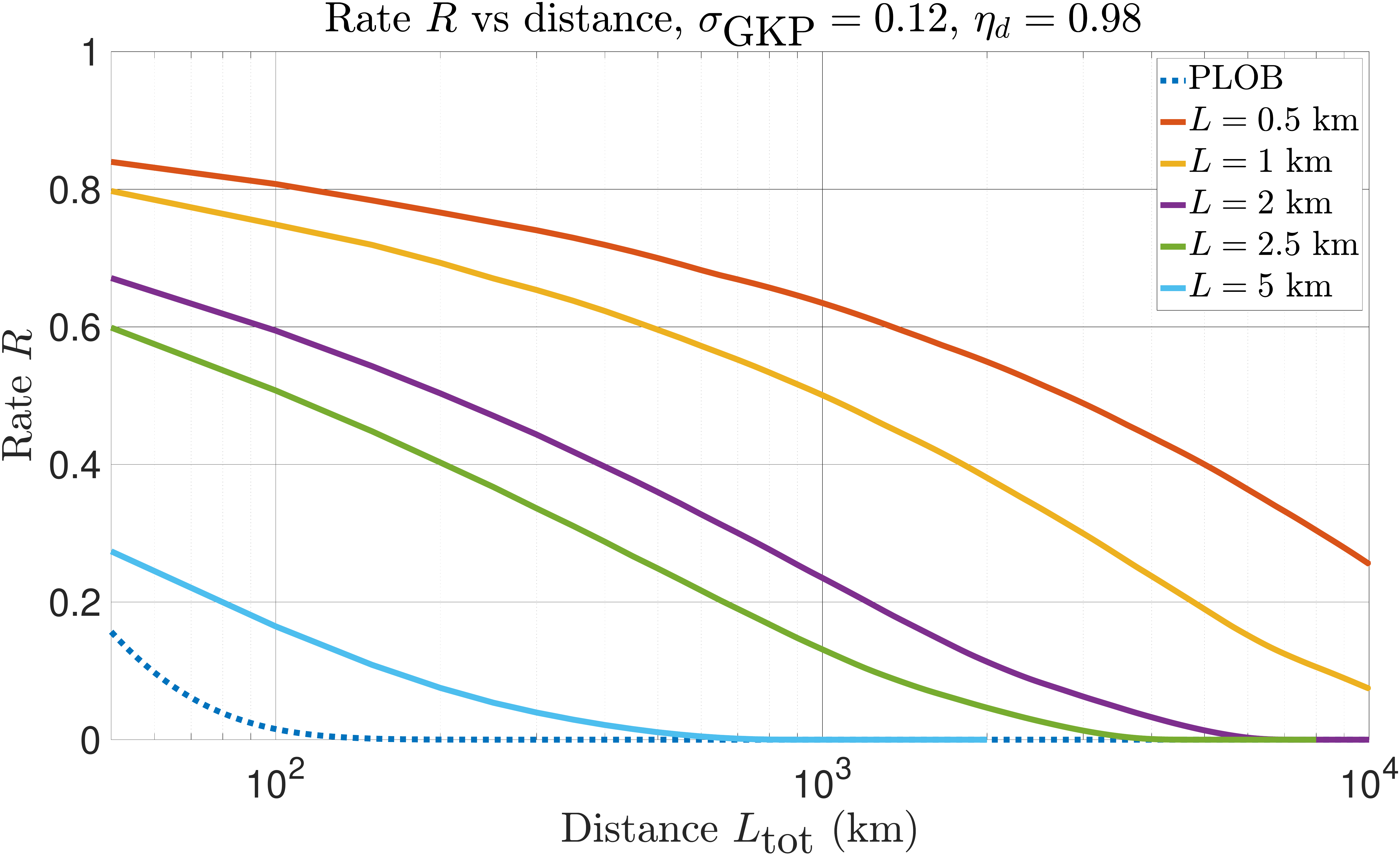}
\includegraphics[width = \columnwidth]{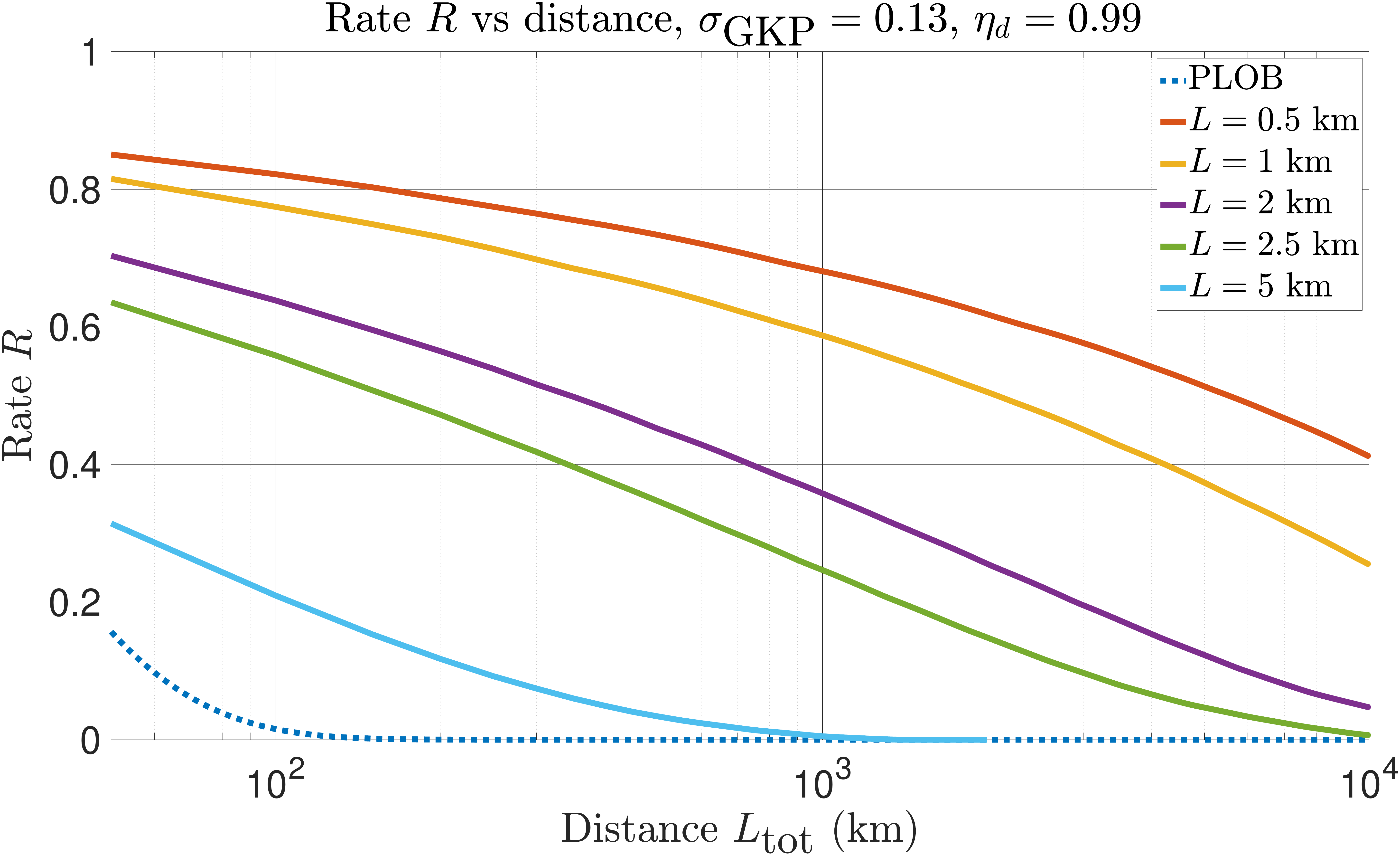}
\caption{Rate as a function of distance for three hardware parameter sets where for each set we plot all the considered repeater spacing scenarios.}
\label{fig:ratevsdistDifferentL}
\end{figure}

We also choose specific hardware parameter sets for which we plot the rate vs distance for all the considered values of $L$, see FIG.~\ref{fig:ratevsdistDifferentL}. We see that the configuration with $L=0.5$ achieves the highest rate for all the distances. We see that the rate is also very sensitive to the homodyne detection efficiency. In particular, in FIG.~\ref{fig:ratevsdistDifferentL} we chose a specific parameter configuration of $\sigma_{\text{GKP}} = 0.13$ and $\eta_d =0.98$ and then investigated the improvement when we improve GKP squeezing by reducing $\sigma_{\text{GKP}}$ by 0.01 or we increase homodyne detection efficiency by 0.01. We see that the latter leads to more dramatic improvement of the rate than the former.

Finally it is important to comment on the values of the achievable rates. We see on the bottom plot of FIG.~\ref{fig:ratevsdistDifferentL} that with the considered parameters we can achieve the rate per mode as large as 0.7 for all the distances up to 750 km. We discuss in Section~\ref{sec:comparison} how these rates compare to the rates of other repeater schemes.

\subsection{Optimizing resources}
\label{sec:OptResourcesResults}

For the resource state generation based on multiplexed fusions along the various steps of the procedure shown in FIG.~\ref{fig:flowchart}, we analyze the number of GKP qubits required at each repeater for near-deterministic generation of the required number of resource states at all repeaters. 
We observe that the success probability $p^{\textrm{res-gen}}_\textrm{succ}$ as a function of the number of GKP qubit resources per repeater undergoes a phase transition where it sharply turns from 0 to 1. Moreover, for the hardware parameters considered here in our investigation, we find that we need around $10^3 - 10^{4}$ GKP qubits per repeater. The results are elaborated below.

\subsubsection{Resource count}

We note that our resource count consists of three components. The first component relates to the number of GKP qubits that each repeater needs in order to build the cube resource state. As building of the cube involves post-selection, to a given number of GKP qubits per repeater one can associate a corresponding probability that all the repeaters simultaneously successfully generate all the required cube resource states $p^{\textrm{res-gen}}_\textrm{succ}$ as described by Eq.~\eqref{multi-res-gen-psucc}. Specifically, for $k$ multiplexing levels, each repeater needs to be able to generate $2k$ such cube resource states. We will see that similarly to the case of the resource generation for the discrete-variable all-photonic scheme of~\cite{pant2017rate}, when increasing the number of GKP qubits per repeater there will be a transition point where provided that a certain amount of GKP qubits is supplied per repeater, the probability that all the repeaters will be able to generate all the cube resource states approaches one. Clearly that required amount of GKP qubits per repeater depends on the chosen discard window $v$. Since in our case the choice of $v$ is associated with a given repeater spacing $L$, the smaller $L$ the more resources we will need per repeater. Of course the total number of resources needed to generate all the cube resource states for the whole protocol run in all the repeaters depends not only on the required resources per repeater but also on the repeater density. Hence we see that from the perspective of number of resources for the generation of the cube states, configurations with larger repeater spacing will be more favorable.

The second component of our resource count are the GKP qubits needed for TEC. The number of these resources is fixed as there is no associated post-selection. Each TEC requires two ancillary GKP qubits. In our scheme we perform TEC on each of the 7 physical GKP qubits of the inner leaf logical qubit every 250 m until the final GKP correction after the last 250 m which is done on the classical level together with the BSM (CC-amplification). Taking into account that for each multiplexing level we store 2 concatenated-coded qubits, the total number of GKP qubits per repeater needed for the implementation of these TECs is $28k \times (4L-1)$. Clearly this number is actually larger the larger $L$ is. Moreover, even for a fixed total end-to-end distance, the total number of GKP qubits in all the repeaters needed to implement all the TECs is slightly larger for the configurations with larger repeater spacings because of the minus one term, accounting for the fact that the final correction in each repeater is always done on the BSM level and hence does not require any ancilla qubits.

The final component contributing only to the end-to-end resource count are the GKP qubits required by the end-nodes of Alice and Bob. As we discuss later in Section~\ref{sec:RequirementsEndNodes} here we assume that Alice and Bob need only to prepare $k$ GKP qubits each so this additional component involves only additional $2k$ GKP qubits. Note that the total number of repeaters is given by $n_{\text{rep}} = \frac{L_{\text{tot}}}{L}-1$.

\begin{figure}
\includegraphics[width =\columnwidth]{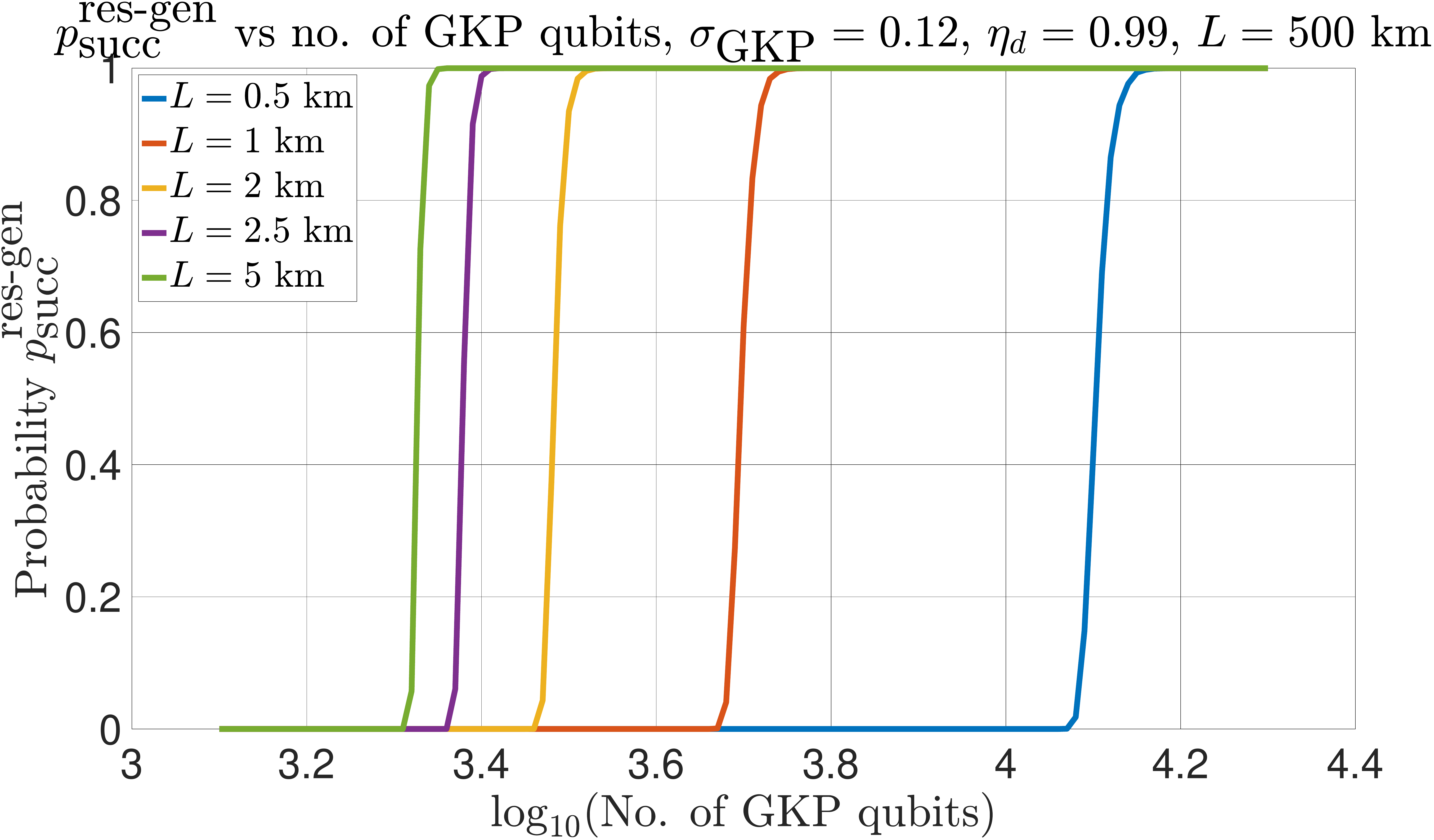}
\caption{Resource generation probability $p^{\textrm{res-gen}}_\textrm{succ}$ as a function of the number of GKP qubits per repeater for different inter-repeater spacings for given hardware parameters and total end-to-end distance.\label{fig:diff_L_res_count}}
\end{figure}

\begin{figure}
\includegraphics[width =\columnwidth]{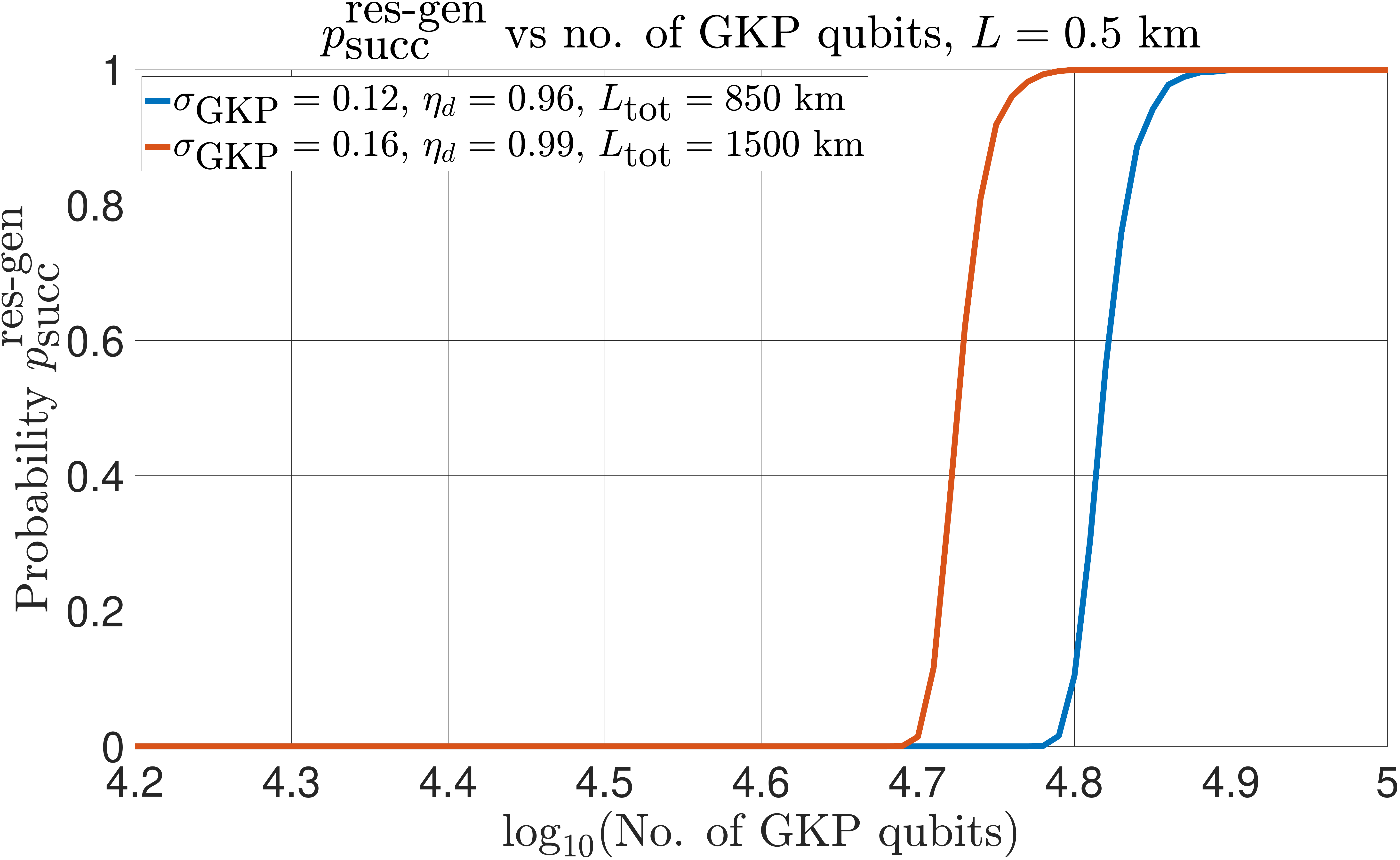}
\caption{Resource generation probability $p^{\textrm{res-gen}}_\textrm{succ}$ as a function of the number of GKP qubits per repeater for two hardware parameter configurations with corresponding total distances for a fixed inter-repeater spacing.  \label{fig:extreme_cases_res_count}}
\end{figure}

Firstly, considering only the first component, in FIG.~\ref{fig:diff_L_res_count} we plot $p^{\textrm{res-gen}}_\textrm{succ}$, the probability that given the specified amount of GKP qubits per repeater, the repeaters are successful in generating all the required cube resource states. The chosen multiplexing value is again $k=20$ and we plot this probability for different repeater spacings for the total distance of $L_{\text{tot}} = 500$ km, GKP squeezing standard deviation $\sigma_{\textrm{GKP}}=0.12$ and homodyne detection efficiency $\eta_d = 0.99$. We see that the probability of successfully generating the required number of cube resource states at all the repeaters end-to-end jumps up to 1 when sufficiently large number of GKP qubits are available at each repeater. We notice that the shorter the repeater spacing, the larger the number of GKP qubits per repeater that are required for successful resource generation at all the repeaters.

Additionally in FIG.~\ref{fig:extreme_cases_res_count} we plot the corresponding probability also for two specific hardware parameter configurations. One with very good squeezing and bad homodyne detection efficiency ($\sigma_{\text{GKP}} = 0.12$ and $\eta_d = 0.96$) and the other vice versa ($\sigma_{\text{GKP}} = 0.16$ and $\eta_d = 0.99$). For the two curves we consider the distances equal to the achievable distances from the left table in FIG.~\ref{fig:AchievableDistances} for these parameters. We only plot the scenario with $L=0.5$ km since for these distances a positive rate can only be achieved with this most dense repeater spacing (see the achievable distances tables in Appendix~\ref{sec:AllAchievableDistanceTABLEs}).

By comparing the two curves in FIG.~\ref{fig:extreme_cases_res_count} with the curve corresponding to $L=0.5$ km in FIG.~\ref{fig:diff_L_res_count}, we see that for better hardware parameters the required resources are generally significantly smaller. Moreover, in FIG.~\ref{fig:extreme_cases_res_count} we also see that the effect of hardware parameters is much stronger than the effect of the total distance $L_{\text{tot}}$, since the configuration corresponding to larger total distance actually requires less resources than the configuration corresponding to the almost twice smaller distance.

When the second component, i.e., resources required for TEC at the repeaters, is also included, the necessary resource count per repeater is no longer monotonic in the repeater spacing. This is because the larger the repeater spacing, the larger the resources required for TEC per repeater. To quote numbers corresponding to the different repeater spacing configurations in FIG.\ref{fig:diff_L_res_count} where the success probabilities jump to 1 and adding to them the required TEC resources, the number of GKP qubits per repeater necessary for $L=\{0.5,1,2,2.5,5\}$ km are given by $\{15351, 7434, 7308, 7670, 12931\}$, respectively. We note that for our Monte-Carlo simulation of $(p^{\textrm{res-gen}}_\textrm{succ})^{1/n_{\text{rep}}}$ we generate $10^7$ simulation runs for each considered number of input GKP qubits. This means that our simulation cannot capture failures occurring with probability of order $10^{-8}$. For the largest considered total distances of 10000 km and for repeater spacing of $L=0.5$ km these can correspond to failure probabilities of the order of $10^{-4}$ for the entire repeater chain. Hence here we define the number of resources needed for deterministic resource generation as the smallest number of GKP qubits for which $p^{\textrm{res-gen}}_\textrm{succ} > 1-10^{-3}$, since that threshold should remain unaffected by our simulation inaccuracy.

Now we can establish the total number of GKP qubits needed end-to-end by multiplying these resources by the corresponding number of repeaters and adding the required 20 GKP qubits for each of the two end-nodes. This gives the total required resources as $\{1.5336\times 10^7, 3.7098\times 10^6, 1.8198\times 10^6, 1.5264\times 10^6, 1.2802\times 10^6\}$, restoring again the monotonic behaviour.

\subsubsection{Resource-performance trade-off}

As we have already seen the densest repeater configurations achieve the best performance but require the most resources. Moreover, for worse hardware parameters the densest repeater configurations might actually be necessary in order to maintain non-zero rate. Here we mathematically quantify the trade-off between the performance and the required resources by introducing a cost function:
\begin{equation}
    C = \frac{\text{no. of GKP qubits used end-to-end}}{R} \, .
    \label{eq:costfunction}
\end{equation}
Here we specifically count the number of GKP qubits per protocol run. For chosen hardware parameters for every distance we minimize this cost function over the five considered repeater spacing configurations in order to optimise the performance-cost trade-off. We perform this task for the three hardware parameter configurations for which we have plotted the rates in FIG.~\ref{fig:ratevsdistDifferentL}. We note that this choice of the cost function can be seen as somehow arbitrary. One could always define a different cost function that puts more weight on minimizing resources by taking resources to some higher power or one that puts more emphasis on maximising the rate, by taking $R$ to some higher power.

\begin{figure}
\includegraphics[width =\columnwidth]{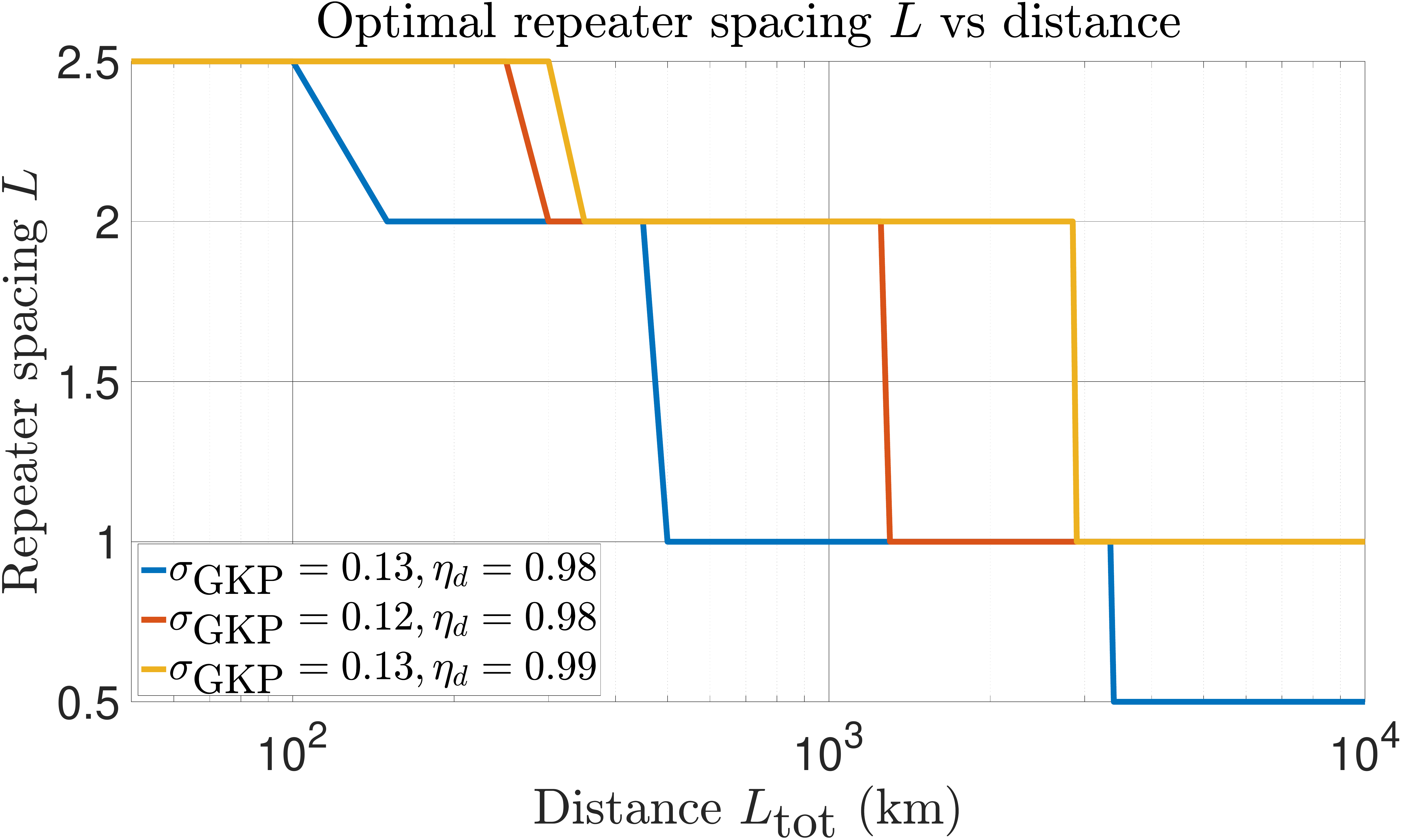}
\caption{Optimal repeater spacing $L$ that minimises cost function $C$ for different hardware parameters.}
\label{fig:optRepSpacing}
\end{figure}

In FIG.~\ref{fig:optRepSpacing} we plot the optimised repeater spacing. We see that the larger the total distance the more densely we should place our repeaters. However, we see that the two better configurations can maintain larger repeater spacing for longer total distances and the worse considered configuration with $\sigma_{\text{GKP}} = 0.13$ and $\eta_d =0.98$ requires actually to reduce repeater spacing to $L=0.5$ km already below 4000 km.

\begin{figure}
\includegraphics[width =\columnwidth]{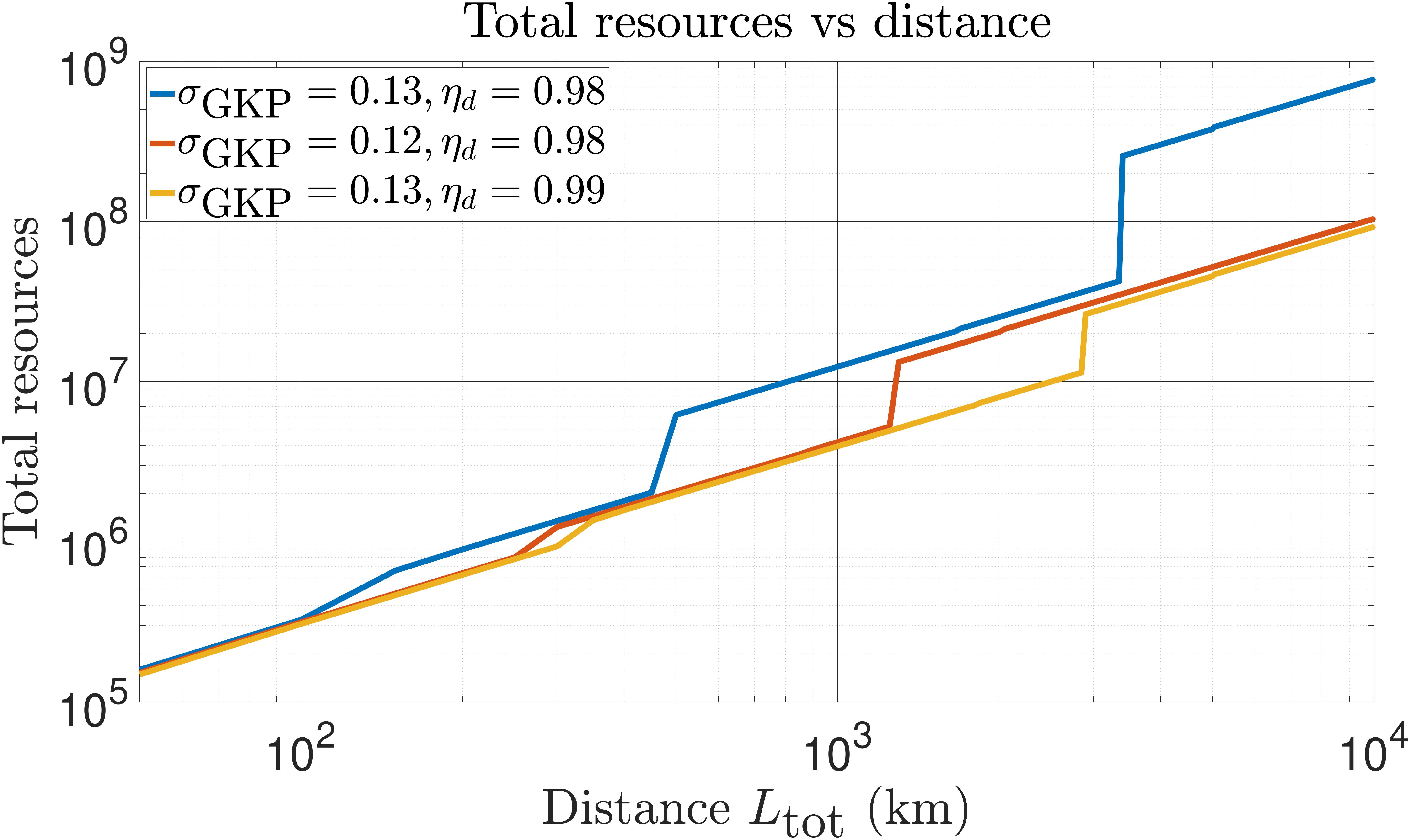}
\caption{Total number of GKP qubits needed for a single protocol run across all the repeater and end-stations that minimises the cost function $C$ for different hardware parameters.}
\label{fig:endtoendresources}
\end{figure}

In FIG.~\ref{fig:endtoendresources} we plot the corresponding end-to-end resources. We see that for smaller distances all the parameter configurations require similar amount of resources. However, for larger distances the configurations with worse hardware parameters require much more resources. In particular the most dramatic increase in the required resources occurs for the worst configuration at the distance where it needs to reduce the repeater spacing to $L=0.5$ km. Overall however, we see that all curves follow a similar pattern and the difference in resources between the three hardware configurations is always less than an order of magnitude.

\begin{figure}
\includegraphics[width =\columnwidth]{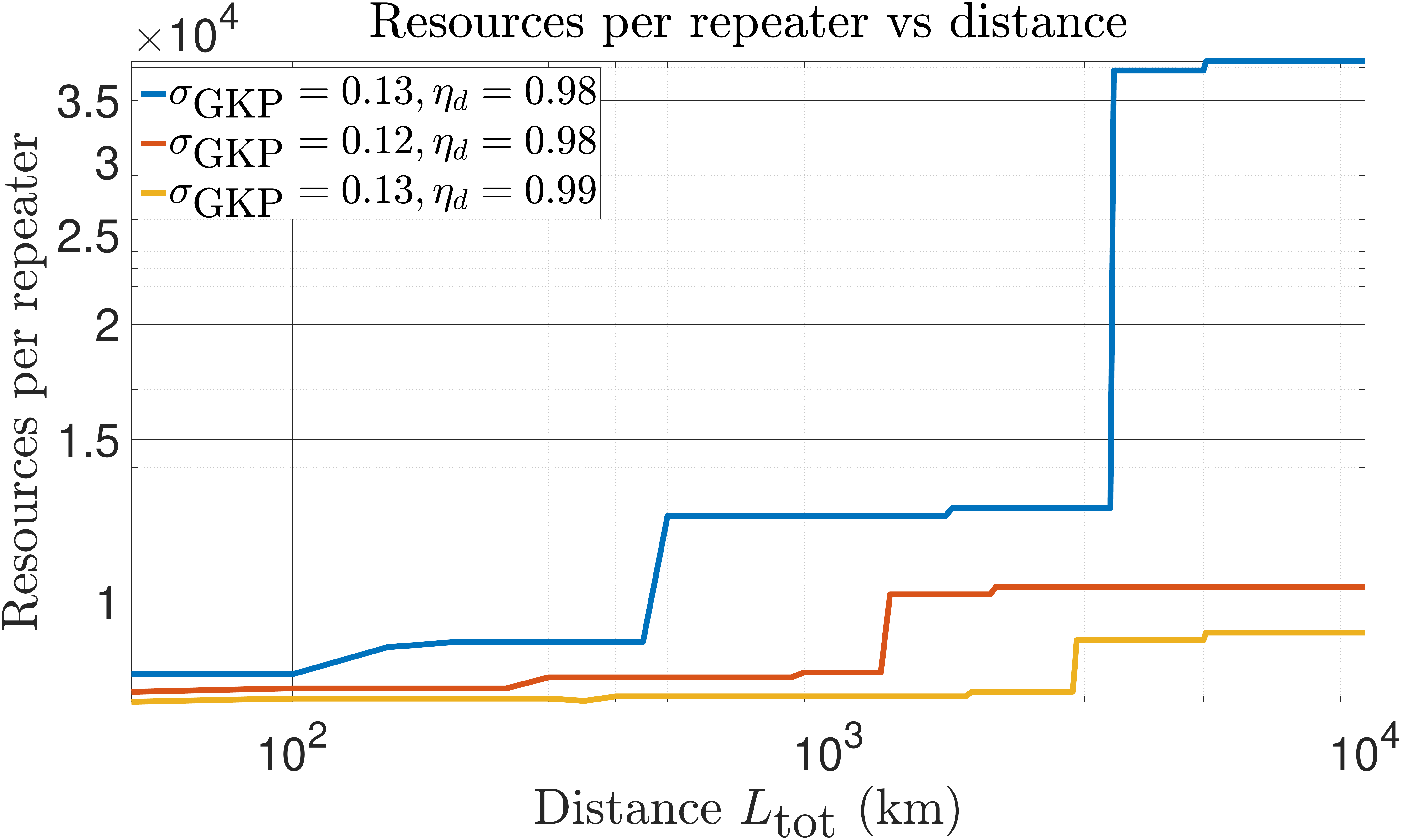}
\caption{Number of GKP qubits needed per repeater, i.e. for near-deterministically building the cube resource states (across all the repeaters) and for TEC on the inner leaves that minimises the cost function $C$ for different hardware parameters.}
\label{fig:resourcesperrepeater}
\end{figure}

Some more interesting conclusions can be drawn from FIG.~\ref{fig:resourcesperrepeater} where we plot the corresponding resources per repeater. The first observation is that the resources per repeater never decrease with distance. This tells us that in the competition between the resources in cube creation, which are larger the smaller the repeater spacing and the resources due to TEC on the inner leaves which increase with $L$, the changes in the resources due to cube generation dominate. Moreover, the most dramatic increase in resources per repeater occurs when decreasing $L$ to $L=0.5$ km for the worst configuration. Furthermore, the plot in FIG~\ref{fig:resourcesperrepeater} also shows that when the repeater spacing $L$ remains fixed, the number of resources per repeater has almost negligible dependence on the total distance. Specifically, it can be seen that over the intervals where $L$ stays fixed in FIG.~\ref{fig:resourcesperrepeater}, all the curves stay almost flat. This is due to the rapid transition of the probability $p_{\text{succ}}^{\text{res-gen}}$ as observed in FIG.~\ref{fig:diff_L_res_count} and FIG.~\ref{fig:extreme_cases_res_count}. Finally, we see in FIG.~\ref{fig:resourcesperrepeater} that the number of needed GKP qubits per repeater for our scheme is generally of the order of $10^3-10^4$. This is around four orders of magnitude less than the number of single-photons needed for the discrete-variable all-photonic scheme of~\cite{pant2017rate}.

To make a more detailed comparison with the single photon resource count of the scheme of~\cite{pant2017rate} such that we also take into account the inter repeater spacing, let us consider a specific total communication distance of 5000 km. In~\cite{pant2017rate} the authors report that they can achieve communication over that distance with $1.2 \times 10^8$ single-photons per repeater with 12411 repeaters in total. This gives us $1.5\times 10^{12}$ single photons needed in total for a single protocol run. On the other hand we see in FIG.~\ref{fig:endtoendresources} that for 5000 km our scheme requires of the order of $10^7-10^8$ GKP qubits in total per protocol run. We also refer the reader to the discussion in Section~\ref{sec:GBS} where we estimate the resources in terms of single-mode squeezers needed to generate this number of GKP qubits.

\begin{figure}
\includegraphics[width =\columnwidth]{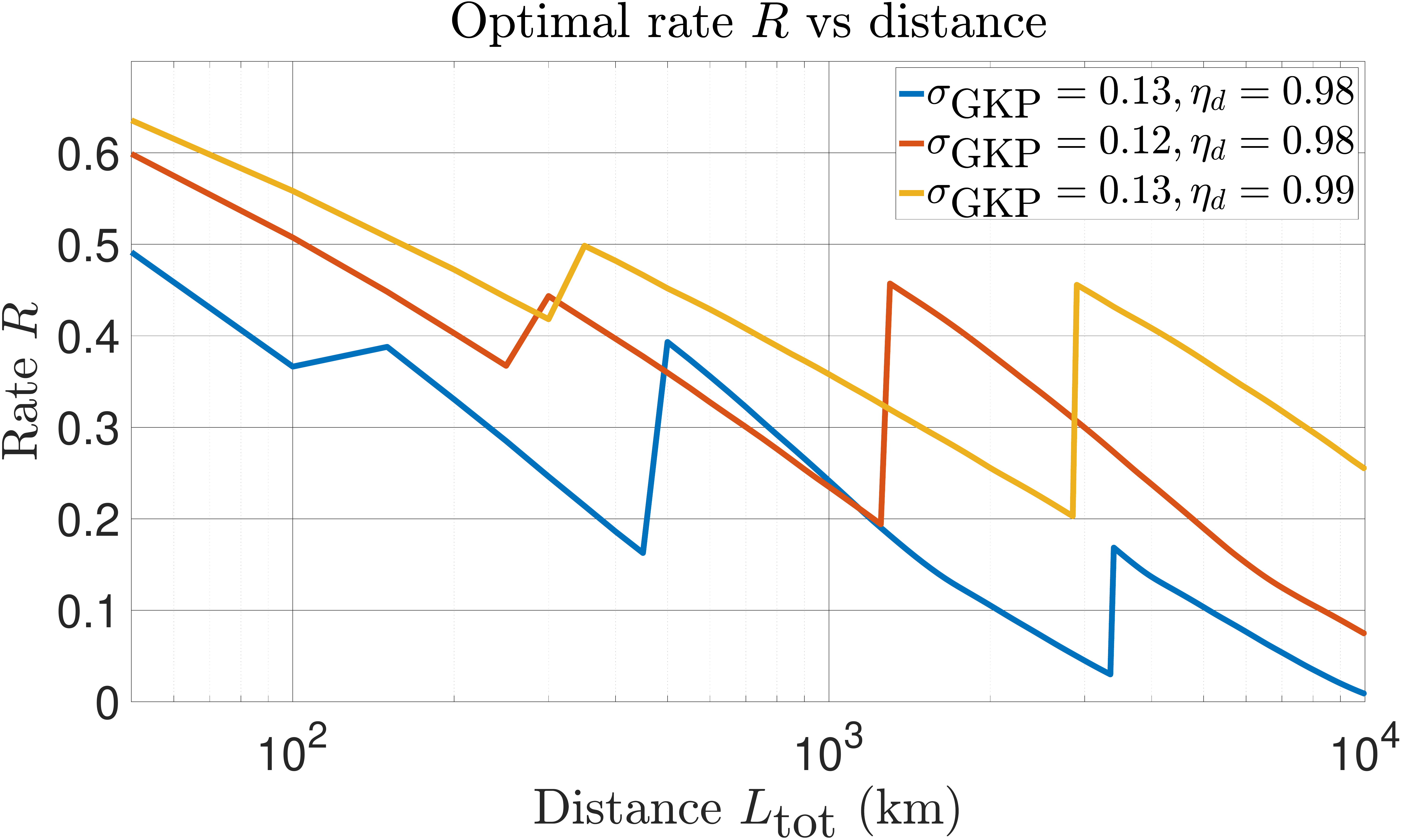}
\caption{Optimised rate $R$ that minimises the cost function $C$ for different hardware parameters.}
\label{fig:optimisedrate}
\end{figure}

Moreover, with the mentioned number of single-photons, the number of secret bits per protocol run that the scheme of~\cite{pant2017rate} can generate drops to 0.14 for 5000 km. On the other hand, we see in FIG.~\ref{fig:optimisedrate} where we plot the achievable rate vs communication distance for the optimised scenario of our scheme, that our scheme is able to generate around 2-8 secret-bits per protocol run for that distance (rate $R$ multiplied by $k=20$ modes per protocol run). In FIG.~\ref{fig:optimisedrate} we also see that for the third weakest configuration the optimised rate decays much faster than for the other two and the switches to denser repeater placing occur for much smaller distances than for the other hardware parameter configurations.

\section{Discussion}
\label{sec:discussion}

In this section we propose an analytical model for the logical error rate of our scheme, we discuss how we envision to generate the optical GKP qubits used in our scheme, we compare the performance of our scheme with other repeater schemes and finally we discuss the resource requirements on the end nodes of Alice and Bob depending on whether the intended task is remote entanglement or secret key generation.

\subsection{Analytical model of the error rate}
\label{sec:AnalyticalModel}

One can see in FIG.~\ref{fig:ratevsdist} that the rate curves corresponding to $L=0.5$ km all appear to have a similar trend. This raises the question whether the effect of changing the amount of GKP squeezing or the efficiency of the homodyne detection on the rate could possibly be modeled just by shifting these horizontally. In this section we attempt to provide a simple analytical error model for our repeater scheme that could explain this behaviour visible in FIG.~\ref{fig:ratevsdist}.

Our repeater scheme has three main sources of errors. These are errors arising during resource state preparation, errors on the inner leaves during storage and errors on the outer links from the channel transmission. As explained, the errors arising at the stage of resource state generation are suppressed below the level of the leading error by using a suitable discard window $v$ and hence have a negligible effect on the rate. The errors due to the outer leaves are most challenging to model analytically as that would require us to model the effect of the ranking, i.e. be able to establish an error probability conditioned on being ranked as i'th link among $k$ multiplexed links. However, we expect that the dominant contribution to the rate will come from the best few links for which the dominant source of error is the inner leaf error. Therefore here we attempt to model and approximate the total error rate of our scheme by just considering the errors arising on the inner leaves.

These inner leaf errors are themselves not easy to model analytically, because the analog information used for the [[7,1,3]] code correction enables us to probabilistically correct most of single and two-qubit errors, but the probability of correcting two-qubit errors seems to have a non-trivial relation to the actual amount of noise in the channel.

Nevertheless, we can still attempt to construct a simple approximate error model as follows. Let $\text{erfc}(x)$ be the complementary error function defined as:
\begin{equation}
    \text{erfc}(x) = 1- \frac{2}{\sqrt{\pi}}\int_0^x e^{-t^2} dt \approx \frac{e^{-x^2}}{x\sqrt{\pi}}
\end{equation}
where the last approximation is valid for large $x$. In our case we can approximate the probability of a logical GKP error when no post-selection is used as:
\begin{equation}
    E_1(v=0;\sigma^2) \approx \text{erfc}\left(\frac{\sqrt{\pi}}{2\sqrt{2}\sigma}\right) \approx \frac{2\sqrt{2}\sigma}{\pi}e^{-\left(\frac{\sqrt{\pi}}{2\sqrt{2}\sigma}\right)^2} \, .
\end{equation}

Although in our case $\frac{\sqrt{\pi}}{2\sqrt{2}\sigma}$ will not necessarily be large, for our estimation purposes we will only consider this leading order term. Moreover, in our approximate error model we will assume that the error rate is the same for both $X$ and $Z$ errors.

In all the considered configurations we perform inner leaf GKP TEC every 250 m and then we perform the final GKP correction through a BSM and CC-amplification on the two GKP qubits from two different resource cubes. Hence, taking into account that there is a single [[7,1,3]]-code correction per two [[7,1,3]]-code encoded qubits, the total number of GKP corrections preceding the [[7,1,3]]-code correction is given by $2 \times (4L-1) + 1 = 8L-1$.

The [[7,1,3]]-code correction with GKP analog information generally suppresses almost all single-qubit errors and most of two-qubit errors as already observed in~\cite{rozpkedek2021quantum}. Nevertheless, our investigation suggests that the leading order errors are still the uncorrected two-qubit errors. There are 21 possible two-qubit errors. The complex behaviour of the correction of these two-qubit errors using analog information is reflected in the fact that the fraction of these errors that can be corrected depends itself on the physical error probability as well as the amount of the analog information available. Based on our investigation we approximate the fraction of uncorrectable two-qubit errors for a noise process with error rate $p$ as $p^{f(p,\lambda)}$, where $\lambda$ denotes the amount of available GKP analog information. Since the leading order errors are still some of the uncorrected two-qubit errors, we have that $f(p,\lambda) \in (0,1)$. In fact here for simplicity we will take $f(p,\lambda)$ to be a constant. 

\begin{figure}
\includegraphics[width =\columnwidth]{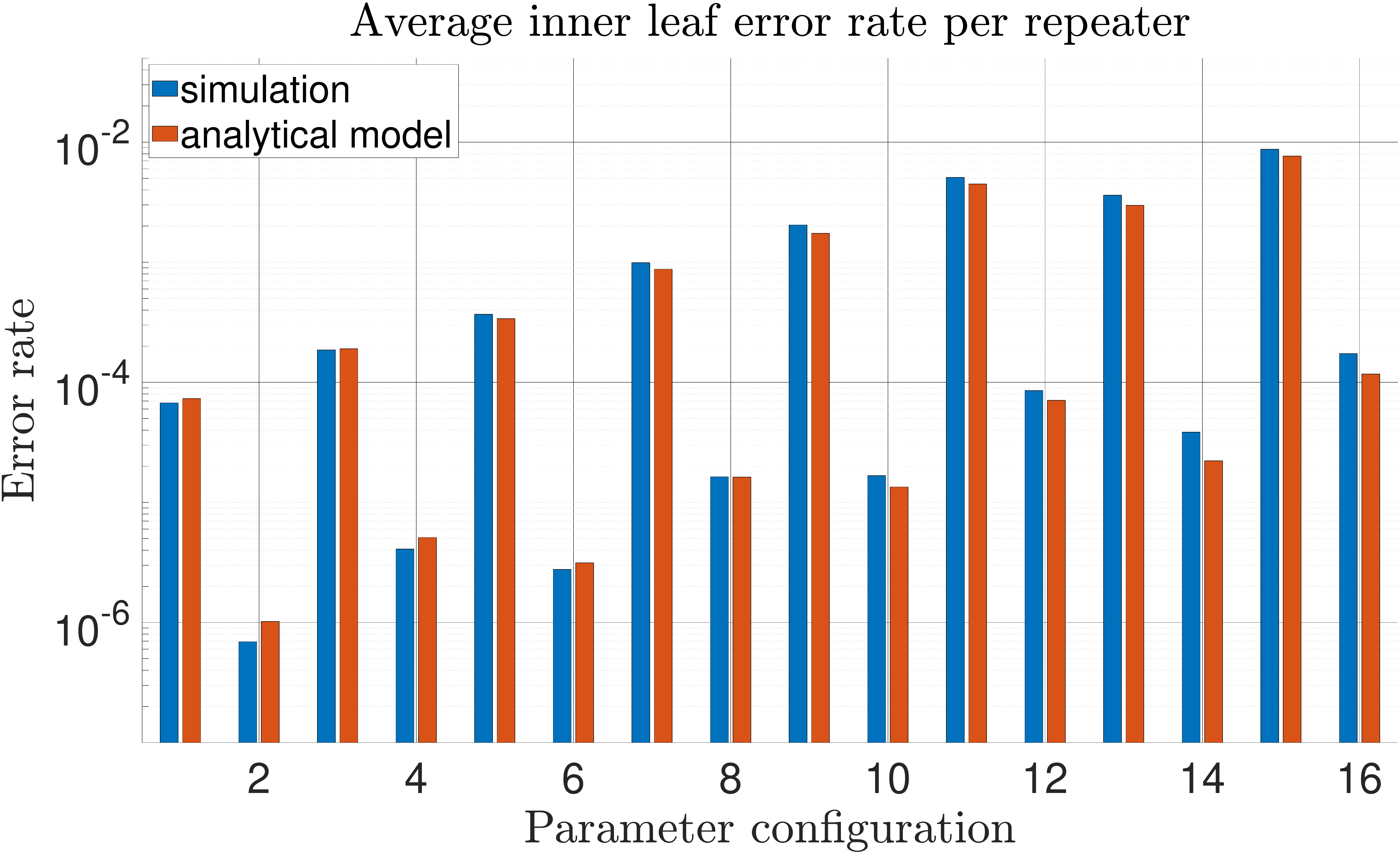}
\caption{Average error rates per a pair of logical inner leaf qubits per repeater obtained through our analytical model and through the simulation. The relevant parameter configurations are listed in Table~\ref{tab:tableParamConfigModelVsSim}. For the analytical model the coefficient $c$ from Eq.~\eqref{eq:errorRatePerRepeaterModel} is set to $c = 2.45$ for all configurations.}
\label{fig:ErrorSimVSModel}
\end{figure}

Putting this together, we can write our approximation for a logical error probability on the pair of inner-leaf logical qubits for each of the $X$ and $Z$ errors for a single repeater as:
\begin{equation}
    \tilde{p}_{\text{error}} =21\left((8L-1) \frac{2\sqrt{2}\sigma}{\pi}e^{-\left(\frac{\sqrt{\pi}}{2\sqrt{2}\sigma}\right)^2}\right)^c \, ,
    \label{eq:errorRatePerRepeaterModel}
\end{equation}
where now $c$ is a constant such that $c \in (2,3)$. The final question is what value of $\sigma$ we should use in our approximation. Here we want to use a single value of $\sigma$ for all the channels between all the inner-leaf GKP corrections. Based on the error channels for which we do not use post-selection from Table~\ref{tab:tableInnerLeaves} in Appendix~\ref{sec:ErrChforInnerLeaves} and including averaging over $X$ and $Z$ channels, the average value of $\sigma$ can be approximated as:
\begin{equation}
    \sigma = \sqrt{a \times \sigma_{\text{GKP}}^2 + 1-\eta + 1-\eta_d} \, ,
\end{equation}
where $\eta = e^{-\frac{1}{4 \times L_{\text{att}}}}$. Here we have used an approximation that $(1-\eta_d)/\eta_d \approx 1-\eta_d$ and $(1-\eta\eta_d)/(\eta\eta_d) \approx 1-\eta\eta_d \approx 1-\eta + 1-\eta_d$ which holds for $\eta$ and $\eta_d$ being close to one. Moreover, the coefficient $a$ corresponds to the average weight of the $\sigma_{\text{GKP}}^2$ term over all the relevant channels and is given by:
\begin{equation}
    a = \frac{3 \times 2 + 2  \times  (4 + 4 \times (4L-2))}{2 + (4 + 4 \times (4L-2))} = \frac{32L - 2}{16L-2} \, .
\end{equation}
We test our numerical model against a simulation data for a couple of configurations. Specifically, we only consider repeater spacing up to $L = 2.5$ km since for these shorter spacing the inner leaf information which we model here is the leading order error for at least the first few best ranked links. For the simulation data we consider the case without the inner leaf information, which as we have seen is only helpful in the regime of low rates. Moreover, we only consider the error rate for the single best of the $k=20$ ranked links as for that link the outer leaf error does not contribute to the overall error. Finally we average the error rate over the $X$ and $Z$ error. The results are presented in FIG.~\ref{fig:ErrorSimVSModel} where the parameter configurations are labelled in Table~\ref{tab:tableParamConfigModelVsSim}. The data from the analytical model have been obtained by setting $c = 2.45$. We see a relatively good agreement between the simulation data and the model, though the visible deviations between the two suggest that a single value of the parameter $c$ for all these configurations might be an oversimplification. Nevertheless, the fact that the model is in principle able to reproduce the simulated error rates shows that it should provide a reliable performance estimate of the repeater protocol of ~\cite{rozpkedek2021quantum}, since all the noise processes and EC operations of that scheme are effectively the same as the corresponding operations on the inner leaf qubits for our scheme.

\begin{table}
\begin{tabular}{ |p{2cm}|p{2cm}|p{2cm}|p{2cm}|}
 \hline
 \multicolumn{4}{|c|}{Parameter configurations for data in FIG.~\ref{fig:ErrorSimVSModel}} \\
 \hline
 Parameter configuration& squeezing $\sigma_{\text{GKP}}$& homodyne efficiency $\eta_d$ & repeater separation $L$\\
   \hline
 1 & 0.13 & 0.97 & 0.5 km \\
 2 & 0.13 & 0.99 & 0.5 km\\
 3 & 0.14 & 0.97 & 0.5 km\\
 4 & 0.14 & 0.99 & 0.5 km\\
 5 & 0.13 & 0.97 & 1 km\\
 6 & 0.13 & 0.99 & 1 km\\
 7 & 0.14 & 0.97 & 1 km\\
 8 & 0.14 & 0.99 & 1 km\\
 9 & 0.13 & 0.97 & 2 km\\
 10 & 0.13 & 0.99 & 2 km\\
 11 & 0.14 & 0.97 & 2 km\\
 12 & 0.14 & 0.99 & 2 km\\
 13 & 0.13 & 0.97 & 2.5 km\\
 14 & 0.13 & 0.99 & 2.5 km\\
 15 & 0.14 & 0.97 & 2.5 km\\
 16 & 0.14 & 0.99 & 2.5 km\\
  \hline
 \end{tabular}
\caption{Parameter configurations for data presented in FIG.~\ref{fig:ErrorSimVSModel}.}
\label{tab:tableParamConfigModelVsSim}
\end{table}

The next question is whether one can use just the error model for the inner leaves to try to capture the performance of the first few best ranked links. Here we investigate this question by making an attempt to interpret our numerical simulation data, which includes the effect of the noise on both the inner and outer leaves, using our inner leaf error model. Specifically, the total end-to-end error rate for $X$ and $Z$ according to our model is given by $e = \frac{L_{\text{tot}}}{L}\tilde{p}_{\text{error}}$. Hence, we could convert the rate vs distance plots from the previous section into rate vs total error rate plots by rescaling the curves horizontally through multiplication of the total distance by $\frac{\tilde{p}_{\text{error}}}{L}$. If the performance for all the considered parameter configurations was to be affected mainly by the noise on the inner leaves we would expect the curves plotted as a function of a universal error rate to have a significant level of overlap.

\begin{figure}
\includegraphics[width =\columnwidth]{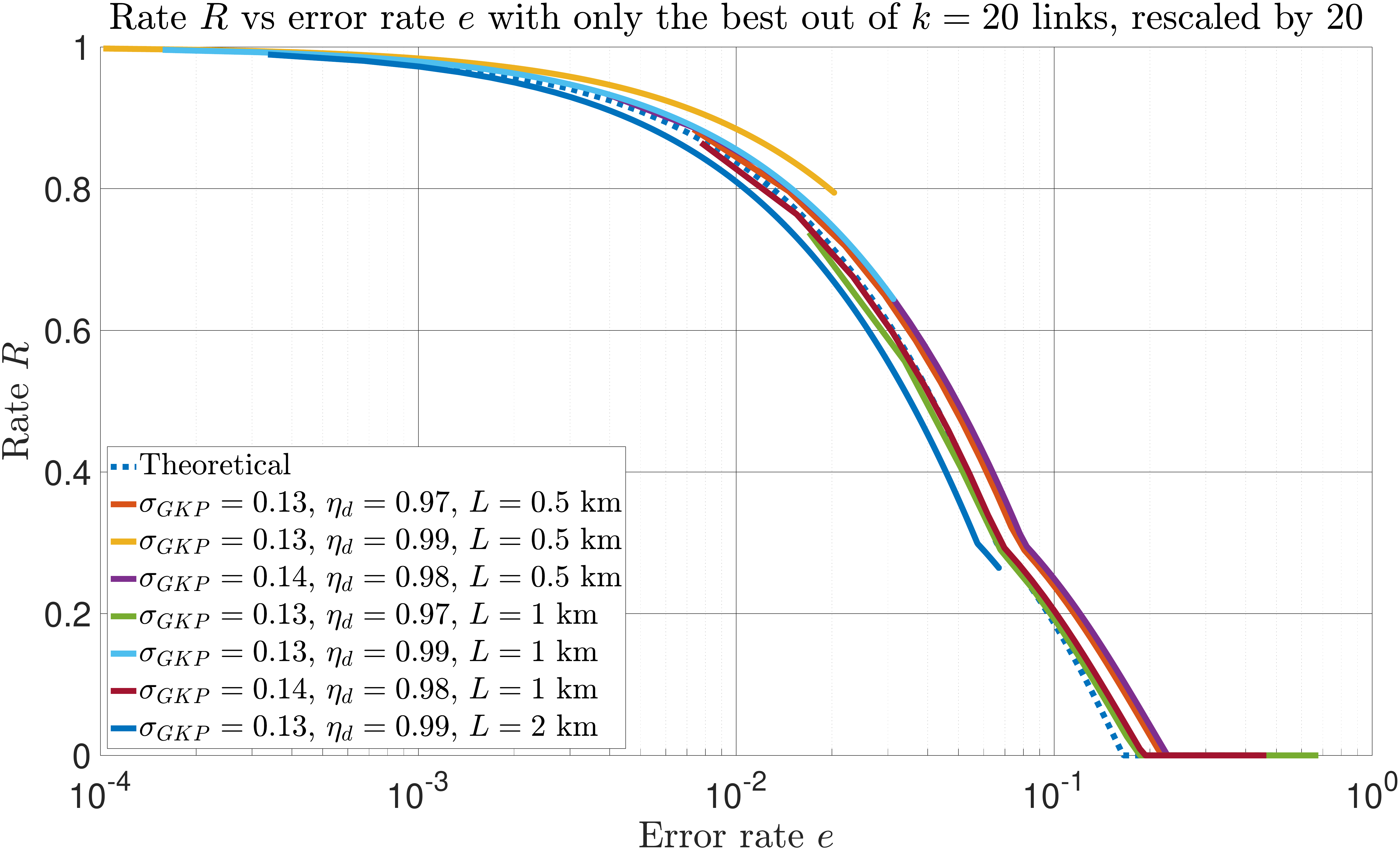}
\caption{Rate $R$ vs. estimated error rate $e$ for various parameter configurations when only the single best out of $k=20$ outer links is post-selected. Due to that post-selection, the rate $R$ would be upper bounded by $1/20$ so we rescale it here by a factor of 20 so that it is again upper bounded by one. The data for rate $R$ are based on the simulation but the corresponding error rate $e$ is based on our analytical model that includes only the errors of the inner leaves. Additionally the dashed curve marked as ``theoretical'' describes the rate that truly corresponds to the given error rate (but without any post-selection). For the analytical model the coefficient $c$ from Eq.~\eqref{eq:errorRatePerRepeaterModel} is set to $c = 2.45$ for all configurations.}
\label{fig:RateVsErrorRateOnlyBestLink}
\end{figure}

We start by considering the scenario when we always post-select only the best outer link out of the $k=20$ links. We have already seen in FIG.~\ref{fig:ErrorSimVSModel} that our model reliably predicts the error rate for this scenario when the outer link does not contribute to the total error. Therefore we expect all the curves on our performance plot for this case to strongly overlap. We plot the results in FIG.~\ref{fig:RateVsErrorRateOnlyBestLink} for various parameter configurations, where for all configurations we keep the coefficient $c$ from Eq.~\eqref{eq:errorRatePerRepeaterModel} set to $c= 2.45$. For clarity we rescale the rate by 20 on the plot so that it is still upper bounded by 1 rather than 1/20. Additionally we also plot the rate $R$ that truly corresponds to the given error rate $e$, i.e. it is the rate obtained by directly substituting the error rate $e$ into the rate formula given in Appendix~\ref{sec:PerformanceMetricsFormulas}.
We see that the curves overlap very well with each other and they also overlap closely with the dashed blue curve as expected. Hence our model can indeed reliably predict the behaviour when only the best outer link is post-selected. When looking at the curves in more detail, we see that the curves that are shifted to the right of the blue dashed curve correspond to the configurations for which the analytical model overestimates the error in FIG.~\ref{fig:ErrorSimVSModel} (though note that not all parameter configurations shown in FIG.~\ref{fig:RateVsErrorRateOnlyBestLink} are also shown in FIG.~\ref{fig:ErrorSimVSModel}). Specifically the yellow curve corresponds to the parameter configuration 2 for which the relative difference between the error due to the analytical model and due to the simulation is very large. The opposite is true for the dark blue curve corresponding to the parameter configuration 10 for which the analytical model underestimates the error. Hence that curve is shifted significantly to the left of the blue dashed curve. We expect that the curves could be made to overlap even better if we allowed to vary the parameter $c$ between the different curves. Finally, we note that the kink occurring around $e = 0.07$ is due to the fact that our rate $R$ corresponds to the maximum of two distinct protocols, see Appendix~\ref{sec:PerformanceMetricsFormulas}.

\begin{figure}
\includegraphics[width =\columnwidth]{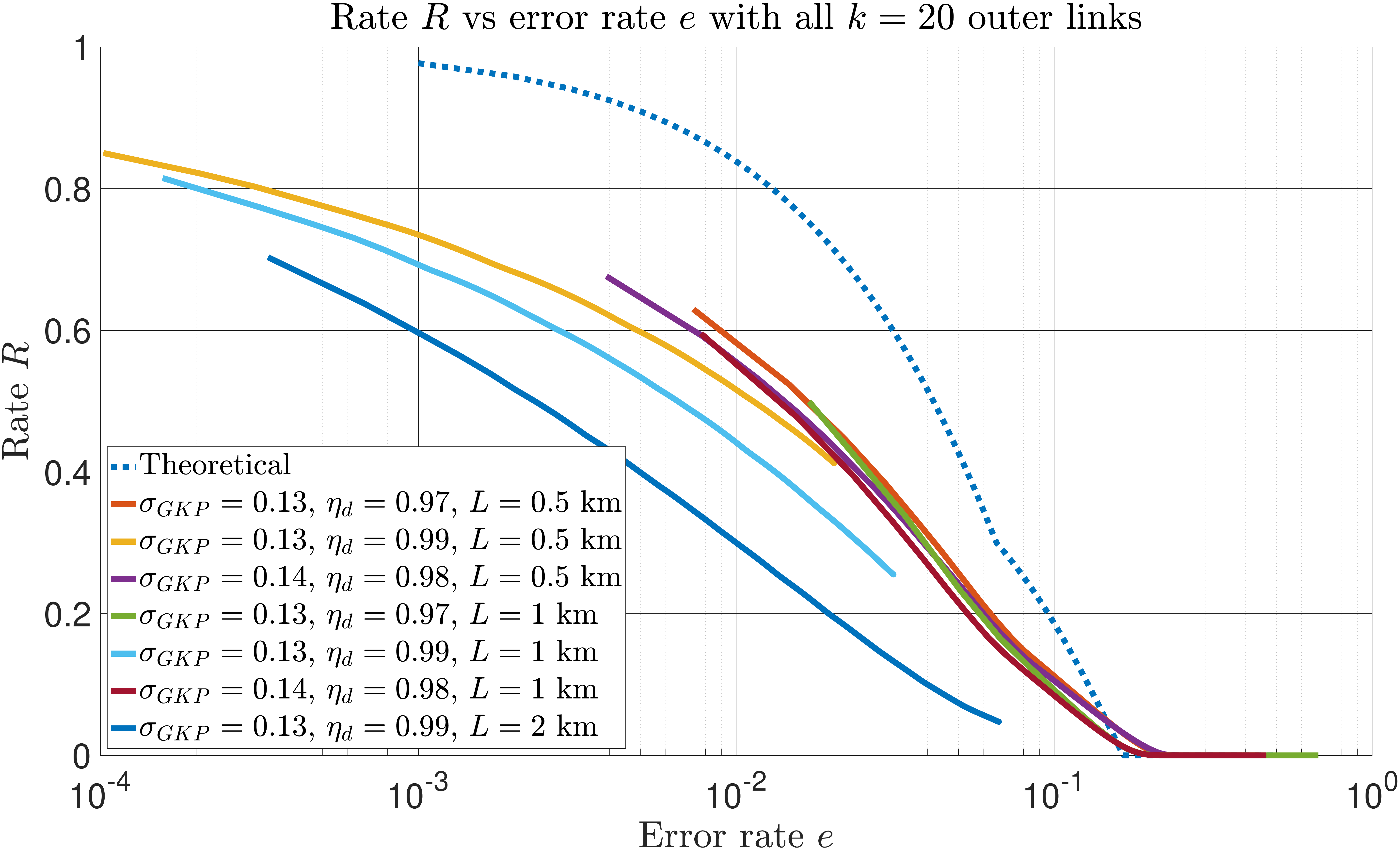}
\caption{Rate $R$ vs. estimated error rate $e$ for various parameter configurations for all the $k=20$ multiplexed links. The data for rate $R$ are based on the simulation but the corresponding error rate $e$ is based on our analytical model that includes only the errors of the inner leaves. Additionally the dashed curve marked as ``theoretical'' describes the rate that truly corresponds to the given error rate. For the analytical model the coefficient $c$ from Eq.~\eqref{eq:errorRatePerRepeaterModel} is set to $c = 2.45$ for all configurations.}
\label{fig:RateVsErrorRate}
\end{figure}

Now, if we start including more outer links, our entanglement/secret-key rate will of course increase relative to the best-link rate upper bounded by 1/20, but as the non-modeled errors of the worse links will start limiting the rate $R$, we no longer expect all the curves to overlap so well. We plot the results for the case when all the $k=20$ multiplexed links are included in FIG.~\ref{fig:RateVsErrorRate}. Indeed we see that the curves do not overlap so well anymore. There is a significant overlap of the curves with moderate homodyne detection efficiency, but the curves corresponding to $\eta_d = 0.99$ clearly do not match the other curves. Clearly the overlap of these specific curves of course only suggests that the outer leaf behaviour for these configurations is similar, but is clearly not modeled here and actually has a significant impact on the rate. This can be seen by the fact that although our secret-key/entanglement rate is now upper bounded by one, all the curves based on the model are significantly below the dashed blue curve. This is because the worse ranked outer leaves experience much higher errors than the modeled error of the inner leaves and many of them do not contribute to the performance rate $R$. Moreover, the number of them that actually contributes significantly to $R$ depends on the actual error rate for the different links. Nevertheless, the plot gives us an indication that the curve corresponding to the configuration with $L=2$ km has much smaller number of outer links with a negligible outer leaf error relative to the inner leaf error. This is because that curve is significantly below all the other curves. That manifests a significant level of outer leaf error, which, if somehow included in the model, would shift that curve to the right to match with all the other ones. All of these complex phenomena cannot be captured by our simple model that only models the effect of the errors in the inner leaves. Yet the fact that many of the plotted curves do overlap suggests that the fraction of the outer links that contribute to the overall performance rate $R$ follows a similar behaviour for many of the considered parameter configurations.

\subsection{GKP resource state preparation via Gaussian Boson Sampling}
\label{sec:GBS}

Bosonic quantum error correction is already an experimentally advanced field. This is reflected by the recent bosonic demonstrations where for the first time the lifetime of a logical qubit has been extended beyond the break-even point~\cite{ni2023beating,sivak2023real}. One of these demonstrations utilised precisely the GKP encoding~\cite{sivak2023real}. Nevertheless, the GKP qubit is still challenging to realize in practice. 
After nearly 20 years of its invention, it has been demonstrated in trapped ion motional modes~\cite{fluhmann2019encoding,de2022error} and in superconducting circuits in microwave cavities~\cite{campagne2019stabilized,sivak2023real}. 
It remains challenging to prepare the GKP qubit in the photonic domain. 
Among different possible proposals to realize GKP qubits in the photonic domain, a promising approach is via the Gaussian Boson Sampling (GBS) paradigm~\cite{GBSorig}, where squeezers, multimode linear optical interferometers and photon number detectors can be used to herald a state that is arbitrarily close to a GKP qubit state in fidelity~\cite{su2019conversion}. 
For example, a GKP qubit in the $|0\rangle$ state with GKP squeezing of $10$ dB ($\sigma=0.224$) can be generated with 99.7\% fidelity using a 3-mode GBS circuit with 3 squeezers each generating $12$ dB squeezed vacuum by heralding 5 and 7 photons in 2 out of the 3 modes, where the probability of heralding the pattern is $\approx 0.11\%$~\cite{tzitrin2020progress}. 
For a total communication distance of 5000 km and repeater hardware configuration given by GKP squeezing of $14.7$ dB and realistic homodyne detection efficiencies around $\eta_d=0.99$, for the repeater spacing that minimizes the cost function of Eq.~\ref{eq:costfunction}, our repeater scheme requires $5\times 10^7$ GKP qubits in total end-to-end for near-deterministic generation of the cube resource graph states at all the repeaters (i.e., 40 graph states of cube topology to support $k=20$ multiplexing on either elementary links associated with each repeater). 
Notwithstanding that the GKP squeezing required is higher than the example GBS-based generation scheme mentioned above, assuming that a similar success probability of heralding similarly high fidelity GKP qubits (in the $|+\rangle$ and $|\oslash\rangle$ states) may be achieved with a better GBS hardware of higher homodyne detection efficiencies, we draw a conservative estimate of the number of squeezers required to near-deterministically generate the necessary number of GKP qubits end-to-end. 
By considering multiplexed 3-mode GBS generation attempts, where each GBS succeeds in heralding the desired GKP qubit with probability $p_0\approx 10^{-3}$, we find that with $\approx 3\times 10^4$ GBS circuits, a single GKP qubit may be successfully generated with a failure probability under $10^{-13}$. If the said number of GBS circuits are dedicated for each GKP qubit generation, which amounts to $\approx 1.5\times 10^{12}$ GBS circuits, i.e., $\approx 4.5\times 10^{12}$ squeezers and $\approx 3\times 10^{12}$ photon number resolving detectors with resolution upto 7 photons, then generating all the required $5\times 10^{7}$ GKP qubits at once would succeed with a probability exceeding $1-3\times10^{-6}$.

We note that GBS is just one way of preparing the optical GKP qubits. There are many other proposals for how this task could be achieved. One possible technique would be to operate on microwave GKP qubits~\cite{campagne2019stabilized,sivak2023real} and then transduce them into the optical regime~\cite{andrews2014bidirectional, han2020cavity, zhong2020proposal, rueda2019electro, lambert2020coherent}. Another proposed strategy is to utilise interactions between light and free electrons to achieve the non-linearity needed to prepare optical GKP qubits~\cite{dahan2023creation}.

\subsection{Comparison with other repeater schemes}
\label{sec:comparison}

Here we compare our scheme to other repeater proposals separately in terms of achievable rates and achievable distances.

\subsubsection{Achievable rates}

We have seen that the rate per mode observed for some of the parameters of our scheme can stay as high as 0.7 for up to 750 km. Up to our knowledge only one other repeater scheme can achieve such a performance~\cite{fukui2021all}. That scheme also makes use of GKP qubits, yet requires significantly more squeezing. Below we provide short justifications why such high rates per mode cannot be achieved by other types of repeater schemes and discuss in more detail the other scheme that can achieve such a rate.

\begin{itemize}
    \item \textbf{Two-way repeaters based on discrete-variable dual-rail encoded photons and heralded entanglement generation.} Generally such schemes require large number of attempts in order to generate the elementary links which makes the rate per channel use and per mode small. Here we use a simple calculation to show that, fundamentally, a rate higher than 0.26 for practically relevant distances is impossible for these schemes. We note that since these schemes use two modes (such as two orthogonal polarisation or two time-bin modes), the rate per mode is upper bounded by 0.5 for these schemes. Moreover, we know that a single attempt over each elementary link will not be enough to establish end-to-end entanglement as that probability corresponds to the probability of direct transmission and decays exponentially with distance. Hence over at least one of the elementary links more than one attempt will be needed. Using more than one attempt over at least one of the elementary links can be counted as having more than one channel use over end-to-end channel since the number of channel uses can be counted as the maximum over the number of channel uses over all the elementary links, see e.g.~\cite{luong2015overcoming,rozpkedek2018parameter}. Hence, assuming perfect devices with perfect memories and assuming arbitrary close repeater placement, the probability of success in a single attempt is upper bounded by $e^{-L_{\text{tot}}/L_0}$. This means that the total rate is upper bounded by:
    \begin{equation}
    \begin{aligned}
        R &< 0.5  \times  (e^{-L_{\text{tot}}/L_0}  \times  1 + (1-e^{-L_{\text{tot}}/L_0})  \times  0.5) \\ &< 0.5  \times  e^{-L_{\text{tot}}/L_0} + 0.25 < 0.26
    \end{aligned}    
    \end{equation}
    for distances above $L_{\text{tot}} > 90$ km. Here the first term in the sum corresponds to the case where all links succeed in the single attempt and so over one end-to-end channel use and the second term corresponds to the case where at least two attempts are needed over at least one elementary link which in the best scenario would correspond to exactly two attempts. Additionally we include a factor of 0.5 due to two modes being used for the encoding. Of course in practice these rates would be much lower as one would generally need much more than 2 attempts to generate elementary link entanglement but the above expression provides a clear upper bound to this rate. We note that this bound also applies to schemes that include any type of multiplexing for these dual-rail-encoding-based schemes. In particular it also applies to the two-way all-photonic repeaters where multiplexing is achieved through local use of the clique-clusters at the repeaters. 
    
    \item \textbf{Two-way repeaters based on discrete-variable single-rail encoded photons and heralded entanglement generation.} A promising alternative to the heralded schemes based on dual-rail encoding are the schemes based on single-rail encoding such as the single photon scheme~\cite{cabrillo1999creation, humphreys2017deterministic}. For these schemes the probability of success for the elementary link generation scales with the square root of the inter-repeater transmissivity i.e. it scales as $e^{-L/2L_0}$. However, this protocol does not produce a perfect Bell state and in order to increase the quality of the generated state one needs to decrease the emission probability which effectively reduces the probability of success of each generation round. Generally, that extra emission probability will need to be reduced below $p_{\text{em}}=0.2$ in order to extract entanglement of reasonable quality~\cite{rozpkedek2019near, humphreys2017deterministic}. This means that the rate for these schemes is also limited. We can see this as follows. Generally, we would expect repeaters to be placed at a distance at which the channel losses are at least as large as the additional reduction of the emission probability. Otherwise repeaters would be adding more effective losses rather than overcoming them. Let us consider the transmissivity of $e^{-L/2L_0} = 0.2$ for around $L=70$ km. Then considering a total distance of 140 km and two elementary links we can bound the rate as follows. With probability $(1-(1-p_{\text{em}} \times e^{-L/2L_0})^3)^2$ both links will succeed within 3 attempts and with the complementary probability at least one link will need at least 4 attempts. Hence we can upper bound the rate as:
    \begin{equation}
    \begin{aligned}
        R &< (1-(1-p_{\text{em}} \times e^{-L/2L_0})^3)^2  \times  1 \\
        &+ \left[1-(1-(1-p_{\text{em}} \times e^{-L/2L_0})^3)^2\right]  \times  0.25 \approx 0.26
    \end{aligned}    
    \end{equation}

    Here the first term assumes the optimistic scenario that within the initial three rounds we actually succeed in the first round while the second terms assumes that with the probability of needing more than three rounds, we will actually only need four rounds. This idealised case already leads to the upper bound of 0.26. Again, if the total distance is larger than 140 km the rate will need to be even smaller. 
    \item \textbf{One-way repeaters based on discrete-variable codes.} For strategies that use discrete-variable codes to overcome transmission losses, up to our knowledge most of the so-far proposed repeater schemes encode one logical qubit in large number of physical qubits, often of the order of few hundred~\cite{muralidharan2014ultrafast,borregaard2019one}. One exception to this are the schemes based on the [[4,2,2]] code, [[7,1,3]] code, [[48,6,8]] code and $[[3,1,2]]_3$ code~\cite{glaudell2016serialized,muralidharan2017overcoming,niu2022all}. For these schemes the individual physical photons are normally also considered to utilise dual-rail encoding (or three-rail encoding for the $[[3,1,2]]_3$ qutrit code). Hence, even in the case of the smallest of these codes, the [[4,2,2]] code, the rate per mode is upper bounded by 25\%. The same bound applies to the code that uses smallest number of modes and can correct photon-loss, which is the [[4,1,2]] code that can approximately correct amplitude damping noise and hence the physical qubits can be single-rail encoded in that case~\cite{leung1997approximate}. Note that the discussed limitations also apply to the concatenated-coded schemes with bosonic codes at the lower level and discrete-variable codes at the higher level, i.e. the proposed one-way schemes based on the GKP code and the [[4,1,2]] or [[7,1,3]] code of~\cite{rozpkedek2021quantum} as well as the clique-cluster based scheme with tree codes of~\cite{fukui2021all}. This again shows that we would not expect rates above 25\% for such one-way repeaters.
    \item \textbf{Two-way all-optical repeaters based on GKP Bell-pair distribution and connection using CC-amplification}
    As discussed, bosonic codes such as the GKP code can make better use of the bosonic nature of the pure-loss channel. One efficient strategy is the scheme of~\cite{fukui2021all} based on individual GKP Bell pairs and CC-amplification. This strategy is up to our knowledge one of the most efficient ways of using single mode GKP qubits without higher level error correction, that has been proposed in the context of quantum repeaters. In this scenario each repeater generates a GKP Bell-pair by preparing two qunaught GKP states (see Eq.~\eqref{eq:qunaughtstate}) and passing them through a beamsplitter. Then each half of the Bell pair is sent towards the two other repeaters in the opposite directions. In the middle between neighbouring repeaters one performs GKP BSMs. These measurements can involve post-selection based on the obtained continuous GKP syndromes and the end-to-end Bell pair is obtained only if the post-selection is passed in both quadratures and at all heralding stations in the same round. Similarly as in the case of our scheme we consider two sources of imperfection for that strategy, the finitely squeezed GKP qubits and lossy homodyne detections. We plot the rates for a few hardware parameter sets in FIG.~\ref{fig:RatesGKPBellpairscheme}, where for each distance we already perform an optimisation over the discard window and repeater spacing with $L=0.5$ km being the minimum allowed value. Here, similarly as for our scheme, $L$ is defined as the distance between the major nodes that generate GKP qubits, i.e. one needs to place additional optical BSM stations half-way in between. Not surprisingly we find that the optimal repeater spacing for all the considered parameters is 0.5 km for all the distances. By comparing these results with the $L=0.5$ curves on the plots in FIG.~\ref{fig:ratevsdistDifferentL}, we see that our scheme significantly outperforms the scheme based on single GKP Bell pairs. However, we find that this scheme can also achieve rates as high as 0.7 for up to around 750 km with the same repeater spacing as for our scheme of $L=0.5$ km. Nevertheless, this requires $\eta_d = 0.99$ and $\sigma_{\text{GKP}} = 0.105$ (16.6 dB) for that scheme, while our scheme can achieve that with $\eta_d = 0.99$ and $\sigma_{\text{GKP}} = 0.13$ (14.7 dB). Hence our scheme can achieve this performance with almost 2 dB of GKP squeezing less than the scheme of~\cite{fukui2021all}. Therefore we see that our scheme becomes especially appealing if one finds that there is a specific hardware limitation with regard to GKP squeezing, such that there is a certain maximum value that cannot be exceeded in an experiment.
\end{itemize}

\begin{figure}
\centering
\includegraphics[width =\columnwidth]{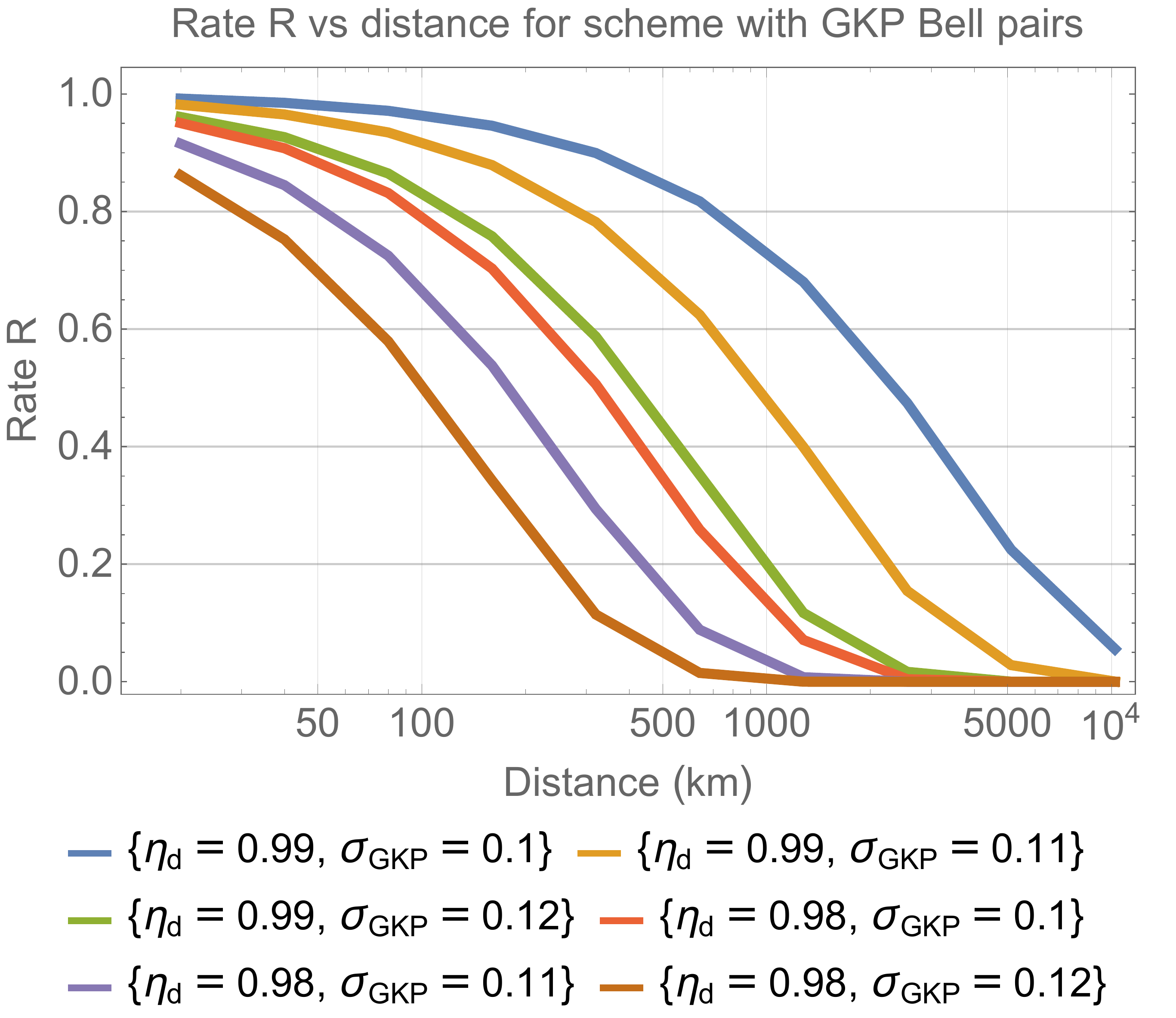}
\caption{Rates vs distance for the all-optical scheme based on distribution of GKP Bell pairs introduced in~\cite{fukui2021all}. The rates are optimised over the discard window.}
\label{fig:RatesGKPBellpairscheme}
\end{figure}

Finally, we note that our discussion above has been focused only on the achievable rate per optical mode. From practical perspective another figure of interest will be throughput, i.e. the number of generated secret-key bits/EPR pairs per second. Here we note that our scheme shares the feature of all all-photonic schemes in that the repetition rate of the protocol is only limited by the repetition rate of the sources. This means that if we can deterministically generate GKP qubits at 1 MHz rate, with $k=20$ we can actually achieve a throughput of 20 entangled pairs/secret bits per second even at distances where $R$ is of the order of $10^{-6}$.

\subsubsection{Achievable distances}
\begin{figure*}
\centering
\includegraphics[width =\textwidth]{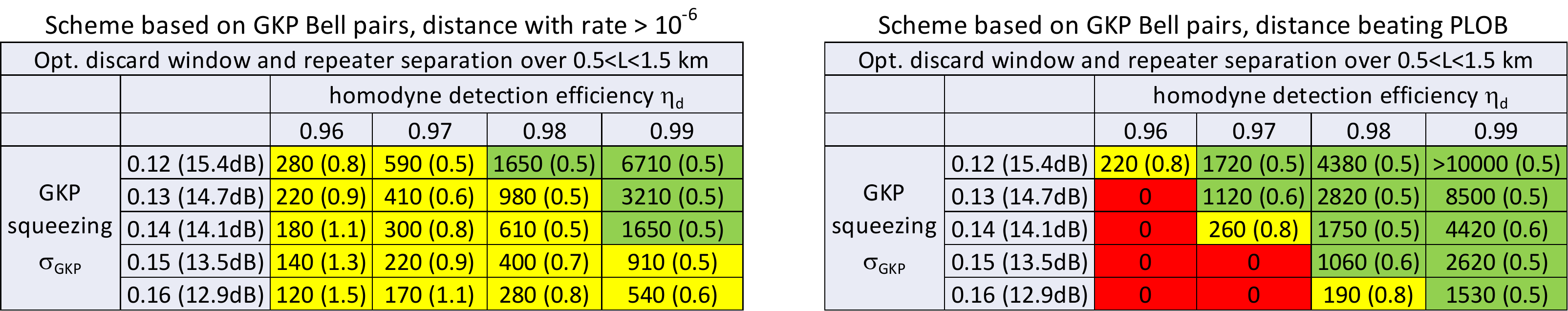}
\caption{Achievable distances for the two-way all-optical repeaters based on GKP Bell-pair distribution and connection using CC-amplification proposed in~\cite{fukui2021all}.}
\label{fig:FukuiAchievableDistances}
\end{figure*}

We can also compare the performance of our scheme in terms of achievable distances. Since hardware parameters for the heralded repeaters based on discrete-variable dual- and single-rail encoding and for the repeaters based on discrete-variable codes are very different than for the GKP-based repeaters, here we will focus only on the comparison with specific other GKP-based repeater schemes:
\begin{itemize}
    \item \textbf{Two-way all-optical repeaters based on GKP Bell-pair distribution and connection using CC-amplification.} We show the achievable distances for that scheme in FIG~\ref{fig:FukuiAchievableDistances} where we perform optimisation over both the discard window and the repeater spacing. We see that for most considered hardware parameters our scheme achieves much larger distances than the simpler GKP repeater chain.

    \item \textbf{One-way repeaters based on concatenated GKP and [[7,1,3]] code.} In~\cite{rozpkedek2021quantum} achievable distances for this scheme are quoted depending on the amount of GKP squeezing and repeater coupling efficiency. While the considered threshold rate there is $10^{-2}$, we have found that increasing the threshold to the value of $10^{-6}$ considered here increases the achievable distances quoted there by about 20-25\%. Moreover, we need to take into account two more differences between the conditions of the two schemes. Firstly, for our new scheme we model lossy homodyne detection, while for the scheme of~\cite{rozpkedek2021quantum} only in-out repeater coupling efficiency is considered with perfect homodyne detection. Clearly the noise model for our new scheme is more stringent as both schemes require large number of homodyne measurements per repeater while in-out coupling efficiency is applied only once per repeater. Secondly the scheme of~\cite{rozpkedek2021quantum} considers the more advanced advantage distillation procedure of~\cite{watanabe2007key} which produces more secret key than the protocol considered here. The reason for using a weaker advantage distillation here is that it is not known whether it is also possible to extract entanglement at the rate corresponding to this more advanced advantage distillation. Taking these into account and comparing achievable distances of our new scheme with $\eta_d = 0.99$ with the achievable distances of the scheme of~\cite{rozpkedek2021quantum} based on the concatenation of the GKP code with the [[7,1,3]] code with repeater in-out coupling efficiency of $\eta_0 = 0.99$, we see that both schemes can achieve similar distances, but our new scheme can achieve them with GKP squeezing reduced by more than 1 dB. The fact that our new scheme has better performance in terms of achievable distances relative to the scheme of~\cite{rozpkedek2021quantum} might seem surprising. After all for that scheme, the multi-qubit Type A repeaters are $L$ distance apart (with $L$ being at least 250 m) and we can place Type B GKP repeaters as dense as every 250 m in between. Then the total elementary link involves transmission of a concatenated-coded state over distance $L$ with the help of those GKP repeaters in between. On the other hand, for our new scheme with repeaters being separated by distance $L$ (where now the smallest considered value is $L=0.5$ km) for a single elementary link we have both the outer leaf qubits and the inner leaf qubits. Specifically, for each pair of the connected outer leaf qubits we then also have the two concatenated-coded inner leaf qubits that are then interfered for error correction through a logical BSM. Hence there seems to be much more losses contributing to the single elementary link of distance $L$ in this case. This is because even if we choose the best ranked outer link, the BSM on the inner leaves effectively performs [[7,1,3]] code correction on a qubit that was stored for distance $2L$ (again with intermediate GKP corrections every 250 m). Hence the total amount of contributed loss on the concatenated-coded qubits corresponds to a channel at least twice longer than in the case of the scheme of~\cite{rozpkedek2021quantum}. However, we believe that the key advantage of our new scheme is that the stabilizers of the [[7,1,3]] code are now measured destructively during BSM, without the need for any ancilla GKP qubits. This contrasts with the way the higher level stabilizers are measured in the scheme of~\cite{rozpkedek2021quantum}. Despite the use of additional techniques such as implementing additional intermediate GKP corrections and repeating the measurement of each stabilizer in~\cite{rozpkedek2021quantum} it seems that obtaining the higher level stabilizer information through a BSM is much more reliable. This also suggests that the stabilizer measurements in the scheme of~\cite{rozpkedek2021quantum} could be improved, e.g.~by using TEC both for the inner and outer code as discussed in~\cite{schmidt2022quantum}. However, that would require us to be able to prepare a logical Bell pair on the [[7,1,3]]-code level, requiring a resource of 14 entangled GKP qubits, so significantly larger than our cube resource of 8 entangled GKP qubits for the proposed scheme. An additional difference between our new scheme and the scheme of~\cite{rozpkedek2021quantum} is that while the latter requires much less resources per repeater, the GKP repeaters need to be placed very densely. In our new scheme the inner leaves are stored locally so we can keep performing GKP correction on them every 250 m even if we would like to place repeaters much further apart. 
    
    \item \textbf{All-optical repeaters based on GKP qubits with clique clusters.} In~\cite{fukui2021all} the authors proposed how to modify the discrete-variable all-photonic scheme of Azuma et al. where the individual qubits are replaced with GKP qubits. The inner leaves are then encoded in a concatenation of the GKP and the tree code. In their scheme, similarly as in the Azuma scheme, the inner leaves are sent out together with the outer leaves. This means that the inner leaves experience overall the losses corresponding to only half of the distance as in our scheme (since if the inner leaves are also sent to the BSM stations, then they can be measured immediately after the BSM without the need for the classical information to come back). However, this also means that it is necessary to pre-generate the clique-cluster even when using GKP qubits on the first level and the inner leaves cannot be periodically corrected which is easy to do in our case where the inner leaves are retained at the repeater nodes. In~\cite{fukui2021all} the authors plot the achievable secret-key rate vs distance for different fixed numbers of repeaters for 15 dB of GKP squeezing and with perfect homodyne detection. This is different from our analysis where we fix the inter-repeater spacing. However, we can still make a comparison by noting that for 500 repeaters the rate of the scheme of~\cite{fukui2021all} drops to zero just below 2100 km. For our scheme with 14.7 dB of squeezing and perfect homodyne detection for $L=5$ km we observe a rate of 0.0014 at 2500 km. This shows us that with 500 repeaters our scheme can definitely achieve a distance larger than 2500 km, hence outperforming the scheme of~\cite{fukui2021all}. 
\end{itemize}

\subsection{Requirements on the nodes of Alice and Bob}
\label{sec:RequirementsEndNodes}

We have already described that the simulation output is the estimate of the error over a single elementary segment and this elementary error can then be put together to establish the end-to-end error probability. This is done by effectively estimating a probability of odd number of $X$ and $Z$ error flips over a chain of $n$ repeaters, taking into account the inner leaf binning information and the outer leaf ranking information. Here we note that every repeater is the same and generates a pair of 8-qubit cubes for each multiplexing level. Specifically the first cube involves an outer leaf qubit sent to the left and the inner leaf GKP + [[7,1,3]] code qubit stored in the repeater. The second cube involves the same leaf structure but now for sending to the right. However, Alice's and Bob's system can in fact be much simpler because they only need to connect with links on one side.

Here we argue that the performance evaluated through our simulation that assumes that all nodes are exactly the same is still a valid lower bound for the real scenario involving Alice's and Bob's modified setups when considering secret-key generation. For it to represent a reliable lower bound on remote entanglement generation we need to impose additional assumptions on the additional memory capabilities of Alice and Bob.

Firstly, we note that as described before, each simulated link can be interpreted as describing errors in a single repeater together with a single outer link, say, to the right of that repeater. We note that the total number of repeaters is one less than the number of links as mentioned in Section~\ref{sec:OptResourcesResults} where we set $n_{\text{rep}} = \frac{L_{\text{tot}}}{L}-1$ for the calculation of the number of required resources. Moreover, we additionally consider there the non-repeater nodes of Alice and Bob. On the other hand the number of repeaters $n$ that we use for the performance simulation, as defined in Section~\ref{sec:performanceMetrics} is actually made equal to the number of links and is given by $n = n_{\text{rep}} + 1$. Here we argue that using the data from a single simulated link to establish performance over $n$ such homogeneous hops can be seen as a lower bound on the actual scenario where we have $n$ links, $n_{\text{rep}}$ repeaters and two additional stations of Alice and Bob. We will argue here that for the purpose of QKD each of the stations of Alice and Bob do not require to produce any inner leaf qubits and so do not introduce any noise at all apart from the finite squeezing of the outer leaf GKP qubit. For the purpose of entanglement generation we will make additional assumptions about quantum memory capabilities of Alice's and Bob's stations. Within these frameworks, the errors introduced by $n_{\text{rep}}$ repeaters together with Alice's and Bob's stations can be upper bounded by the errors introduced by $n$ repeaters for both QKD and entanglement generation. Let us examine these cases separately for key and entanglement generation.

\subsubsection{Key generation}

\begin{figure*}
\centering
\includegraphics[trim={0 0 0 0}, clip, width = \textwidth]{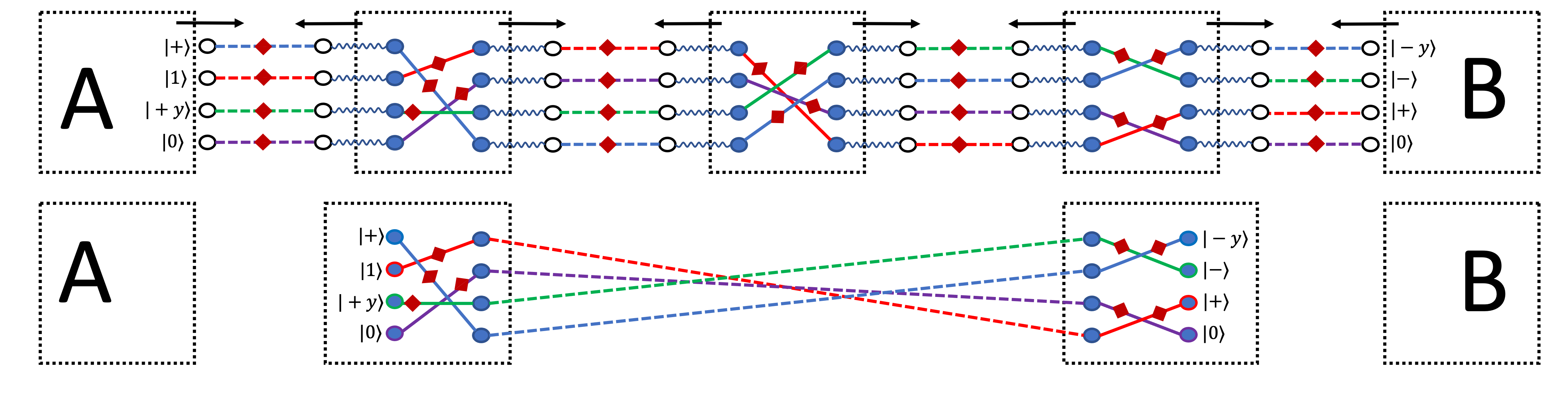}
\caption{Repeater setup for QKD with minimal resource on the sides of Alice and Bob. The scheme can effectively be viewed as Measurement Device Independent (MDI) QKD setup where Alice and Bob only require sources of GKP qubits and do not need to perform any measurements. Specifically Alice and Bob send GKP qubits in protocol basis states towards the neighbouring BSM stations where they become effectively teleported to their neighbouring repeaters. These two repeaters effectively establish entanglement with each other through the repeater chain. Moreover, they hold ranking information about the long-distance links that they share and about the elementary-link teleported states from Alice and Bob. They then perform BSMs between the teleported states of Alice and Bob and their two respective qubits of the long distance entangled pair according to the ranking information. This effectively implements a BSM between the two states sent by Alice and Bob as in MDI QKD protocols.}
\label{fig:Rep_QKD}
\end{figure*}

In the case of key generation the required capabilities of Alice's and Bob's stations are very limited. Specifically it is sufficient for Alice and Bob to just send the individual outer leaf GKP qubits in one of the six-state QKD protocol basis states. One can then think that effectively through the whole repeater chain the two most outer repeaters generate $k$ ranked Bell pairs that they share between each other, see FIG.~\ref{fig:Rep_QKD}. At the same time they locally generate Bell pairs whose outer leaves are sent for a BSM with the qubits sent by Alice and Bob. These BSMs effectively teleport the qubits of Alice and Bob into those outermost repeaters. The teleportation BSM has also generated analog information which ranks the teleported states according to their expected quality and hence determines which of the teleported qubits should be connected to which of the long distance links. This connection of the qubits of Alice and Bob through BSMs to a long-distance entangled resource effectively realises a BSM between the qubits of Alice and Bob. Hence the whole setup can be seen as the Measurement Device Independent (MDI) QKD implementation~\cite{lo2012measurement} where Alice and Bob need to trust their preparation of the six-state protocol basis states, but all the resources generated inside the repeaters as well as all the measurements taking place in this protocol can be untrusted. 

As mentioned, in this case Alice and Bob do not require to generate, store and measure any of the inner leaf qubits. Since no resource cube generation on their part is needed not only will there be no inner leaf errors (since there are no inner leaves for them) but the outer leaf errors are also smaller. Clearly in that case simulating errors over a chain of $n$ links consisting of $n$ outer links and $n$ repeaters upper bounds the errors arising in our real scenario from $n$ outer links with two of those links having smaller errors and $n_{\text{rep}}$ repeaters.

\subsubsection{Entanglement generation}

In the case of entanglement generation the capabilities of the nodes of Alice and Bob need to be larger than in the case of key generation. Specifically in this case Alice and Bob need to be able to error-protect their qubits while attempting to entangle them with each other through the repeater chain. One issue that is of crucial importance is for how long Alice and Bob need to store their qubits. Our repeater is based on Bell measurements so in order to know which Bell pair Alice and Bob hold, they would need to wait for the information from all repeaters about the overall Pauli frame. Fortunately, the information about end-to-end link ranking based on all the rankings of all the outer links is available to Alice and Bob immediately with the information about the local rankings from the nearest outer link connections. This is because all the repeaters preserve the local rankings by always connecting the links of the same rank. However, the ranking based on inner leaf syndromes, i.e. based on the value of $\abs{\vec{s}_{X/Z}}$ requires also for additional classical communication time from all repeaters to Alice and Bob. The ranking information is needed in order for Alice and Bob to be able to group their links into bins where from each bin they later extract/distill entanglement separately. Hence these two requirements, the need for communication of the Pauli frame and the need for communication of the inner leaf syndromes means that Alice and Bob need to be able to preserve their qubits for the duration of the end-to-end classical communication. Regarding the Pauli frame information, we note that one could argue that depending on the application of that raw end-to-end entanglement one can already start distilling/processing that entanglement before the Pauli frame information arrives. This could be possible if the initial part of the application only requires certain deterministic operations, since any adaptive step of the protocol that requires any kind of measurements and interpretation of the measured data can only be performed when the Pauli frame is already known to Alice and Bob. However, the second part related to inner-leaf ranking imposes more constraints in this case as e.g.~in order to distill better Bell pairs within the bins, Alice and Bob need to know which pairs belong to which bins before even starting the distillation protocol. This implies that the waiting requirement cannot be alleviated in our case. Therefore in this paper we assume that our performance metric provides a reliable lower bound on the entanglement generation rate only under the assumption that Alice and Bob are each in possession of $k$ (multiplexing levels) perfect quantum memories that are able to perfectly store the state for the duration of the entire repeater protocol i.e. for the time scaling linearly with and lower bounded by the total communication time over the whole repeater chain. Then they locally generate Bell pairs where one of the qubit of each Bell pair is the outer leaf qubit encoded in the GKP code and sent over for entanglement swapping, while the other qubit of that Bell pair is simply loaded into that perfect memory where it is stored until the Pauli frame and the inner leaf ranking information is communicated. In that scenario we can, similarly to the QKD scenario, argue that simulating errors over a chain of $n$ links consisting of $n$ outer links and $n$ repeaters upper bounds the errors arising in our real scenario from $n$ outer links with two of those links possibly having smaller errors (as Alice and Bob do not need to generate full cubes as they have perfect memories), $n_{\text{rep}}$ repeaters and 2 end-nodes with perfect memories.

\section{Conclusion}
\label{sec:conclusion}

We have proposed a novel type of multiplexing for quantum repeaters based on the GKP code. We have then analysed the performance of the corresponding repeater scheme using a numerical Monte-Carlo simulation whose code is freely available at~\cite{Note2}. Our model includes a detailed analysis of the preparation of the resource state used for the repeater protocol and takes into account imperfect experimental hardware, i.e. imperfect finitely squeezed GKP qubits as well as lossy homodyne detectors. We have shown that our repeater scheme can sustain high entanglement and secret-key rates per mode as large as $R=0.7$ for distances as large as 750 km. Since this is the rate for $k=20$ multiplexed modes, it corresponds to the rate of 14 ebits/secret-key bits per protocol run. Due to the all-photonic nature of our scheme, its throughput is only limited by the protocol repetition rate which in our case is dictated by the rate of generating the GKP qubits. Assuming our deterministic GKP sources can operate at a rate of at least 1 MHz, this corresponds to $1.4 \times 10^7$ ebits/secret-key bits per second. Our scheme can also achieve high distances both in terms of the rate per mode and in terms of overcoming the PLOB bound, the fundamental limit of repeater-less quantum communication. The performance of our scheme can be achieved with significantly relaxed hardware requirements relative to the previously proposed repeater schemes based on GKP qubits. We have also performed a detailed analysis of the number of GKP qubits needed to realise our scheme. Though that number is significantly higher than for the GKP-based schemes with only one level of encoding and one-way GKP-DV concatenated-coded schemes, the much higher performance and relaxed requirement on GKP squeezing make our scheme particularly promising if due to experimental reasons one encounters certain squeezing thresholds establishing the maximum amount of squeezing attainable in practical scenarios.

We also need to mention that apart from the requirement of GKP qubits, most operations required for our protocol involve only beamsplitters, homodyne detectors, and rotations and displacements in phase space. However, during the resource state preparation one also requires CZ gates for GKP qubits, which require additional squeezing resource. In principle these operations can be performed using only offline squeezing, as demonstrated experimentally in~\cite{yoshikawa2008demonstration}. Our model could be extended to provide a more detailed description of such two-mode gates as in this paper we have assumed those gates to be perfect.

Another issue related to squeezing is the possibility of using this resource or some form of phase sensitive amplification to overcome homodyne detection inefficiency. The benefit of such a strategy which aims to amplify/antisqueeze the measured quadrature with adding no or minimal amount of noise has already been demonstrated experimentally in the scenarios where the intrinsic homodyne detection efficiency was low~\cite{frascella2021overcoming,nehra2022few}. This technique could potentially increase homodyne detection efficiencies beyond 99\%.

Finally, we note that although the CC-amplification strategy of~\cite{fukui2021all} offers significant gain over the before considered pre-amplification strategy, we know that much better strategies of using the GKP code against photon loss exist.~\cite{albert2018performance,noh2018quantum}. However, the existence of these improved strategies have been proven using semidefinite programs which make it hard to extract the corresponding decoding procedure. More research is needed to extract the optimal decoding strategies for GKP code against photon loss as the results of~\cite{albert2018performance} suggest that if the optimal strategy could potentially correspond to a scheme realizable experimentally, then long distance communication with GKP qubits with relaxed squeezing requirement could potentially be possible without the need for code concatenation.

\section{Acknowledgements}
 We would like to thank Rafael N. Alexander, Debayan Bandyopadhyay, Prajit Dhara, Kosuke Fukui, Kenneth Goodenough, Michael Hatridge, Eneet Kaur, Gideon Lee, Gl\'{a}ucia Murta, Changhun Oh, Ashlesha Patil, Nithin Raveendran, Narayanan Rengaswamy, Hassan Shapourian, Qian Xu, Pei Zeng, and Changchun Zhong for helpful discussions. We would also like to thank Prajit Dhara, Kenneth Goodenough and Gl\'{a}ucia Murta for useful feedback on the manuscript. We acknowledge support from the ARO (W911NF-23-1-0077), ARO MURI (W911NF-21-1-0325), AFOSR MURI (FA9550-19-1-0399, FA9550-21-1-0209), AFRL (FA8649-21-P-0781), DoE Q-NEXT, NSF (OMA-1936118, ERC-1941583, OMA-2137642, CCF-2204985), ONR (N00014-19-1-2189), NTT Research, and the Packard Foundation (2020-71479). The authors are also grateful for the support of the University of Chicago Research Computing Center for assistance with the numerical simulations carried out in this work.  

\bibliography{library}{}
\onecolumngrid
\appendix

\section{Error propagation during fusion}
\label{sec:ErrorPropagationFusion}

In this appendix we use properties of graph states to describe the rules of error propagation during graph state BSM (fusion). Specifically, we describe how Pauli $X$ and $Z$ errors on the qubits involved in the BSM propagate onto the outcome graph state.

The BSM or fusion for graph states is implemented by first applying a CZ gate between the relevant qubits and then measuring them both in the $X$ basis. Note that the graph state BSM is rotated with respect to the standard BSM in the sense that if $\ket{\phi_{ij}}$ denotes the four canonical Bell states, then the graph state fusion projects on the two-qubit basis defined as:
$H \otimes \id \ket{\phi_{ij}}$. In order to see that the BSM for graph states can be implemented as described above and in order to see the rules of error propagation during such a graph state BSM we will use here the framework of graph state operations. In particular, the $Z$ and $X$ basis measurements affect the graph as follows~\cite{hein2006entanglement}:
\begin{equation}
    \begin{aligned}
    P_{Z,\pm}^{(a)}\ket{G} &= \ket{Z,\pm}^{(a)} \otimes U_{Z,\pm}^{(a)} \ket{G - {a}} \, , \\
    P_{X,\pm}^{(a)}\ket{G} &= \ket{X,\pm}^{(a)} \otimes U_{X,\pm}^{(a)} \ket{\tau_{b_0}(\tau_a \circ \tau_{b_0}(G) - {a})} \, ,
    \end{aligned}
    \label{fusion_rules_1}
\end{equation}
where

\begin{equation}
    \begin{aligned}
    U_{Z,+}^{(a)} &= \id \, ,\\
    U_{Z,-}^{(a)} &= \prod_{b \in N_a} Z^{(b)} \, ,\\
    U_{X,+}^{(a)} &= \sqrt{iY^{(b_0)}}\prod_{b \in N_a - N_{b_0}-b_0} Z^{(b)} \, ,\\
    U_{X,-}^{(a)} &= \sqrt{-iY^{(b_0)}}\prod_{b \in N_{b_0} - N_a-a} Z^{(b)} \, ,
    \end{aligned}
\end{equation}
and $P_{i,\pm}^{(a)}$ for $i=\{Z,X\}$ denotes a projection onto one of the $i$-basis eigenstates $\ket{i,\pm}$ corresponding to the two outcomes $+$ and $-$ on the qubit $a$. Here, $G$ denotes the original graph, $a$ the measured vertex of the graph, $b_0$ any chosen neighboring vertex of $a$ (different choice of $b_0$ result in different outcome graphs that are equivalent up to local Clifford operations), and $\tau_i$ denotes local complementation at vertex $i$. A local complementation at vertex $i$ is a graph operation which connects all the unconnected neighbours of $i$ and disconnects all its connected neighbours. Moreover, note that:
\begin{equation}
    \sqrt{\pm i Y} = e^{\pm i \frac{\pi}{4} Y}\, .
\end{equation}

\begin{figure}
\centering
\includegraphics[width = 0.5\columnwidth]{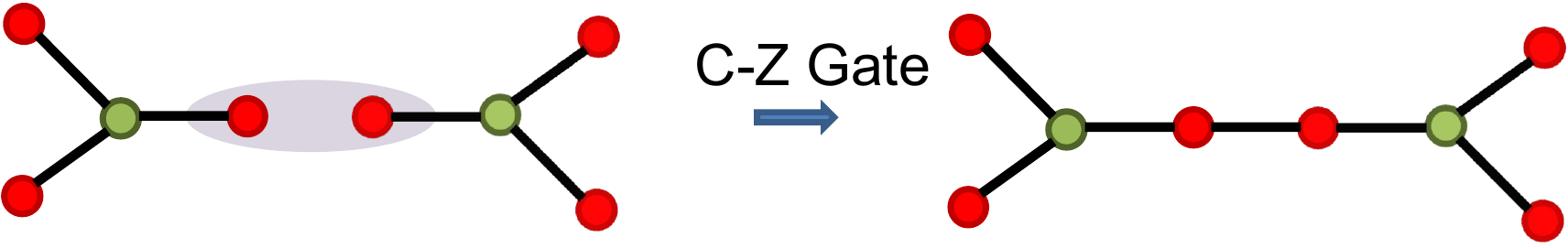}
\caption{Fusing two 4-tree clusters through their leaves. The first step involves a $C_Z$ gate between the leaves.}
\label{fig:4Trees_LL_fusion}
\end{figure}

\begin{figure}
\centering
\includegraphics[width =0.6\columnwidth]{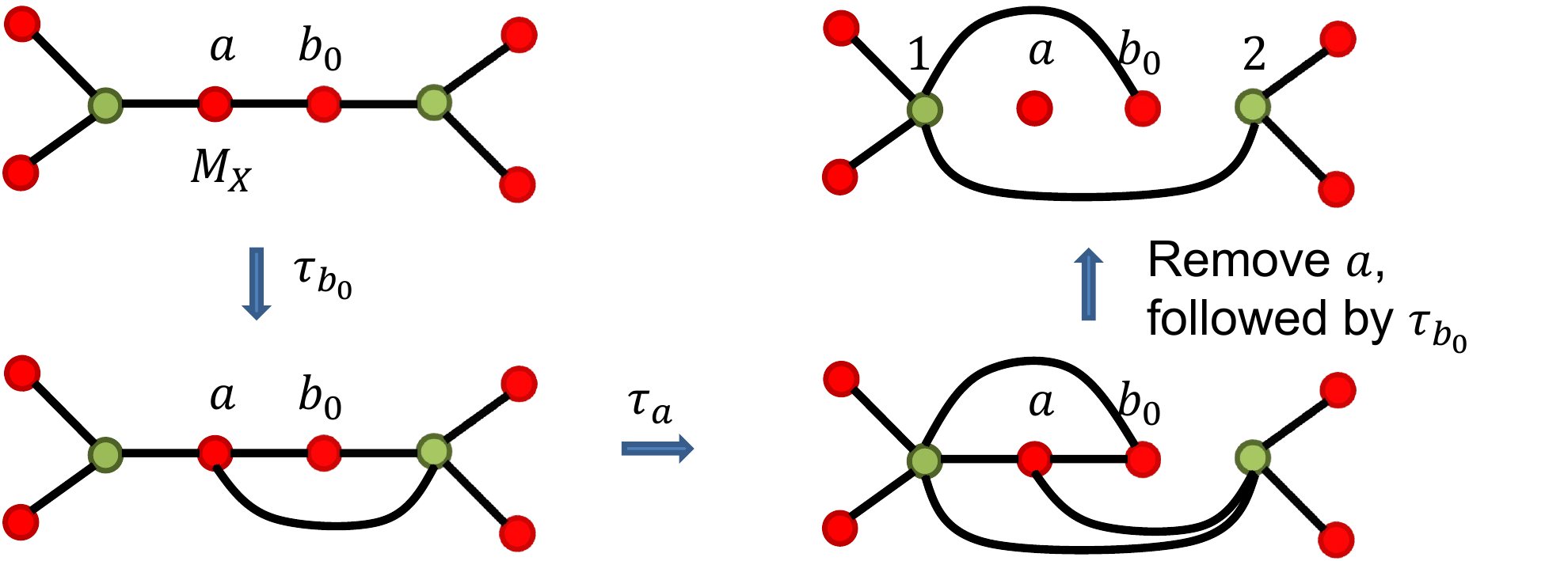}
\caption{Fusing two 4-tree clusters through their leaves. Following the $C_Z$ gate between the leaves, the vertex labeled $a$ is measured in the $X$ basis. The outcome graph state is depicted graphically by tracking the measurement as described in Eq.~\ref{fusion_rules_1}.}
\label{fig:4Trees_LL_MX_lcomps}
\end{figure}

To see the propagation of errors, let us illustrate the action of such a graph BSM in a simple scenario occurring during the generation of our resource state when we fuse two 4-trees through their leaves. Firstly we apply a CZ gate between the leaves as shown in FIG.~\ref{fig:4Trees_LL_fusion}.
Then we shall perform the $X$ measurement on the leaf qubit marked $a$, with $b_0$ being the chosen neighbour of $a$ as shown in FIG.~\ref{fig:4Trees_LL_MX_lcomps}. Following the rules of graph state evolution, after the measurement the state of the graph evolves as shown in FIG.~\ref{fig:4Trees_LL_MX_lcomps} up to the unitary $U_{X,\pm}^{(a)}$, depending on the measurement outcome.
Now we need to apply $(U_{X,\pm}^{(a)})^\dag$ to bring the state to the graph state shown. This requires us to apply $Z$ on qubits labeled 1 or 2 in the outcome graph of FIG.~\ref{fig:4Trees_LL_MX_lcomps}, depending on the measurement outcome and $(\sqrt{\pm iY})^\dag = \sqrt{\mp iY} = \exp(\mp \frac{\pi}{4} Y)$ on qubit $b_0$.

\begin{figure}
\centering
\includegraphics[width =0.6\columnwidth]{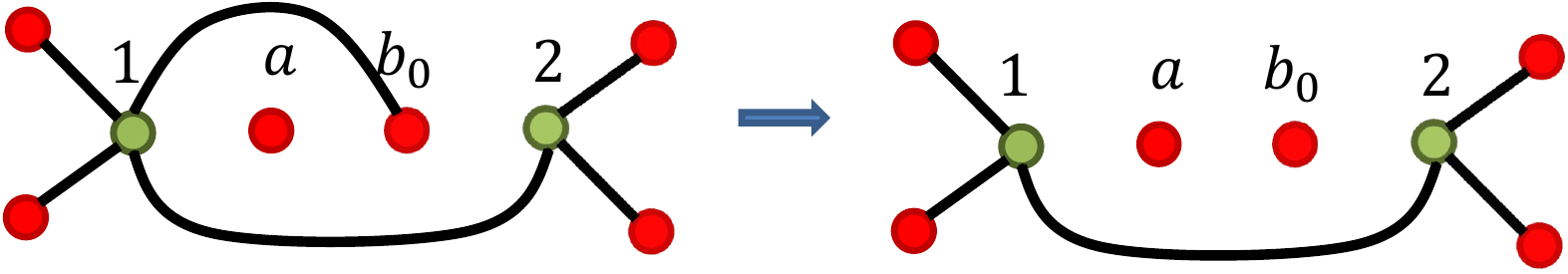}
\caption{Fusing two 4-tree clusters through their leaves. Following the $C_Z$ gate between the leaves, $X$ measurement of vertex $a$ and conditional correction operations, the vertex $b_0$ is measured out.}
\label{fig:4Trees_LL_MX_remove_b0}
\end{figure}

Clearly after bringing the state to the above form the final step would be to remove the vertex $b_0$ by measuring it in the $Z$ basis and applying $U_{Z,\pm}$ to bring the state to the final desired fused form as shown in FIG.~\ref{fig:4Trees_LL_MX_remove_b0}. However, note that:
\begin{equation}
\begin{aligned}
 P_{Z,\pm}^{(b_0)} \sqrt{-iY^{(b_0)}} &= \sqrt{-iY^{(b_0)}} P_{X,\mp}^{(b_0)} \, \\
 P_{Z,\pm}^{(b_0)} \sqrt{iY^{(b_0)}} &= \sqrt{iY^{(b_0)}} P_{X,\pm}^{(b_0)} \, .
 \label{remove_b0_measurement}
 \end{aligned}
\end{equation}
Hence, since we do not care about the final state of the $b_0$ qubit, we can actually perform an $X$ measurement on that qubit and then interpret the outcome as if we have applied the $\sqrt{\pm iY^{(b_0)}}$ correction first followed by a $Z$ measurement after. Therefore after that we need to undo the $U_{Z,\pm}^{(b_0)}$ to go back to the graph state.

Now let us consider an error on the $b_0$ qubit. Specifically, we will consider an $X$ or $Z$ error arising after the CZ gate, i.e. after the GKP quadrature residual shifts from qubit $a$ have already propagated onto $b_0$ through that gate. We note that a $Y$ error can be expressed as a simultaneous $X$ and $Z$ error. Clearly an $X$ error has no effect on our state as qubit $b_0$ is then measured in the $X$ basis, i.e. that error can be commuted through the measurement and therefore it does not affect the final state of the remaining graph state. A $Z$ error on the other hand will flip the measurement outcome, which will lead us to the wrong decision whether the final correction to be applied should be $(U_{Z,+}^{(b_0)})^\dag$ or $(U_{Z,-}^{(b_0)})^\dag$. These two unitaries differ by $Z$ operation on the neighbours of $b_0$ hence leading in our case to a $Z$-error on qubit 1 in FIG.~\ref{fig:4Trees_LL_MX_remove_b0}.

\begin{figure}
\centering
\includegraphics[width =0.6\columnwidth]{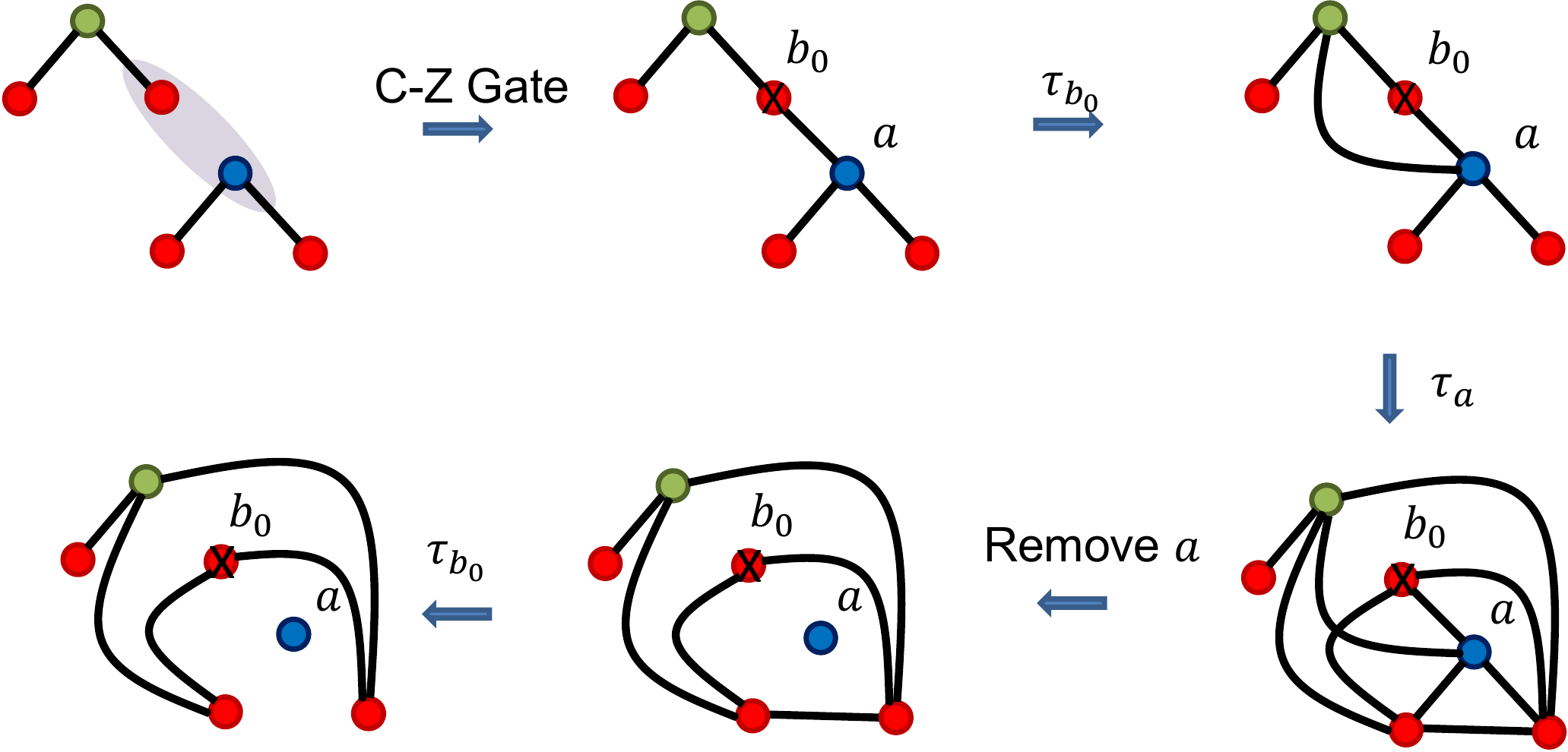}
\caption{Fusion between two 3-tree clusters by connecting a leaf of one (vertex $b_0$) with the node of the other (vertex $a$). Transformation of the graph following the $X$ measurement on $a$. Here $b_0$ has a $Z$ error marked by a cross after the CZ gate. This error could have been there from the beginning, it could originate from the $X$ error on qubit $a$ that has propagated to $b_0$ through the CZ gate or in the case of GKP qubits, it could arise due to the accumulation of the continuous shifts in $p$ quadrature beyond the critical value on qubit $b_0$ after the CZ gate.}
\label{fig:fusion_2_3trees}
\end{figure}

\begin{figure}
\centering
\includegraphics[width =0.4\columnwidth]{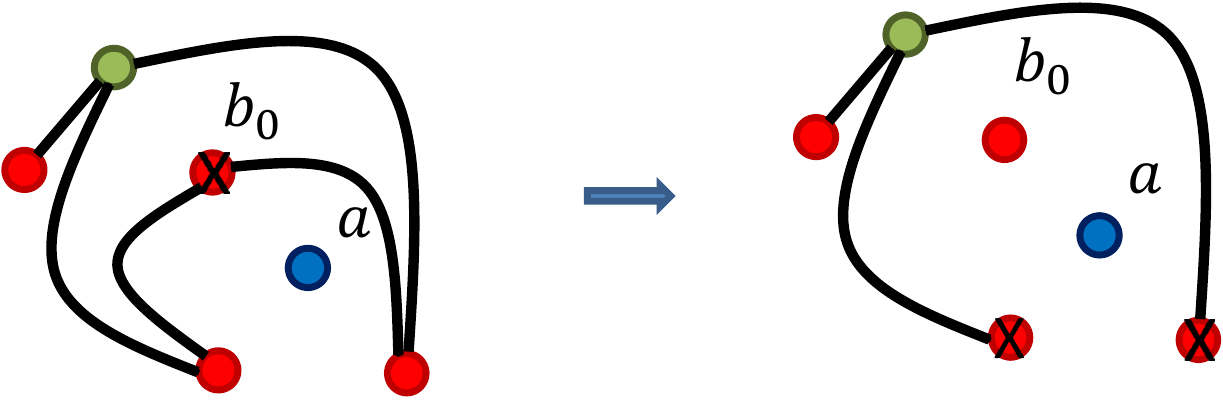}
\caption{Fusion between two 3-tree clusters by connecting a leaf of one with the node of the other. Final measurement of vertex $b_0$ in accordance with Eq.~\ref{remove_b0_measurement} followed by conditional correction operations. A $Z$-type error in vertex $b_0$ results in correlated $Z$ errors on vertices that were originally neighbors of $a$.}
\label{fig:fusion_2_3trees_remove_b0}
\end{figure}

In FIG.~\ref{fig:fusion_2_3trees} and FIG.~\ref{fig:fusion_2_3trees_remove_b0} we illustrate the error propagation for the other case occurring in our resource state creation, where we fuse a leaf of a 3-tree with a node of another 3-tree. We see, as before, than a $Z$ error on $b_0$ propagates as a $Z$ error to all the initial neighbours of qubit $a$. In FIG.~\ref{fig:fusion_2_3trees} we illustrate the transformation of the graph corresponding to the $X$ measurement on qubit $a$. Since $a$ had originally two neighbours, after that $X$ measurement $b_0$ has two neighbours. In FIG.~\ref{fig:fusion_2_3trees_remove_b0} we see the effect of removing the $b_0$ qubit, which leads to the correlated $Z$-error on two neighbours of $b_0$ which were original neighbours of $a$.

This result can be generalised to a rule, according to which a $Z$ error after the $CZ$ gate on one of the two qubits involved in the graph state BSM, will propagate as a correlated $Z$ error to all the initial neighbours of the other qubit involved. An $X$ error after the $CZ$ gate will not affect the resulting state.

\section{Secret-key fraction and entanglement-distillation rate}
\label{sec:PerformanceMetricsFormulas}

In this appendix we provide the formulas for achievable entanglement and secret-key rates that we use to quantify the performance of our scheme. The lower bound on the secret-key rate that we consider is based on the six-state QKD protocol~\cite{bruss1998optimal}. We consider two variants of this protocol. In the low noise regime we consider the fully asymmetric (i.e. one of the three bases is chosen for key generation and that basis is chosen with a probability approaching one) one-way six-state protocol for which the achievable secret fraction is given by~\cite{lo2001proof,renner2008security,kraus2005lower}:

\begin{equation}
    r_{\text{six-state,one-way}} = 1 - H(\rho).
\label{eq:hashing}
\end{equation}
Here $H$ is the von Neumann entropy and $\rho$ is the unique Bell diagonal state consistent with the observed Quantum Bit Error Rate (QBER). We note that as our repeater protocol distributes logical Bell diagonal states (as on the logical level we only have Pauli errors), in our case the state $\rho$ is exactly the state distributed by the repeater protocol, i.e. the parameter estimation step of the QKD protocol enables Alice and Bob to exactly reconstruct the density matrix of the shared state. We note that this achievable rate of the one-way six-state QKD protocol is independent of which basis is chosen to generate the key.

In the high noise regime we supplement our six-state protocol with 2-bit advantage distillation~\cite{maurer1993secret,gottesman2003proof,kraus2007security}. In this case before proceeding with information reconciliation and error correction, Alice and Bob divide their raw key-bits in groups of two and publicly announce and compare the xor of the two bits from each group. If the bits match they keep the first bit in each group, otherwise they discard both bits from the group. This procedure enables for significant suppression of errors at the expense of the 2-way classical communication, which we can assume to be free, and at the expense of needing to discard at least 50\% of the initial raw bits.

It has been shown in~\cite{murta2020key} that the largest amount of key with advantage distillation can be extracted if the key basis is the basis with largest QBER. For CSS codes this is the $Y$ basis and therefore we consider extraction of key in the $Y$ basis. To provide an expression for the corresponding secret key we need to introduce some notation.
Let
\begin{equation}
\ket{\psi(\san{x},\san{z})} =
\frac{1}{\sqrt{2}}(
\ket{0}\ket{0 + \san{x}} + (-1)^{\san{z}}
\ket{1} \ket{1 + \san{x}\,\,(\textrm{mod} \, 2)}
),
\end{equation}
for $\san{x},\san{z} \in \{0,1\}$.
We can then write the Bell-diagonal state as
\begin{eqnarray}
\rho_{AB} = \sum_{\mathclap{\san{x},\san{z} \in \{0,1\} }}\; p_{\san{xz}}
\ket{\psi(\san{x},\san{z})}\bra{\psi(\san{x},\san{z})}\ .
\label{eq:belldiag}
\end{eqnarray}
The four coefficients $p_{\san{xz}}$ define the probability distribution $P_{\san{XZ}}$. These Bell diagonal coefficients can be expressed in terms of $Q_{X,\text{end}}(\abs{\vec{s}_X} = m_X,j), Q_{Z,\text{end}}(\abs{\vec{s}_Z} = m_Z,j)$ as stated in Eq.~\eqref{eq:FlipsToBellDiagCoeffs}. For the fully asymmetric six-state protocol with key extraction in the $Y$ basis, together with the two-way advantage distillation, the corresponding secret-key fraction is~\cite{kraus2007security,renner2008security}:
\begin{eqnarray} 
r_{\text{six-state,AD}} = 
\frac{P_{\bar{\san{X}}}(0)}{2} [
1 - H(P_{\san{XZ}}^\prime) ],
\label{eq-key-rate-vs-error-rate}
\end{eqnarray} 
where
\begin{equation}
\begin{aligned}
P_{\bar{\san{X}}}(0) &=& (p_{00} + p_{11})^2
+ (p_{10} + p_{01})^2\ , \\
P_{\bar{\san{X}}}(1) &=& 2 (p_{00} + p_{11})
(p_{10} + p_{01})\ , \\
p_{\san{00}}^\prime &=&
\frac{p_{00}^2 + p_{11}^2}{(p_{00} + p_{11})^2 + (p_{10}+p_{01})^2}, \\
p_{\san{01}}^\prime &=& 
\frac{2 p_{00} p_{11}}{(p_{00} + p_{11})^2 + (p_{10}+p_{01})^2}, \\
p_{\san{10}}^\prime &=&
\frac{p_{10}^2 + p_{01}^2}{(p_{00} + p_{11})^2 + (p_{10}+p_{01})^2}, \\
p_{\san{11}}^\prime &=&
\frac{2 p_{10}p_{01} }{(p_{00} + p_{11})^2 + (p_{10}+p_{01})^2} \ ,
\end{aligned}
\label{eq:ParamForSecKey}
\end{equation}
and $H(P_{\san{XZ}})$ is the Shannon entropy of the distribution $P_{\san{XZ}}$. Here $P_{\bar{\san{X}}}(0)$ is the probability that the advantage distillation succeeds and the factor $1/2$ takes into account the fact that even in the case of success Alice and Bob keep only the first bit from each group.

In our protocol we calculate the secret-key fraction as:
\begin{equation}
    r_{\text{six-state}} = \max\{r_{\text{six-state,one-way}}, r_{\text{six-state,AD}},0\} \, .
\end{equation}

Now we note that the formula in Eq.~\eqref{eq:hashing} also describes the achievable rate of distilling entanglement through one-way hashing from a Bell diagonal state $\rho$~\cite{bennett1996mixed}. Moroever, it was shown in~\cite{murta2020key} that performing advantage distillation in the most noisy basis is equivalent to performing the DEJMPS entanglement distillaton protocol~\cite{deutsch1996quantum} with additional pre-rotations that maximise the fidelity of the output state conditioned on success. The formula in Eq.~\eqref{eq-key-rate-vs-error-rate} hence provides the rate of distilling entanglement from a Bell diagonal state $\rho$ through a procedure involving first the DEJMPS protocol checking for the $Y \otimes Y$ parity followed by one-way hashing distillation. The distribution $P'_{\san{XZ}}$ describes the improved Bell diagonal state which we obtain conditioned on the success of the DEJMPS protocol which succeeds with probability $P_{\bar{\san{X}}}(0)$. This output state described by $P'_{\san{XZ}}$ is the one to which we now apply the one-way hashing. Similarly the factor $1/2$ describes the fact that DEJMPS distillation requires us to sacrifice the target copy. Hence the figure of merit that we use here is:
\begin{equation}
    r = r_{\text{entdist}} = r_{\text{QKD}} = r_{\text{six-state}} \, .
\end{equation}

\section{Strategies for connecting links}
\label{sec:strategiesConnectingLinks}

In this appendix we justify our specific strategy of connecting the links in the repeaters based on the ranking information. Specifically we will show that connecting best links with each other, second best links with each other etc. is the optimal strategy, while the strategy of connecting the links randomly without using the ranking information would lead to the worst performance. We first prove this statement for the depolarising channel and then extend the proof to our case of independent noise in $X$ and $Z$ errors under the constraint that the errors remain small, even for the end-to-end links.

The secret-key rate generated through different QKD protocols is convex with respect to Quantum Bit Error Rate (QBER). The same applies to different bounds on distillable entanglement. As an example let us consider the entanglement hashing rate, which also describes the secret-key fraction of the one-way six-state QKD protocol. As a function of the QBER $e$ for a depolarising channel, it can be written as:
\begin{equation}
    r(e) = \max \left[1 - e -(1-e)h\left(\frac{1-\frac{3}{2}e}{2-e}\right) - h(e),0 \right] \, .
\end{equation}
Here $h(x) = -x \log_2(x) - (1-x)\log_2(1-x)$ is the binary entropy function. Clearly, in the physically meaningful regime of $e \in [0,1/2]$, $r(e)$ is a convex function of $e$.

The QBER for a depolarised EPR pair:
\begin{equation}
    \rho = (1-q) \dyad{\Phi} + q\frac{\mathbb{I}}{4} 
\end{equation}
is $e = q/2$. The end-to-end depolarised EPR pair with depolarising parameter $q_{\text{end}}$ would be established by connecting elementary depolarised pairs for which we will use the parameter $q$. By applying entanglement swapping to $n$ depolarised EPR pairs we obtain a final state with $q_{\text{end}} = 1-(1-q)^n \approx nq$, assuming that $q$ was small.

Now, let us assume that over the chain of $n$ elementary links that we want to connect together, we actually consider $k$ multiplexing levels. Hence let $q_i$ denote a depolarising parameter over a single elementary link ranked as $i$'th for $i \in \{1,...,k\}$ according to the analog information. Let $j \in \{1,...,k\}$ denote the $j$'th end-to-end link established by connecting $n$ elementary links according to some chosen strategy. Also, let $\alpha_i^j$ be a fraction in $[0,1]$ associated with each end-to-end link $j$, which tells us what fraction of the $n$ connected elementary links consists of elementary links ranked as $i$'th. Then the $j$'th end-to-end link or $j$'th path between Alice and Bob through different multiplexed links will consist of $\alpha_i^j n$ $i$'th ranked links. Since the total number of elementary links in each end-to-end path is $n$ we have that:
\begin{equation}
    \sum_{i=1}^k \alpha_i^j = 1 \, .
\end{equation}
Moreover, since each elementary segment contributes one multiplexed link of each ranking $i$, the total number of links of ranking $i$ over all paths must also sum up to $n$. Hence:
\begin{equation}
    \sum_{j=1}^k \alpha_i^j = 1 \, .
\end{equation}
Therefore a total hashing-rate $r$ over a single path or single end-to-end link labelled $j$ with the corresponding QBER $e^j$ and established using a specific swapping strategy defined by $\{\alpha_i^j\}_{i,j}$ is:
\begin{equation}
    r(e^j) = r\left(\frac{q_{\text{end}}^j}{2}\right) \approx r\left(\sum_{i=1}^k \alpha_i^j n \frac{q_i}{2}\right) \, ,
\end{equation}
where we have assumed $q_i$ to be small for all $i$, so that we could use:
\begin{equation}
    q_{\text{end}}^j = 1- \prod_{i=1}^{k}(1-q_i)^{\alpha_i^j n} \approx \sum_{i=1}^k \alpha_i^j n q_i \, , 
\end{equation}
for all $j$. Now the total rate per mode established over $k$ multiplexed links is:
\begin{equation}
    R = \frac{1}{k}\sum_{j=1}^k r(e^j) \approx \frac{1}{k}\sum_{j=1}^k r\left(\sum_{i=1}^k \alpha_i^j n \frac{q_i}{2}\right) \, .
\end{equation}
Now by convexity of $r(e)$:
\begin{equation}
    R \leq \frac{1}{k}\sum_{j=1}^k \sum_{i=1}^k \alpha_i^j r\left(n \frac{q_i}{2}\right) = \frac{1}{k} \sum_{i=1}^k r\left(n \frac{q_i}{2}\right) \, .
    \label{eq:rateUpperBound}
\end{equation}
Clearly the upper bound can be achieved by a strategy corresponding to $\alpha_i^j = \delta_{i,j}$. This is a strategy where a path $j$ corresponds to connecting all elementary links of rank $j$ i.e. best with the best, second best with the second best etc. This indicates that this strategy maximises the hashing rate, one-way secret-key fraction obtained through a six-state protocol but in fact also other distillable entanglement and secret-key bounds which are also convex with respect to QBER as mentioned before.

As a side note we note that by convexity:
\begin{equation}
    R \geq r\left(\frac{1}{k}\sum_{j=1}^k \sum_{i=1}^k \alpha_i^j n \frac{q_i}{2}\right) =  r\left(n \frac{1}{2} \sum_{i=1}^k\frac{q_i}{k}\right) \, .
    \label{eq:rateLowerBound}
\end{equation}
Note that this value can be achieved by setting $\alpha_i^j = 1/k$, i.e. by making all the paths the same and each consisting of equal number of elementary links of each ranking. This is in fact equivalent to the strategy of randomly connecting the links, i.e. without using the analog information at all. This can be seen by noting that $\sum_{i=1}^k\frac{q_i}{k} = q_{\text{av}}$, where $q_{\text{av}}$ is the average depolarising parameter for a single elementary link averaged over all the multiplexed ranked links with equal distribution. Hence the strategy of randomly connecting the links without using the analog information leads to the worst performance, as one might intuitively expect.

It is important to note here that the depolarising parameters $q_i$ are statistical properties of the ranked links that effectively arise as a consequence of the ranking. This means that it is not the case that each link has an a priori defined depolarising parameter $q_i$ but we just do not know their order over the $k$ links and the goal of analog information is to just discover which link has which parameter. Rather depending on what information is used to rank the links, the values of the arising depolarising parameters will be different. This can be best explained through an example. Let us consider only a bit flip channel for simplicity and let us assume that in total we have data from $N$ attempts to generate elementary link entanglement. In each protocol run we have assigned one link to the class ``best link'' based on some specific type of analog information called $analog_1$. We find that out of those links in the ``best link'' bin, $N_{\text{flip}}$ links have a bit flip. This means that for ranking based on $analog_1$, the bit flip probability for the channel describing the ``best link'' is given by $N_{\text{flip}}/N$. If we rank the links according to some other information, say $analog_2$, we might find a different number of links having a flip in that bin, say $N'_{\text{flip}}$ leading to a different bit-flip rate $N'_{\text{flip}}/N$. The bit-flip rate $N_{\text{flip}}/N$ could be the smallest out of all $k$ bit-flip rates when using $analog_1$ and the bit-flip rate $N'_{\text{flip}}/N$ could be the smallest out of all $k$ bit-flip rates when using $analog_2$ if both $analog_1$ and $analog_2$ are reliable. However, the fact that they are different depending on the information that we use to rank the links shows that ranking itself effectively creates the depolarising parameters.

Hence, clearly it is important that the ranking itself is done based on reliable information. One could imagine ranking based on completely unreliable information where in every protocol run the links are assigned distinct identifiers $i$ completely at random leading to their statistical property $q_i$ being the same for every $i$ so that $q_i = q$ for every $i$. In that case the two bounds in Eq.~\ref{eq:rateUpperBound} and Eq.~\ref{eq:rateLowerBound} would become the same. Hence the high performance of our scheme is an evidence that the GKP analog information reliably predicts the uncorrectable errors.

Clearly the argument presented here applies to a depolarising channel where the ranking and the hashing rate is a function of a single parameter $e = \frac{q_{\text{end}}}{2}$. In the case of our concatenated CSS codes, we do not have a depolarising channel, but rather our error rate is characterised by two parameters $Q_X$ and $Q_Z$ describing the probability of an independent $X$-flip and $Z$-flip respectively. Similarly as before we can define $Q_{X/Z,i}$ for an elementary link and:
\begin{equation}
    Q_{X/Z,\text{end}}^j = \frac{1-\prod_{i=1}^k (1-2 Q_{X/Z,i})^{\alpha_i^j n}}{2} 
\label{eq:EndToEndQ}
\end{equation}
as the total probability of $X/Z$ flip over the end-to-end link for a path $j$ where an end-to-end flip occurs for and odd number of elementary flips. Then we can use $Q_{X/Z,\text{end}}^j$ to define the probability of $X$, $Z$, and $Y$ error for the $j$'th path as:
\begin{equation}
    \begin{aligned}
    p_{10,\text{end}}^j &= Q_{X,\text{end}}^j (1-Q_{Z,\text{end}}^j) \, , \\
    p_{01,\text{end}}^j &= Q_{Z,\text{end}}^j (1-Q_{X,\text{end}}^j) \, , \\
    p_{11,\text{end}}^j &= Q_{Z,\text{end}}^j Q_{X,\text{end}}^j \, ,
    \end{aligned}
    \label{eq:FlipsToBellDiagCoeffs}
\end{equation}

where $X$ error occurs if there is an $X$-flip but not a $Z$-flip and vice versa for the $Z$ error. The $Y$ error occurs if both $X$- and $Z$-flip occur and hence is quadratically suppressed for the CSS codes. Here the coefficients $p_{01}, p_{10}, p_{11}$ and $p_{00} = 1 - p_{01} - p_{10} - p_{11}$ are the Bell diagonal coefficients of the resulting state according to Eq.~\eqref{eq:belldiag}.

As mentioned before the probabilities $Q_{X/Z,i}$ arise as a consequence of our ranking. Let us also assume here that we are in a regime in which both $Q_{X,\text{end}}^j$ and $Q_{X,\text{end}}^j$ are small for all $j$. We can then perform the following approximation:

\begin{equation}
    \begin{aligned}
    p_{10,\text{end}}^j &\approx Q_{X,\text{end}}^j \, , \\
    p_{01,\text{end}}^j & \approx Q_{Z,\text{end}}^j \, , \\
    p_{11,\text{end}}^j &= Q_{Z,\text{end}}^j Q_{X,\text{end}}^j \, .
    \end{aligned}
    \label{eq:FlipsToBellDiagCoeffs2}
\end{equation}

Now, the secret-key fraction/entanglement hashing rate for the corresponding Bell diagonal state satisfies:

\begin{equation}
\begin{aligned}
    &r - p_{00}\log_2(p_{00})= 1- H(\rho)- p_{00}\log_2(p_{00}) \\
    &= 1 + Q_{X,\text{end}}^j \log_2(Q_{X,\text{end}}^j) + Q_{Z,\text{end}}^j \log_2(Q_{Z,\text{end}}^j) + Q_{X,\text{end}}^j Q_{Z,\text{end}}^j\log_2(Q_{X,\text{end}}^j Q_{Z,\text{end}}^j) \\
    &= 1 + Q_{X,\text{end}}^j \log_2(Q_{X,\text{end}}^j) + Q_{Z,\text{end}}^j \log_2(Q_{Z,\text{end}}^j) + Q_{X,\text{end}}^j Q_{Z,\text{end}}^j\log_2(Q_{X,\text{end}}^j) + Q_{X,\text{end}}^j Q_{Z,\text{end}}^j\log_2(Q_{Z,\text{end}}^j) \\
    &\approx 1 + Q_{X,\text{end}}^j \log_2(Q_{X,\text{end}}^j) + Q_{Z,\text{end}}^j \log_2(Q_{Z,\text{end}}^j) \, .
    \end{aligned}
\end{equation}
Also:
\begin{equation}
\begin{aligned}
p_{00}\log_2(p_{00}) &= (1 - Q_{X,\text{end}}^j - Q_{Z,\text{end}}^j - Q_{X,\text{end}}^jQ_{Z,\text{end}}^j) \log_2(1 - Q_{X,\text{end}}^j - Q_{Z,\text{end}}^j - Q_{X,\text{end}}^jQ_{Z,\text{end}}^j) \\ &\approx (1 - Q_{X,\text{end}}^j - Q_{Z,\text{end}}^j) \log_2(1 - Q_{X,\text{end}}^j - Q_{Z,\text{end}}^j) \, .
\end{aligned}
\end{equation}
Hence:
\begin{equation}
    r \approx 1 + Q_{X,\text{end}}^j \log_2(Q_{X,\text{end}}^j) + Q_{Z,\text{end}}^j \log_2(Q_{Z,\text{end}}^j) + (1 - Q_{X,\text{end}}^j - Q_{Z,\text{end}}^j) \log_2(1 - Q_{X,\text{end}}^j - Q_{Z,\text{end}}^j) \, .
\end{equation}
Therefore we see that our secret-fraction/entanglement rate can be approximated by the corresponding rate extractable from a Bell diagonal state with
\begin{equation}
    \begin{aligned}
    p_{10,\text{end}}^j &= Q_{X,\text{end}}^j \, , \\
    p_{01,\text{end}}^j & = Q_{Z,\text{end}}^j \, , \\
    p_{11,\text{end}}^j &= 0 \, .
    \end{aligned}
\end{equation}
Moreover, note that:
\begin{equation}
    Q_{X/Z,\text{end}}^j \approx \sum_{i=1}^k \alpha_i^j n Q_{X/Z, i} \, .
\end{equation}
Hence we have effectively linearised all the relations between the probabilities of Pauli flips over elementary links and end-to-end links as well as the Bell diagonal coefficients of the resulting Bell diagonal states. Therefore we can define $p_{10,\text{end,} i} = n Q_{X, i}$, $p_{01,\text{end,} i} = n Q_{Z, i}$, $p_{11,\text{end,} i} = 0$ and $p_{00,\text{end,} i} = 1-p_{10,\text{end,} i} - p_{01,\text{end,} i} - p_{11,\text{end,} i}$ as approximate Bell diagonal coefficients of the end-to-end Bell diagonal state obtained by connecting all the links of the same ranking.
Then we see that
\begin{equation}
    \begin{aligned}
    p_{10,\text{end}}^j &\approx \sum_{i=1}^k \alpha_i^j p_{10,\text{end,} i}  \, , \\
    p_{01,\text{end}}^j &= \sum_{i=1}^k \alpha_i^j p_{01,\text{end,} i} \, , \\
    p_{11,\text{end}}^j &= \sum_{i=1}^k \alpha_i^j p_{01,\text{end,} i} \, , \\
    p_{00,\text{end}}^j &= 1 - p_{10,\text{end}}^j - p_{01,\text{end}}^j - p_{11,\text{end}}^j = \sum_{i=1}^k \alpha_i^j (1 - p_{10,\text{end,} i} - p_{01,\text{end,} i} - p_{11,\text{end,} i}) = \sum_{i=1}^k \alpha_i^j p_{00,\text{end,} i} \, .
    \end{aligned}
\end{equation}
Hence we have that:
\begin{equation}
\begin{aligned}
    R &= \frac{1}{k} \sum_{j=1}^k r(\{p_{00,\text{end}}^j,p_{01,\text{end}}^j,p_{10,\text{end}}^j,p_{11,\text{end}}^j\}) = \frac{1}{k} \sum_{j=1}^k r\left(\sum_{i=1}^k \alpha_i^j\{p_{00,\text{end,} i},p_{01,\text{end,} i},p_{10,\text{end,} i},p_{11,\text{end,} i}\}\right) \\
    &\leq \frac{1}{k} \sum_{j=1}^k \sum_{i=1}^k \alpha_i^j r\left(\{p_{00,\text{end,} i},p_{01,\text{end,} i},p_{10,\text{end,} i},p_{11,\text{end,} i}\}\right) =\frac{1}{k} \sum_{i=1}^k r\left(\{p_{00,\text{end,} i},p_{01,\text{end,} i},p_{10,\text{end,} i},p_{11,\text{end,} i}\}\right) \, .
    \end{aligned}
\end{equation}
We again see that the upper bound is achievable for the ranking strategy described by $\alpha_i^j = \delta_{i,j}$ corresponding to connecting links of the same ranking.

We also find a lower bound:

\begin{equation}
    R \geq r\left(\frac{1}{k} \sum_{j=1}^k \sum_{i=1}^k \alpha_i^j\{p_{00,\text{end,} i},p_{01,\text{end,} i},p_{10,\text{end,} i},p_{11,\text{end,} i}\}\right) = r\left(\frac{1}{k} \sum_{i=1}^k\{p_{00,\text{end,} i},p_{01,\text{end,} i},p_{10,\text{end,} i},p_{11,\text{end,} i}\}\right) \, ,
\end{equation}
which is achievable by setting $\alpha_i^j = 1/k$ which corresponds to uniform connections such that each end-to-end link is the same. Hence we see that the same ranking strategy is also optimal for the noise channel with independent $X$ and $Z$ errors in the regime where not only the elementary-link error probabilities, but also the end-to-end error probabilities are small.

Again, it is important to rank the links based on reliable information. Here the situation is slightly more challenging than for depolarising noise. This is because in our case ranking generates two statistical properties of the ranked links given by the error probabilities $Q_{X, i}$ and $Q_{Z, i}$. Nevertheless, we see from our repeater performance that the GKP analog information serves as a good measure for establishing different bins of different quality links which leads to high overall performance rate. Clearly the GKP analog information is also generated separately for $X$ and $Z$ errors. In Appendix~\ref{sec:errorsOuterLeaves} and Section~\ref{sec:rankingOuter} we describe how we combine these two pieces of information into a single quantity dictating the ranking of the links.

\section{Elementary link postselection strategy}
\label{sec:postselectionstrategy}
\subsection{With multiplexing}

While the multiplexing strategy considered in this work based on ranking of the elementary links is hard to model analytically, the reference strategy based on post-selection depicted in FIG.~\ref{fig:RateVsMultiplexing} can be easily modelled analytically. The important difference in this case is that the links are not generated deterministically. Therefore we need to first establish the probability distribution for the number of end-to-end links that can be generated.

Let us assume that $p$ is the probability of passing the post-selection test for a single link in a single quadrature, i.e. $p=P_\textrm{PS}(v;\sigma^2)$ as defined in Eq.~\eqref{eq:PostselectionProbAndPostselectedError} for some noise variance $\sigma^2$ and discard window $v$. We attempt on $k$ multiplexed levels and therefore the probability that we successfully generate $X=m$ elementary links out of $k$ is given by:
\begin{equation}
    p_{\text{elem}}(X=m) = \binom{k}{m} p^{2m} (1-p^2)^{k-m} \, .
\end{equation}
Here we include $p^2$ as a single-link is successfully generated only if we pass the post-selection test in both quadratures for that link.
Now the probability of generating exactly $Y=m$ end-to-end links corresponds to the probability that over all $n$ elementary links we generated at least $m$ links and over at least one of the elementary links we generated exactly $m$ links:
\begin{equation}
    p_{\text{end}}(Y=m) = \sum_{l=1}^n \binom{n}{l} p_{\text{elem}}(X=m)^l p_{\text{elem}}(X>m)^{n-l} \, .
\end{equation}
Then we can calculate the average number of successfully generated end-to-end links as:
\begin{equation}
    \expval{Y} = \sum_{m=0}^k m p_{\text{end}}(Y=m) \, .
\end{equation}
After establishing the average QBER for the postselected links, we can calculate the rate as:
\begin{equation}
    R = \frac{\expval{Y}}{k} r
\end{equation}
where $r$ is the secret-key fraction/lower bound on distillable entanglement for a single successfully post-selected end-to-end link.
Clearly both $\expval{Y}$ and $r$ depend on the post-selection window during GKP syndrome measurement leading to a trade-off between the two.

\subsection{Without multiplexing}

Here we describe the rate corresponding to the single-mode strategy without multiplexing based on CC-amplification, which up to our knowledge is the best known strategy of using only bare square-lattice, single-mode GKP qubits for repeaters. We note that this strategy effectively corresponds to the above described strategy based on multiplexing with post-selection for $k$ being restricted to 1 in the case where we only consider the noise on the outer leaves. Then the rate just becomes $R = p^{2L_{\text{tot}}/L}  \times  r$. Here the secret-key fraction/entanglement rate $r$ can be calculated in terms of the flip probability $Q_{X/Z,\text{end}} = Q_{\text{end}}$ where $Q_{\text{end}}$ can be calculated using Eq.~\ref{eq:EndToEndQ} with $k=1$ and $j=1$, $Q_{X/Z,1} = Q = E_\textrm{PS}(v;\sigma^2)$ and $\alpha_1^1 = 1$. Moreover, the probability of passing post-selection in each quadrature is given by $p = P_\textrm{PS}(v;\sigma^2)$. We note the dependence on the discard window $v$ while the noise standard deviation $\sigma$, including lossy channel, imperfect GKP qubits and lossy detectors is then given by:
\begin{equation}
    \sigma = \sqrt{\frac{1-\eta_{\text{det}}\exp(-L/(2L_0))}{\eta_{\text{det}}\exp(-L/(2L_0))} + 2\sigma_{\text{GKP}}^2}
\end{equation}

\section{GKP Bell State Measurement}
\label{sec:GKPBSMs}

In our repeater architecture three types of BSM occur. The first one is the GKP BSM which occurs either during TEC on the GKP qubits or during GKP entanglement swapping on the outer leaves. We already know from~\cite{walshe2020continuous, fukui2021all} that such a BSM can be implemented using a beamsplitter followed by a homodyne measurement of one mode in the $q$ quadrature and the other in the $p$ quadrature. Finally the outcomes need to be rescaled by $\sqrt{2}$ to compensate for the rescaling due to the beamsplitter. This procedure effectively corresponds to measuring $q_1+q_2$ on one of the modes and $p_1-p_2$ on the other. The logical GKP outcomes are read by establishing whether the measured value is closer to an even (logical zero) or odd (logical one) multiple of $\sqrt{\pi}$.

The second BSM is a logical BSM on the Steane code level used for entanglement swapping on the inner leaves. Since the Steane code admits transversal CNOT gates, this BSM amounts to seven parallel CNOT gates followed by measurement of seven modes in the $q$ quadrature and the other seven modes in the $p$ quadrature. Again, we can then use the argument from~\cite{fukui2021all} to implement each of these seven blocks using a beamsplitter followed by a homodyne measurement of one output mode in the $q$ quadrature, the other in the $p$ quadrature and then rescaling the outcomes by $\sqrt{2}$ as before.

The final BSM is again on the GKP level, but it is a graph state BSM which is rotated relative to the standard BSM. As already discussed it corresponds to a $CZ$ gate followed by the measurement of both qubits in the $X$ basis. This procedure can also be decomposed into a purely optical implementation using again the result of~\cite{fukui2021all} that:
\begin{equation}
    (M_p \otimes M_q) \circ CNOT = \sqrt{2}(M_q \otimes M_p) \circ BS
\end{equation}
where $BS$ denotes the 50-50 beamsplitter. Specifically, we can then show that for the GKP encoding:
\begin{equation}
\begin{aligned}
    (M_X \otimes M_X) \circ CZ &= (M_X \otimes M_X) \circ (\mathbb{I} \otimes H) CNOT (\mathbb{I} \otimes H)^\dag = (M_X \otimes M_Z) \circ CNOT (\mathbb{I} \otimes H)^\dag \\
    &= (M_p \otimes M_q) \circ CNOT \circ (\mathbb{I} \otimes H)^\dag = \sqrt{2}(M_q \otimes M_p) \circ BS \circ (\mathbb{I} \otimes H)^\dag
\end{aligned}
\end{equation}
Hence we see that this measurement can be implemented in the same way as the standard BSM using a beamsplitter and two homodyne measurements with an initial extra Hadamard transformation on qubit 2.

To see the effect of this measurement note that:
\begin{equation}
    \begin{aligned}
    H\p H^\dag &= - \q \, , \\
    H\q H^\dag &= \p \, , \\
    H^\dag \p H &= \q \, , \\
    H^\dag \q H &= - \p \, . \\
    \end{aligned}
\end{equation}
Hence $H^\dag$ on mode 2 followed by the the CNOT leads to the following transformation:
\begin{equation}
\begin{pmatrix}
\q_1 \\ \p_1 \\ \q_2 \\ \p_2 
\end{pmatrix}
\rightarrow (\id \otimes H^\dag) \rightarrow
\begin{pmatrix}
\q_1 \\ \p_1 \\ -\p_2 \\ \q_2 
\end{pmatrix}
\rightarrow CNOT \rightarrow
\begin{pmatrix}
\q_1 \\ \p_1 - \q_2 \\ \q_1-\p_2 \\ \q_2 
\end{pmatrix} \, .
\end{equation}

Finally, the two homodyne measurements read off $\p_1 - \q_2$ on the first mode and $\q_1-\p_2$ on the second mode.

\section{Modelling noisy GKP Bell measurements}
\label{sec:NoisyGKPMeasurements}

Now that we have established how to perform different types of BSMs using only beamsplitters and homodyne detectors, we shall now discuss how we can model the effect of noisy BSMs on our encoded qubits. While the main type of noise arising in the context of quantum communication is photon loss in the fiber, here we consider additional two imperfections, namely imperfect finitely squeezed GKP ancillas and lossy detectors.

\subsection{Teleportation-based error correction (TEC)}
\begin{figure}
\includegraphics[width =\columnwidth]{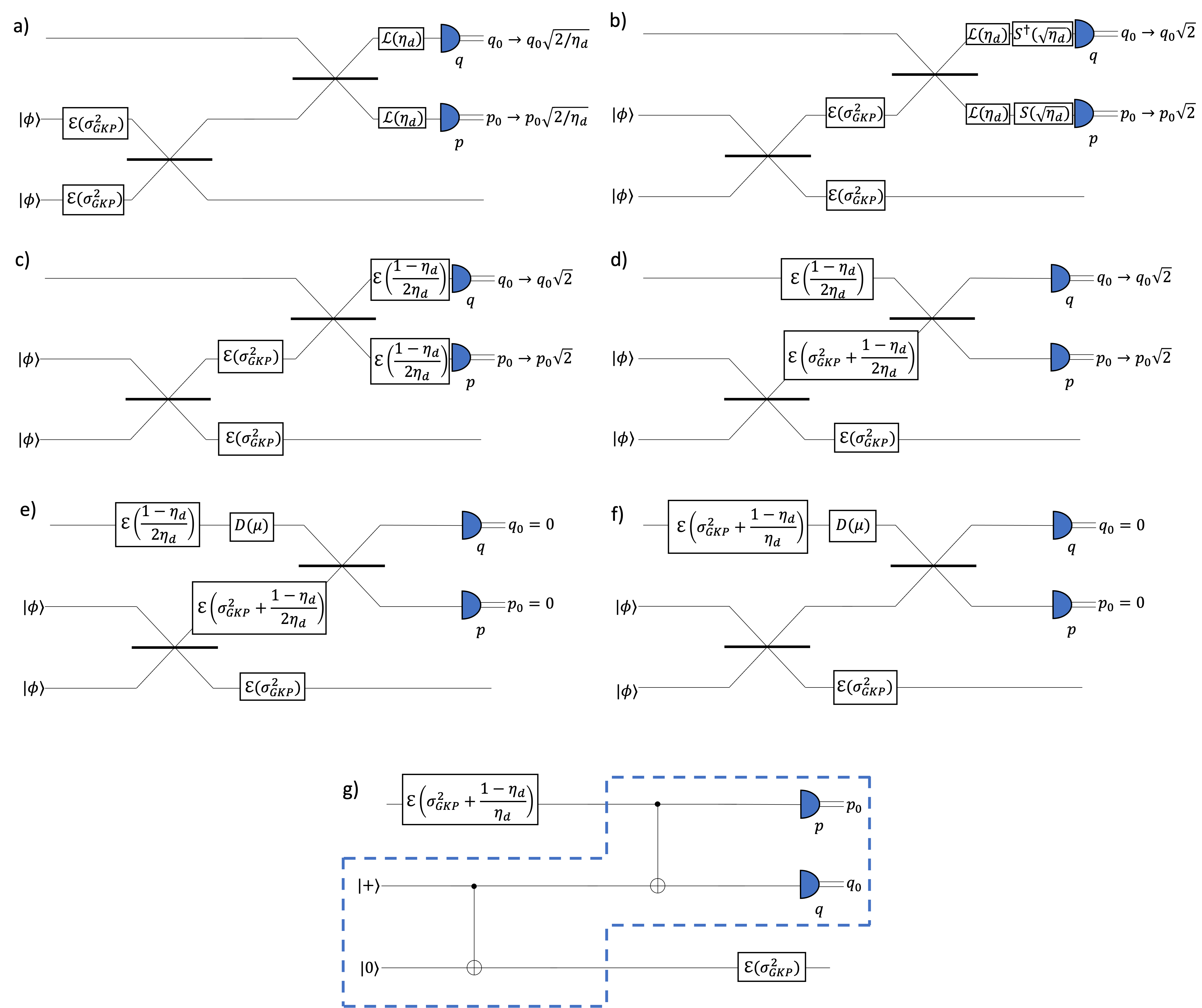}
\caption{Optical circuit for TEC showing how the noise due to finitely squeezed GKP ancillas and due to lossy detectors can be moved into the communication channel.}
\label{fig:TECcircuits}
\end{figure}

Let us start by considering the teleportation based GKP error correction~\cite{walshe2020continuous, fukui2021all}. We start by noting that under the twirled-GKP model, the ancilla GKP states can be modelled as initially perfect and then subjected to a Gaussian random displacement channel with standard deviation $\sigma_{GKP}$.
The effect of lossy detectors on the other hand can be modelled by lossy channel before perfect detection, where we try to compensate for the loss by rescaling the measurement outcomes by the inverse of the square root of the detector efficiency $\eta_d$ as shown in FIG.~\ref{fig:TECcircuits} a).

As shown in FIG.~\ref{fig:TECcircuits} b) the two Gaussian random displacement channels can be commuted through the beamsplitter. Moreover, rescaling the measurement outcomes to compensate for detector losses is equivalent to actually assuming that before perfect detection we apply single mode squeezing by $\sqrt{\eta_d}$, where the quadrature to be measured is the anti-squeezed one~\cite{fukui2021all}. Specifically, we define here the action of squeezing operator $\hat{S}(\lambda)$ as:
\begin{equation}
\begin{aligned}
    \hat{S}(\lambda)\hat{q}\hat{S}^\dag(\lambda) &= \lambda \hat{q} \, , \\
    \hat{S}(\lambda)\hat{p}\hat{S}^\dag(\lambda) &= \frac{1}{\lambda} \hat{p} \, .
\end{aligned}
\end{equation}

A lossy channel followed by the corresponding anti-squeezing of a quadrature to be measured immediately after, is equivalent to a Gaussian random displacement channel with $\sigma = \sqrt{\frac{1-\eta_d}{2\eta_d}}$ as shown in FIG.~\ref{fig:TECcircuits} c). This shows that by rescaling the measurement outcome rather than supplementing the lossy channel by phase insensitive amplification we can roughly decrease the variance by a factor of 2. This is because we only have the contribution of Gaussian noise from the lossy channel, while the rescaling, equivalent to a single-mode squeezing, does not introduce extra Gaussian noise in contrast to amplification. We note that rescaling actually amplifies the Gaussian noise contributed by loss, which is why the above expression for $\sigma$ has a factor $\eta_d$ in the denominator.

It is also worth noting that in principle one could aim to completely remove the effect of lossy homodyne detection by using a strong squeezer/phase sensitive amplification before the homodyne detection as demonstrated experimentally in~\cite{frascella2021overcoming,nehra2022few}. The action of such a squeezer can be decomposed into two effects. Firstly, it would again convert the action of the subsequent lossy channel into a Gaussian random displacement, now with standard deviation $\sigma = \sqrt{\frac{1-\eta_d}{2}}$. Secondly, if the squeezing/antisqueezing factor $\lambda$ is larger than $\sqrt{\eta_d}$, then this process would also transform the square GKP lattice into a rectangular lattice, hence increasing the lattice size along the measured direction. This would provide a further robustness against the effective Gaussian random displacement noise added by the detector and in the limit of large squeezing could completely mitigate the effect of that noise. This is because with a larger lattice the probability of that random displacement leading to a logical error after measurement can be made completely negligible. However, the feasibility of this strategy in the limit where our homodyne detection already has 99\% efficiency is unclear, given that such phase-sensitive amplification could actually also add non-negligible amount of noise in this regime. Therefore we only consider the rescaling strategy. Moreover, we note that with regard to squeezing, in our scheme we make use only of processes that require easier offline squeezing. Specifically the two-mode CZ gate can be implemented using ancillary squeezed vacuum modes. 

The two Gaussian random displacement channels just before the detectors can then be commuted through the beamsplitter as shown in FIG.~\ref{fig:TECcircuits} d). The two Gaussian random displacement channels on the middle mode can be combined into one channel. Moreover, the beamsplitter followed by the two homodyne measurements is equivalent to a displacement on the top mode followed by the beamsplitter and the projection on squeezed vacuum $q=0$ on the top mode and $p=0$ on the middle mode as shown in FIG.~\ref{fig:TECcircuits} e). The beamsplitter followed by the projections on the two vacuums corresponds to a projection onto a two-mode squeezed vacuum, i.e. a CV EPR pair. Such a CV EPR pair exhibits the same feature as its DV counterpart i.e. $U \otimes \id \ket{\Phi} = \id \otimes U^T \ket{\Phi}$. Note that the transpose map of a Gaussian random displacement channel is $\mathcal{E}^T = \mathcal{E}$ and so we can simply bounce the noise from the middle mode to the top one. Then we can commute it through the displacement operator and combine with the other noise on the top mode as shown in FIG.~\ref{fig:TECcircuits} f). Hence we obtain a perfect GKP error correction which is preceded by the Gaussian random displacement channel with $\sigma^2 = \sigma_{GKP}^2 + \frac{1-\eta_d}{\eta_d}$ and followed by another such channel with $\sigma = \sigma_{GKP}$ as shown in FIG.~\ref{fig:TECcircuits} g). Now, let $\eta$ be the transmissivity of the fiber between two consecutive GKP corrections. Hence, if we transform this fiber loss into Gaussian random displacement using pre-amplification, then the total Gaussian random displacement channel between two consecutive TECs, as occuring on the inner GKP qubits while being stored inside the repeater, has $\sigma^2 = 1-\eta + 2\sigma_{GKP}^2 + \frac{1-\eta_d}{\eta_d}$.

Finally, let us describe briefly how the measurement outcomes $q_0$ and $p_0$ are processed to obtain the GKP logical BSM outcomes. The two classical bits  corresponding to the $ZZ$ and $XX$ parity can be obtained according to the following rule:

\begin{equation}
S(q_0) = \left.
\begin{cases}
        0, & \text{for } \abs{R_{2\sqrt{\pi}}(q_0)}< \sqrt{\pi}/2 \, ,\\ 
        1, & \text{for } \abs{R_{2\sqrt{\pi}}(q_0)}\geq \sqrt{\pi}/2 \, .
\end{cases}
\right.
\label{eq:multiSyndromeRule}
\end{equation}
for the bit corresponding to the $ZZ$ parity and similarly for the $S(p_0)$ denoting the bit describing the $XX$ parity. Moreover, the two GKP stabilizer values which we store for [[7,1,3]] code correction (inner leaves) correspond to $R_{\sqrt{\pi}}(q_0)$ and $R_{\sqrt{\pi}}(p_0)$ respectively.

\subsection{CC-amplification}

\begin{figure}
\includegraphics[width =\columnwidth]{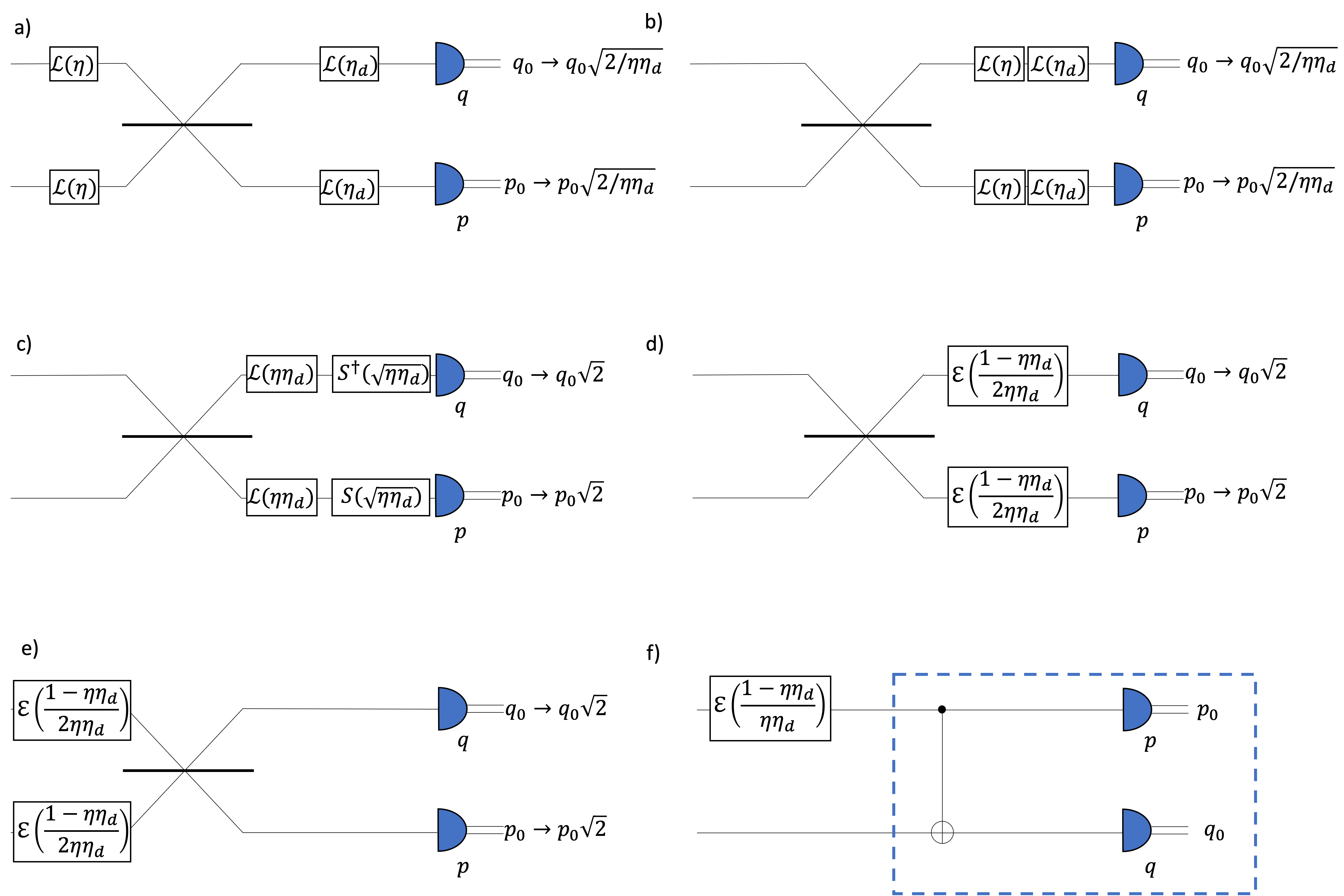}
\caption{Optical circuit for the CC-amplification showing how the noise due to lossy detectors can be moved into the communication channel.}
\label{fig:CCamp}
\end{figure}

The second BSM process that we want to consider is the CC-amplification with lossy detectors~\cite{fukui2021all}. In this case each mode is affected before the beamsplitter by the lossy channel of transmissivity $\eta$ modelling the fiber and by a lossy channel with transmissivity $\eta_d$ after the beamsplitter, which models lossy detectors. Now the measurement outcomes are rescaled by $1/\sqrt{\eta \eta_d}$. By a similar argument all the lossy channels followed by the rescaling of the measurement outcomes can be mapped onto a Gaussian random displacement channel on one mode with $\sigma^2 = \frac{1-\eta \eta_d}{\eta \eta_d}$, no noise on the other mode and BSM with loss-less detectors as shown in FIG.~\ref{fig:CCamp}. The rules for extracting the two classical BSM outcomes and the GKP stabilizer values are the same as for the TEC.

\section{\label{sec:resource_generation} Resource state generation}
\begin{figure}
\begin{center}
\begin{tikzpicture}

\node[draw,circle,fill=white!30] (n5) at (0,0) {$8$};
\node[draw,circle,fill=white!30] (n2) at (2,0) {$3$};
\node[draw,circle,fill=white!30] (n3) at (0,2) {$2$};
\node[draw,circle,fill=white!30] (n8) at (2,2) {$5$};

\node[draw,circle,fill=white!30] (n4) at (1,1) {$1$};
\node[draw,circle,fill=white!30] (n7) at (3,1) {$6$};
\node[draw,circle,fill=white!30] (n6) at (1,3) {$7$};
\node[draw,circle,fill=white!30] (n1) at (3,3) {$4$};

\draw (n2) -- (n5);
\draw (n3) -- (n5);
\draw (n4) -- (n5);
\draw (n2) -- (n7);
\draw (n4) -- (n7);
\draw (n1) -- (n7);
\draw (n1) -- (n6);
\draw (n3) -- (n6);
\draw (n4) -- (n6);
\draw (n1) -- (n8);
\draw (n2) -- (n8);
\draw (n3) -- (n8);
\end{tikzpicture}
\end{center}
\caption{Resource state used in the proposed repeater scheme is a Bell state between a bare physical GKP qubit and a [[7,1,3]] logical qubit comprised of 7 physical GKP qubits. Upto Hadamard gates on 4 out of the 8 qubits it corresponds to the depicted cube graph state.}
\label{fig:ResourceState}
\end{figure}
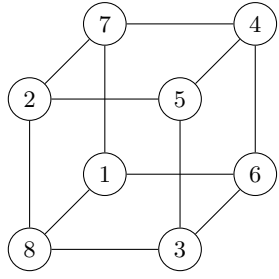

In this appendix, we describe the entangled resource state and its generation. 
The resource state is a maximally entangled state $(1/\sqrt{2})(\ket{00} + \ket{11})$ between a concatenated-coded $[[7,1,3]]$-GKP logical qubit and a physical GKP qubit. 
That is, for qubits numbered $1-8$, with the first 7 qubits in the Steane code and the $8^\textrm{th}$ qubit as the physical qubit, the generators of the stabilizer group associated with the resource state are given by
\begin{align}
    \begin{matrix}
        S_1=X_1X_3X_5X_7 & S_5=Z_1Z_3Z_5Z_7\\
        S_2=X_2X_3X_6X_7 & S_6=Z_2Z_3Z_6Z_7\\
        S_3=X_4X_5X_6X_7 & S_7=Z_4Z_5Z_6Z_7\\
        S_4=X_1X_2X_3X_8 & S_8=Z_1Z_2Z_3Z_8,
    \end{matrix}
\end{align}
where the generators $S_4$ and $S_8$ are the stabilizers of the Bell state, while the others are of the $[[7,1,3]]$-Steane code. 
The resource state is local-Clifford equivalent to the graph state~\cite{hein2006entanglement} associated with a cube topology as shown in FIG.~\ref{fig:ResourceState}. 
To see this, consider the stabilizer generators of the graph state, namely,
\begin{align}
    \begin{matrix}
        G_1=X_1Z_6Z_7Z_8 & G_5=X_5Z_2Z_3Z_4\\
        G_2=X_2Z_5Z_7Z_8 & G_6=X_6Z_1Z_3Z_4\\
        G_3=X_3Z_5Z_6Z_8 & G_7=X_7Z_1Z_2Z_4\\
        G_4=X_4Z_5Z_6Z_7 & G_8=X_8Z_1Z_2Z_3.
    \end{matrix}
\end{align}
When qubits $1-4$ are rotated by a Hadamard, the above stabilizer generators transform as
\begin{align}
    \begin{matrix}
        G_1\rightarrow G_1'= Z_1Z_6Z_7Z_8 & G_5\rightarrow G_5'=X_2X_3X_4X_5\\
        G_2\rightarrow G_2'=Z_2Z_5Z_7Z_8 & G_6\rightarrow G_6'=X_1X_3X_4X_6\\
        G_3\rightarrow G_3'=Z_3Z_5Z_6Z_8 & G_7\rightarrow G_7'=X_1X_2X_4X_7\\
        G_4\rightarrow G_4'=Z_4Z_5Z_6Z_7 & G_8\rightarrow G_8'=X_1X_2X_3X_8,
    \end{matrix}
\end{align}
which can easily be seen to be equivalent to products of the generators of the stabilizer group of the resource state. 
For example, $G_1'=S_6S_8$, $G_2'=S_5S_8$, $G_3'=S_5S_6S_8$, $G_4'=S_7$.

The procedure we use to generate the resource state is depicted in FIG.~\ref{fig:resource_prep}. It involves a suite of operations acting on finitely squeezed GKP qubits prepared in the qunaught and $\ket{+}$ states that include the $CZ$ gate, GKP fusion operation, post-selected homodyne measurements, and rotations in phase space (Hadamard gate), which all have been discussed before in this manuscript.
We characterize the quality of the resource states in terms of the GKP squeezing variance and the logical error probabilities on the individual GKP qubits that constitute the resource state. 
The latter arise from the homodyne detections that are performed as part of the fusion operations and depend on both the GKP squeezing as well as the homodyne detection inefficiency.

The squeezing variances of the GKP qubits involved in the generation procedure in FIG.~\ref{fig:resource_prep} at different stages of the procedure are indicated by the color / shade of filling of the qubits. 
The qubits that finally constitute the resource state have their squeezing variances given by $\left(\sigma_{\text{GKP}}^2,2\sigma_{\text{GKP}}^2\right)$. 
This is because these qubits underwent a $CZ$ gate (that transforms the gray/ lighter shade filled qubit to the red/darker shade fill qubit), and in the case of the purple qubits, additionally, Hadamard gates that cause the $q$ and $p$ variances to be interchanged. These so-called native shifts will later add up with the shifts due to transmission through lossy fiber and lead to potential logical GKP errors after subsequent GKP EC. 

The logical error probabilities associated with the resource state GKP qubits reflect the errors propagated and accrued from the measurements involved in the fusion operations in the resource state preparation procedure. Below, we describe the probabilities of $X$ and $Z$ errors on the red qubits of the resource state, which also correspond to probabilities of $Z$ and $X$ errors, respectively, for the purple qubits.

The fusion operation in step c involves two red qubits, i.e., both with GKP squeezing variance of $(\sigma_{\text{GKP}}^2,2\sigma_{\text{GKP}}^2)$ (denoted as fusion 1 in the main text). 
The homodyne measurements as part of the fusion, which are measurements of $ \hat{p}_1-\hat{q}_2$ and $ \hat{q}_1-\hat{p}_2$ are characterized by noise with variance $3\sigma_{\text{GKP}}^2+\frac{1-\eta_d}{\eta_d}$ each.
Since these are $X$ measurements of the two participating qubits, they introduce $Z$ errors in the immediate neighborhood of the qubits. 
To be precise, the measurement of the node qubit of the one 3-tree graph state produces a $Z$ error on the node qubit (red) of the other 3-tree graph state (as derived in Appendix~\ref{sec:ErrorPropagationFusion}), which make up one of the qubits in the final resource state, with probability given by
\begin{align}
P_Z^{\textrm{Fusion,c}}=E_\textrm{PS}\left(v;3\sigma_{\text{GKP}}^2+\frac{1-\eta_d}{\eta_d}\right).
\label{eq:PzFusionC}
\end{align}
Likewise, the measurement of the leaf qubit of the second 3-tree graph state produces correlated $Z$ errors on the two leaf qubits of the first 3-tree graph state, also with probability given by
\begin{align}
P_Z^{'\textrm{Fusion,c}}=E_\textrm{PS}\left(v;3\sigma_{\text{GKP}}^2+\frac{1-\eta_d}{\eta_d}\right).\label{eq:step_c_propagated_error}
\end{align}

\begin{figure}
\includegraphics[width =0.9\columnwidth]{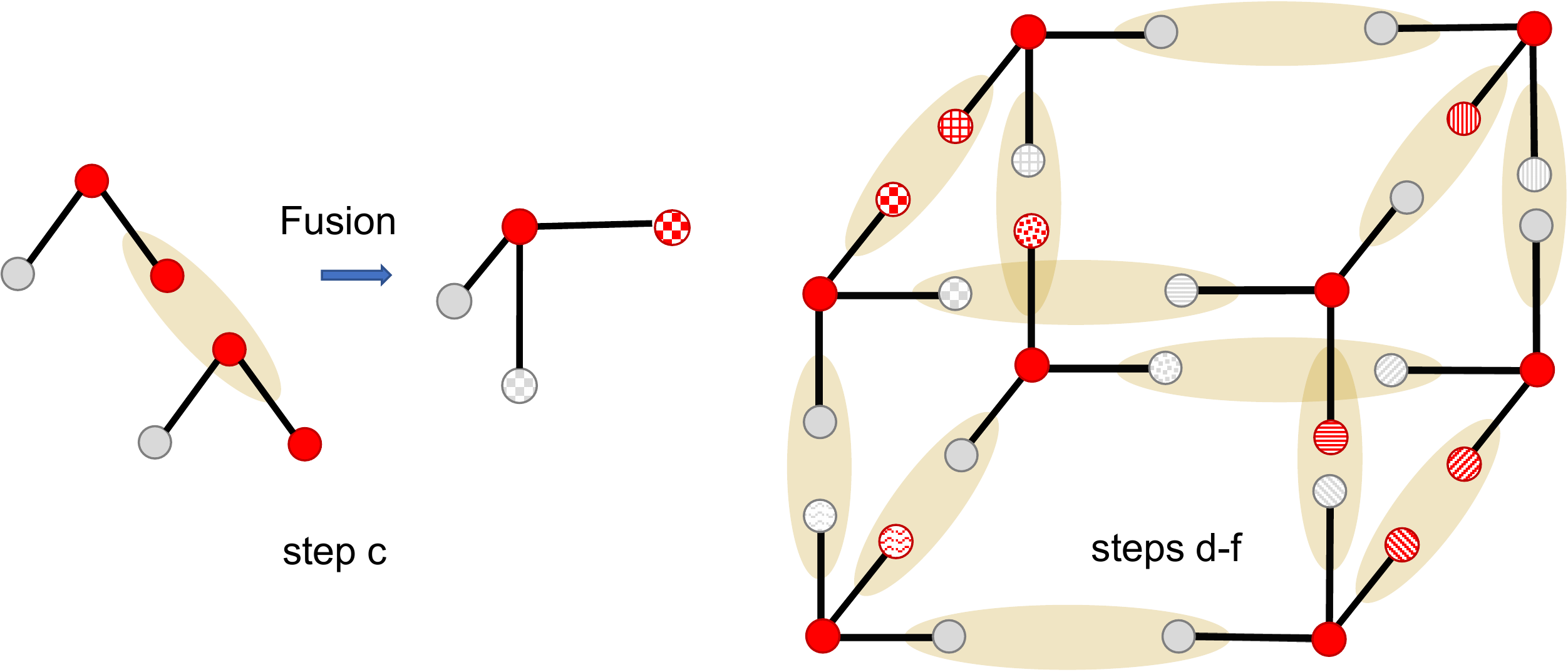}
\label{fig:fusion-steps-de}
\caption{(Left) Fusion step c of the resource state generate procedure described in FIG.~\ref{fig:resource_prep} and the origin of possible correlated $Z$-type errors on the leaf qubits of the lower 3-tree graph state. The pairs of similar pattern fillings (between a red and a gray qubit) indicate correlated $Z$-type errors. Absence of tiling indicates absence of errors from step c. (Right) The correlated $Z$-type errors from step c may then propagate to the resource state qubits via the subsequent fusion operations that are pictured here.}
\end{figure}

The fusion operation in step d involves fusion between two red qubits (fusion-1), and thus introduces $Z$-type errors on their immediate neighboring qubits with probability given by
\begin{align}
P_Z^{\textrm{Fusion,d}}=E_\textrm{PS}\left(v;3\sigma_{\text{GKP}}^2+\frac{1-\eta_d}{\eta_d}\right).
\end{align}
These are the node qubits of the participating 4-tree graph states, which eventually end up in the resource state. 
On the other hand, the fusion operation in step d' involves one red and one gray qubits (fusion-2), and thus introduces $Z$-type errors on their immediate neighboring qubits, i.e., on the node qubits of the participating 4-tree clusters, which similarly to step d eventually end up in the resource state. The probability of these errors occurring however are not the same on the two qubits, but instead one of them has a probability given by:
$P_Z^{\textrm{Fusion,d'}}=E_\textrm{PS}\left(v;3\sigma_{\text{GKP}}^2+\frac{1-\eta_d}{\eta_d}\right)$, and the other by $P_Z^{'\textrm{Fusion,d'}}=E_\textrm{PS}\left(0.7v;2\sigma_{\text{GKP}}^2+\frac{1-\eta_d}{\eta_d}\right)$.
The fusion operations in steps e and f similarly result in $Z$-type errors on the resource state qubits with similar probabilities associated with either noise variance $3\sigma_{\text{GKP}}^2+\frac{1-\eta_d}{\eta_d}$ or $2\sigma_{\text{GKP}}^2+\frac{1-\eta_d}{\eta_d}$.
For the numbering of the resource state qubits in FIG.~\ref{fig:ResourceState}, we tabulate in TABLE~\ref{tab:uncorr_Z_errors} the maximum number of possible uncorrelated logical $Z$-type errors on the individual qubits before the final Hadamard gates (i.e. for the red qubits) along with the associated probabilities of each error occurring that may arise from all the fusion operations in steps c through f. 
Additionally, recall that the fusion operation in step c gave rise to correlated $Z$ errors between a red and a gray leaf qubits. 
These errors propagate through subsequent fusion operations and settle as possible correlated errors between the resource state qubits. 
We tabulate these additional propagated $Z$ errors on the resource state qubits (red qubits before the action of the final Hadamards) in TABLE~\ref{tab:correlated_Z_errors}.

\begin{table}
\begin{tabular}{ |c|c|c|c|c| } 
\hline
\multirow{2}{4em}{Qubit Number} & \multicolumn{2}{|c|}{$Z$-type errors with $P_{\textrm{PS}}(v;3\sigma_{\text{GKP}}^2+\frac{1-\eta_d}{\eta_d})$} & \multicolumn{2}{|c|}{$Z$-type errors with $P_{\textrm{PS}}(0.7v;2\sigma_{\text{GKP}}^2+\frac{1-\eta_d}{\eta_d})$}\\
\cline{2-5}
& contributing steps & Max. total count & contributing steps & Max. total count\\
\hline
1&c,d'&2&e,f&2\\
2&c,d&2&e,f&2\\
3&c,d,e&3&f&1\\
4&c&1&d',e,f&3\\
5&c,d'&2&e,f&2\\
6&c,d&2&e,f&2\\
7&c,d,e&3&f&1\\
8&c&1&d',e,f&3\\
\hline
\end{tabular}
\caption{Possible independent $Z$ errors arising from fusion steps c through f on the resource state qubits at the end of step f (before the final Hadamard gates) of the resource state generation procedure of FIG.~\ref{fig:resource_prep}. For the two possible error probabilities, the table shows the steps that may contribute to the corresponding error, and maximum total count of each type of error.}
\label{tab:uncorr_Z_errors}
\end{table}

\begin{table}
\begin{tabular}{ |c|c|c| } 
\hline
Correlated $Z$ errors & Max. total count (on each pair)\\
\hline
(1,2); (5,6) & 2  \\ 
(1,3); (2,3); (5,7); (6,7) & 1 \\  
\hline
\end{tabular}
\caption{Possible correlated $Z$ errors in the resource state at the end of step f of the resource state generation procedure of FIG.~\ref{fig:resource_prep} arising and propagated from step c. The table shows the possible qubit pairs (numbered as in FIG.~\ref{fig:ResourceState}) that could carry these correlated errors and their maximum counts, where each correlated error occurs with probability $P_Z^{'\textrm{Fusion,c}}$ of Eq.~\eqref{eq:step_c_propagated_error}.}
\label{tab:correlated_Z_errors}
\end{table}

\section{Error channel for the outer leaves}
\label{sec:errorsOuterLeaves}

\begin{figure}
\includegraphics[width =\columnwidth]{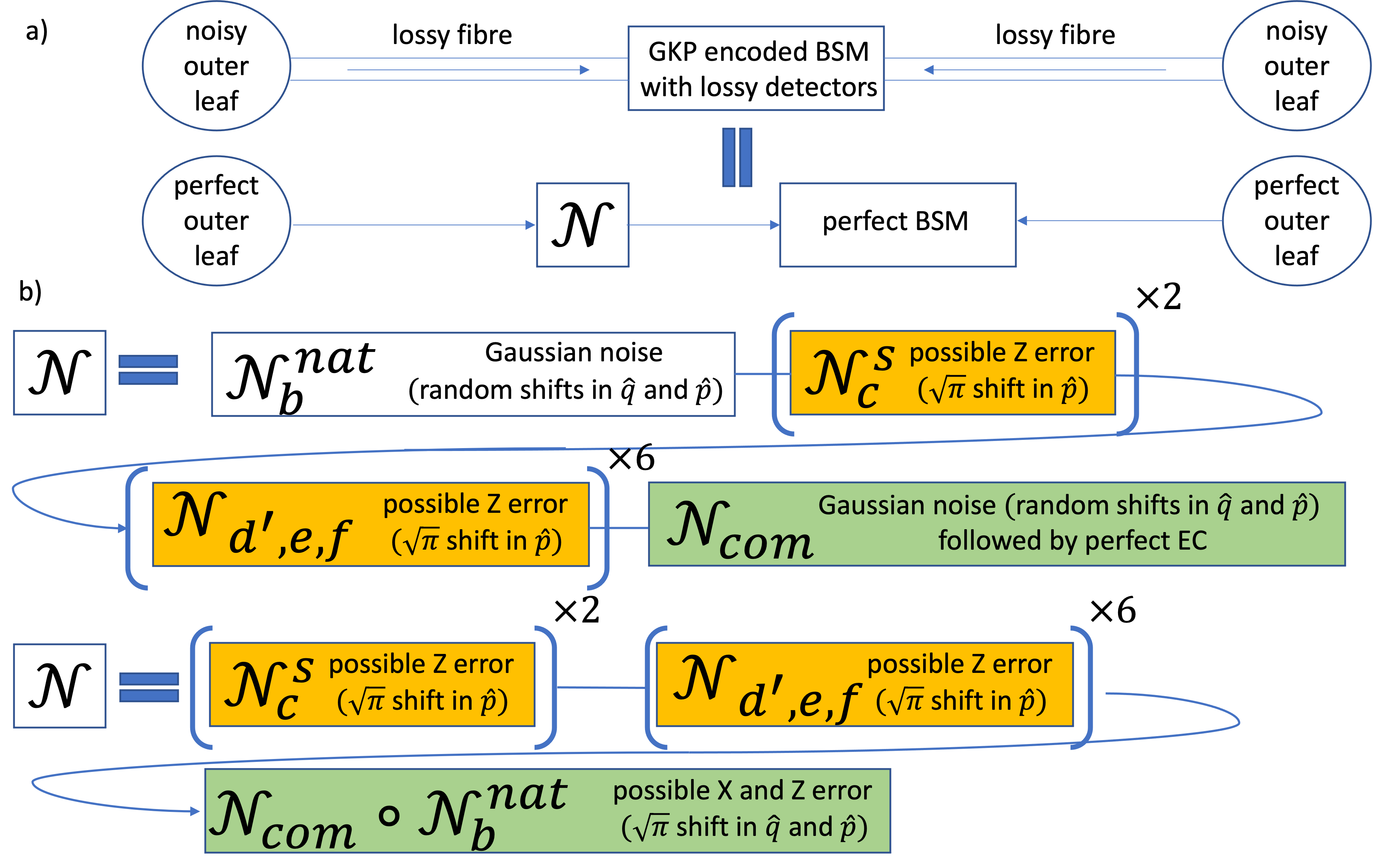}
\caption{Description of the error processes affecting the outer leaves. a) Outer leaves are affected by possible errors during the preparation of the cube resource state. The noisy outer leaves are then sent through lossy fibers towards a BSM station which performs a GKP encoded BSM using lossy homodyne detection modules. As described and depicted in FIG.~\ref{fig:TECcircuits} all the noise channels can be bounced through the BSM onto one qubit, so that effectively we can model the entire process as perfect with a single noise channel affecting the left qubit. b) Decomposition of the noise channel affecting the left outer leaf qubit. The orange and green boxes describe noise channels which involve GKP stabilizer measurement and GKP error correction. This stabilizer GKP analog information is used in orange channels for post-selection. The analog information from green channels is used for the ranking of the outer links. The Gaussian channel $\mathcal{N}_b^{\text{nat}}$ can be combined with the Gaussian channel in $\mathcal{N}_{\text{com}}$ followed by GKP EC to obtain a single GKP logical Pauli channel.}
\label{fig:OuterLeaves}
\end{figure}

In this section we discuss all the noise processes that lead to errors when creating outer links. The noisy processes affecting the GKP qubits sent from the left and right to meet in the middle heralding station are marked in FIG~\ref{fig:OuterLeaves}. Again using the properties of the GKP optical BSM, which can be decomposed into a displacement on one mode followed by a projection on a two-mode squeezed vacuum, we can bounce all the noise processes from the right GKP qubit onto the left. This enables us to see the whole process as having a perfect GKP qubit from the left, which then undergoes all the noise processes corresponding to noisy state preparation, noisy transmission and noisy BSM followed by a perfect BSM with a perfect GKP qubit from the right as also shown in FIG.~\ref{fig:OuterLeaves}.

\begin{table}
\begin{tabular}{ |p{2cm}||p{4.5cm}|p{1cm}|p{5.5cm}|p{1cm}| p{1.25cm}|p{1.25cm}| }
 \hline
 \multicolumn{7}{|c|}{Logical error channels per outer leaf pair} \\
 \hline
 Logical channel & description& $Z$ or $X$ error &standard deviation $\sigma$ & no of times & post-selection& analog info for ranking\\
 \hline
 $\mathcal{N}^s_{c}$&   single qubit error propagating from the fusion in step c)   & Z   &$\sigma^2_{Z} = 3\sigma_{GKP}^2 +\frac{1 - \eta_d}{\eta_d}$ & 2 &yes & no\\
 $\mathcal{N}_{d',e,f}$ &single qubit errors propagating from fusions in steps d'), e) and f) & Z&  $\sigma^2_{Z} = 2\sigma_{GKP}^2 +\frac{1 - \eta_d}{\eta_d}$ & 6 & yes & no \\
 $\mathcal{N}_{com} \circ \mathcal{N}^{\text{nat}}_{b}$    &outer leaf BSM EC, correcting native errors after CZ gate in step b) and the communication channel noise & Z and X&  $\sigma^2_{Z} = 4\sigma_{GKP}^2 + \frac{1 - \eta\eta_d}{\eta\eta_d}$, $\;\;\;\;$ $\sigma^2_{X} = 2\sigma_{GKP}^2 + \frac{1 - \eta\eta_d}{\eta\eta_d}$, where ${\eta=\exp\left(-L/(2L_0)\right)}$ & 1 & no & yes   \\
 \hline
\end{tabular}
\caption{Error processes affecting outer leaves.}
\label{tab:tableOuterLeaves}
\end{table}

Now we will describe in detail all the noise channels contained in the overall noise process $\mathcal{N}$. These noise processes can be divided into two categories. The first one consists of the error processes for which we collect analog information during error correction. This analog information is then used for ranking the outer links. The second error processes are those for which we use post-selection during error correction. For these processes we do not store the analog information if the post-selection is passed. This means that we do not store any information about the likelihood of a possible logical error, that can arise even if the post-selection is passed, for the later ranking of the links. All these error channels are described in TABLE~\ref{tab:tableOuterLeaves} and represented graphically in FIG.~\ref{fig:OuterLeaves}.

We start with errors arising during the preparation of the ``cube state'' resource, as discussed in Appendix~\ref{sec:resource_generation} and affecting the outer leaf qubit. Here we follow the labelling of the specific operations using the steps marked in FIG.~\ref{fig:resource_prep} and the outer leaf qubit is the qubit located in the position marked as ``8'' in FIG~\ref{fig:ResourceState}. Firstly, after the CZ gate in step b) our root qubit also ends up with a native error. In our case of combining the errors from both the left and right qubit onto the left qubit, the native error in $p$ quadrature corresponds to a Gaussian channel with $\sigma_n^2 = 4\sigma_{\text{GKP}}^2$ while in $q$ quadrature it corresponds to a Gaussian channel with $\sigma_n^2 = 2\sigma_{\text{GKP}}^2$.

Then during the fusion in step c) the logical error when measuring the red root qubit will lead to the $Z$-error on our outer leaf. This is modeled by the channel $\mathcal{N}_c^s$ with logical error probability given in Eq.~\eqref{eq:PzFusionC}. Again this channel appears twice corresponding to the possible error on both the left and the right qubit. We note that the possible logical error when measuring the other red leaf qubit in that step does not propagate to the outer-leaf qubit but in step d') propagates to one of the inner-leaf qubits.

Then we have the possible errors in steps d'), e) and f) marked by the $\mathcal{N}_{d',e,f}$ channel. This corresponds to the $Z$-error that can arise on the outer leaf qubit if there is a $Z$-error during one of the three fusions of the adjacent grey qubits. This is because a logical error during the measurement of the grey qubit as part of one of these three fusions, specifically the grey qubit that was not originally part of the graph with the outer leaf root, will cause a logical $Z$-error on the outer leaf qubit as derived in Appendix~\ref{sec:ErrorPropagationFusion}. In total this channel $\mathcal{N}_{d',e,f}$ adds this error six times because of three fusions neighbouring the left outer leaf and three fusions neighbouring the right outer leaf. The logical error probability for each of these measurements is $E_\textrm{PS}\left(0.7v;2\sigma_{\text{GKP}}^2+\frac{1-\eta_d}{\eta_d}\right)$. For these steps that involve post-selection, in the simulation we simply sample from the Bernoulli distribution and then either do nothing or add a shift by $\sqrt{\pi}$ to the $p$ quadrature of the outer leaf.

The final channel to consider is the $\mathcal{N}_{c-c}$ channel describing the errors due to the fiber losses in long-distance communication when bringing the two outer leaves for swapping using CC-amplification. We proceed by sampling the $q$ and $p$ shifts from the Gaussian distribution with $\sigma_{cc}^2 = \frac{1-\eta \eta_d}{\eta \eta_d}$, where $\eta = \exp(-L/2L_0)$ with the repeater separation $L$. We then add the shifts to the accumulated native error shifts. Finally the BSM is performed between our GKP qubit from the left that has accumulated all the noise and the perfect GKP qubit from the right. A logical $X$ $(Z)$ error happens in the form of a flip of a BSM outcome at this stage if the accumulated $q$ $(p)$ quadrature error $q_0$ $(p_0)$ has an odd multiple of $\sqrt{\pi}$ shift relative to $R_{\sqrt{\pi}}(q_0)$ $(R_{\sqrt{\pi}}(p_0))$.

We then proceed to rank the links according to the analog information obtained during the last step. We note that the analog information i.e. the information about the likelihood of error is always contained in the GKP stabilizer value, i.e. the measurement outcome modulo $\sqrt{\pi}$. To evaluate the likelihood of error we additionally need the information about the standard deviation of the Gaussian random displacement channel that resulted in the given syndrome. Then the error likelihood for that syndrome measurement $R_{\sqrt{\pi}}(q_0)$ is~\cite{noh2020fault}:

\begin{equation}
p[\sigma](R_{\sqrt{\pi}}(q_0)) = \frac{\sum_{n \in \mathbb{Z}} \exp[- (R_{\sqrt{\pi}}(q_0) - (2n+1)\sqrt{\pi})^2/(2\sigma^2)]}{\sum_{n \in \mathbb{Z}} \exp[- (R_{\sqrt{\pi}}(q_0) - n\sqrt{\pi})^2/(2\sigma^2)]}.
\label{eq:analogInfoprob}
\end{equation}

We note that in our scheme we have two categories of channels. Channels $\mathcal{N}_{c}^a$ and $\mathcal{N}_{d',e,f}$ are effectively logical channels acting on our outer leaf, i.e. these channels either do nothing or add a $\sqrt{\pi}$ shift. Hence the final stabilizer measurement occurring as part of the final BSM does not carry any additional information about the likelihood of errors during those channels. The second category are channels $\mathcal{N}_{b}^{\text{native}}$ and $\mathcal{N}_{c-c}$ which introduce Gaussian displacement errors. Hence the syndrome obtained during the final BSM can provide us with information about a logical error arising as a consequence of an overall shift occurring after the convolution of these two channels. This means that for the evaluation of the error likelihood for that final syndrome $R_{\sqrt{\pi}}(q_0)$ $(R_{\sqrt{\pi}}(p_0))$ we will use $\sigma^2 = \sigma_n^2 + \sigma_{cc}^2$, where as we have seen $\sigma_n$ is different for the $q$ and $p$ quadrature.

Having the error likelihood in both quadratures at hand we can rank all the $k$ links according to the likelihood that the final BSM correctly reads the values in both quadratures, as given in Eq.~\eqref{eq:PnoErrorOuterLeaf} and described in Section~\ref{sec:rankingOuter}.

\section{Error channel for the inner leaves}
\label{sec:ErrChforInnerLeaves}

Here we describe the evolution, noise process, error correction and entanglement swapping on the inner leaf qubits. The logical inner leaf qubits are encoded in the [[7,1,3]] code consisting of seven GKP qubits. These qubits are stored inside the repeater. After the outer leaves have been connected and the outer links have been ranked according to the expected error likelihood, the ranking information determines which inner leaf logical qubits to connect with each other. We have already discussed how the BSM for the concatenated GKP + [[7,1,3]] code qubits is performed. Here we will focus on the error analysis.

In the preparation of the cube, the first seven out of eight GKP qubits in FIG.~\ref{fig:ResourceState} form the inner leaf qubit. The uncorrelated logical errors that can arise on each of these inner leaf qubits at this stage are given in TABLE~\ref{tab:uncorr_Z_errors} (note that while for the outer leaves we moved all the relevant errors in the preparation of both the left and the right cube onto the left outer leaf qubit, for the inner leaves we separately simulate every qubit). The native shift errors on the inner leaf qubits are the same as for the outer leaf qubit, namely they have variances $(\sigma_{\text{GKP}}^2,2\sigma_{\text{GKP}}^2)$ in the $q$ and $p$ quadrature respectively. Moreover, one extra error possibility relates to the correlated errors arising in step c) when the logical error on the measured red leaf qubit of the top 3-tree would propagate as the correlated error on the grey and red leaf qubits of the bottom 3-tree. These correlated errors further propagate to the qubits forming the cube in steps d), d'), e) and f) as described in Appendix~\ref{sec:resource_generation}. As described in TABLE~\ref{tab:correlated_Z_errors}, six out of the seven qubits forming the encoded inner leaf qubit are susceptible to the correlated errors. As these errors result from the BSM in step c) we consider post-selection here and do not store analog information. Hence in our simulation we sample from the Bernoulli distribution with the error probability variable given in Eq.~\eqref{eq:step_c_propagated_error}.

\begin{figure}
\includegraphics[width =\columnwidth]{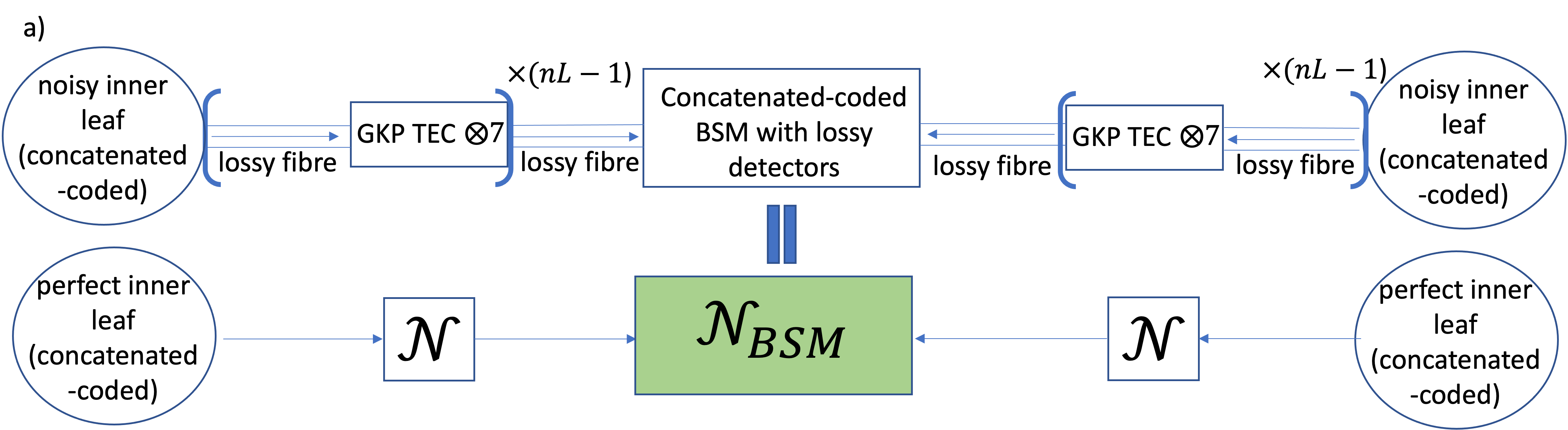} \\
\includegraphics[width =\columnwidth]{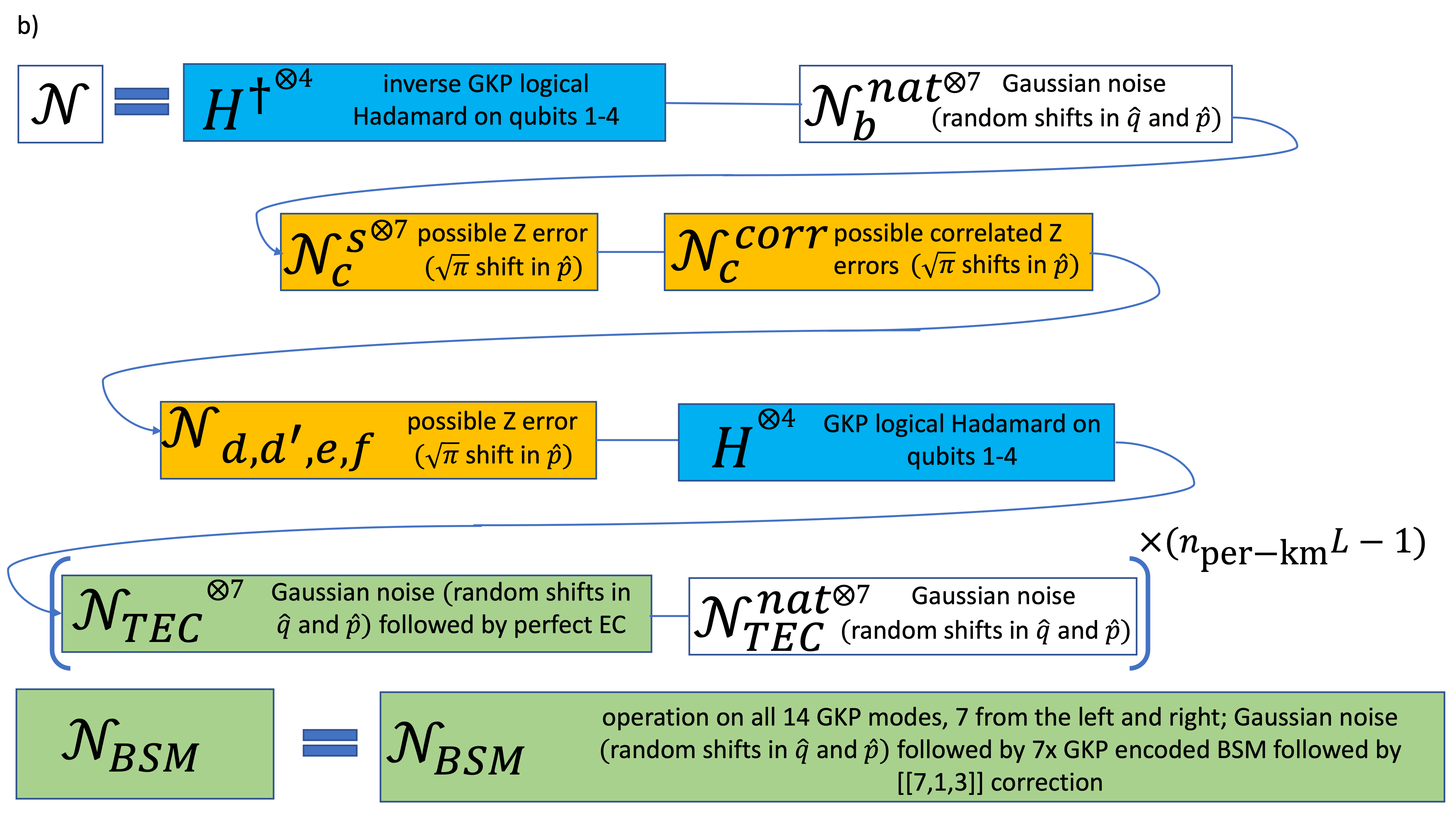}
\caption{Description of error processes affecting inner leaves. a) The imperfect preparation of the inner leaves is followed by their local storage in the lossy fiber with intermediate GKP TEC to overcome storage losses. Finally a concatenated-coded error-protected BSM is performed on the two concatenated-coded qubits. This can be modelled as having perfect inner leaf qubits passing through noisy channels $\mathcal{N}$ followed by the noisy BSM. b) Decomposition of $\mathcal{N}$ into individual error channels. The orange and green boxes describe noise channel which involve GKP stabilizer measurement and GKP error correction. This stabilizer GKP analog information is used in orange channels for post-selection. The analog information from green channels is used for boosting the performance of the [[7,1,3]] code decoding after the BSM. The blue channels denote additional Hadamard operations on qubits 1-4 which swap the noise and errors between the $q$ and $p$ quadratures on these qubits. Since in the presented model we assume that a perfectly prepared concatenated-coded inner leaves are sent through noisy channels $\mathcal{N}$ we need to apply the Hadamards twice as part of the noisy channel. Firstly we apply $H^\dag$ to undo the Hadamards from the perfect state preparation, then we apply the noise and then restore the Hadamards again.}
\label{fig:InnerLeaves}
\end{figure}

After the cube preparation, we first apply a Hadamard on qubits 1-4 to transform the cube to the desired Bell pair between the physical GKP qubit and the concatenated GKP + [[7,1,3]] qubit. These Hadamards exchange the native shift errors on these qubits and transfer the $Z$ logical errors onto $X$ logical errors on these qubits.  Then we store the 7 GKP qubits in a fiber loop until the ranking information from the outer leaves arrives. In the meantime, to counteract loss errors we periodically implement GKP TEC and store the corresponding generated analog information. To overcome channel losses through this TEC we consider the standard pre-amplification strategy. Hence the effective noise corresponds to a Gaussian random displacement channel with $\sigma^2 = 1-\eta$, where $\eta = \exp(-\frac{L}{n L_0})=\exp(-\frac{1}{n_{\text{per-km}} L_0})$. Here $L$ is the repeater separation and the total number of GKP corrections on a given GKP qubit during its storage is given by $n = n_{\text{per-km}}L$ (including the final classical GKP correction during the BSM) with $n_{\text{per-km}}$ being the number of GKP corrections during storage per km of fiber. As discussed in Appendix~\ref{sec:NoisyGKPMeasurements} the TEC can be modeled as ideal GKP correction with additional Gaussian random displacement channels, one with $\sigma^2 = \sigma_{\text{GKP}}^2 + \frac{1-\eta_d}{\eta_d}$ preceding the correction and one with $\sigma^2 = \sigma_{\text{GKP}}^2$ after the correction. The final channel directly preceding the logical BSM does not involve a pre-amplfication but rather the cc-amplification, i.e. each of the GKP qubits passes through a Gaussian random displacement channel with $\sigma^2 =  \frac{1-\eta\eta_d}{2\eta\eta_d}$. We list all the error channels in TABLE~\ref{tab:tableInnerLeaves} and graphically illustrate them in FIG.~\ref{fig:InnerLeaves}.

The final logical BSM implements both GKP and [[7,1,3]] code correction. After the 14 homodyne measurement outcomes from the 7 GKP Bell measurements are decoded into binary outcomes, the [[7,1,3]] code correction is implemented for both $X$ and $Z$ errors. For each of these, the classical parity check matrix corresponding to the 3 $X$ ($Z$) stabilizers is:

\begin{equation}
H = 
\begin{pmatrix}
0 & 0 & 0 & 1 & 1 & 1 & 1\\
0 & 1 & 1 & 0 & 0 & 1 & 1\\
1 & 0 & 1 & 0 & 1 & 0 & 1
\end{pmatrix}
\end{equation}
and it is applied to the string of 7 outcome bits. We interpret the obtained syndrome using the analog information. In this way we can correct not only single but also most of the two-qubit errors and increase the reliability of our encoded BSM~\cite{rozpkedek2021quantum}. Since the stabilizers in $X$ and $Z$ for the [[7,1,3]] code are the same, the procedure is the same for both quadratures. For each quadrature each error syndrome is consistent with one single-qubit error and three two-qubit errors. The analog information allows us to infer which one of these 4 cases is most likely. Specifically we evaluate the error likelihood for each case by finding the likelihood that the expected erroneous qubit(s) had an error and the other qubits did not have an error. We note that the three two-qubit error cases are equivalent to each other up to the stabilizers, but due to numerical precision and simulation efficiency issues we find that treating them separately leads to a better performance. Note that in total we have 14 GKP qubits involved in this measurement but for each quadrature we have only 7 bits of classical outcomes. This is because in our BSM we only care about the reliability of the final outcome, hence we want to identify which of the qubit pairs had an error. Within a pair of GKP qubits that was transversally connected through a BSM, we do not care whether the error that led to the $X$ ($Z$) flip of the outcome occurred on the left or the right qubit. We need to only identify on which pair(s) the error happened to identify which of the 7 bits are erroneous. Therefore for each pair $j \in \{1,...,7\}$ we evaluate the likelihood of error as:

\begin{equation}
    P_j = \frac{1 - \Pi_i^m (1-2p^j_i)}{2} \, ,
\end{equation}

where $p^j_i$ corresponds to a likelihood of GKP error at GKP correction $i$ on one of the two qubits belonging to pair $j$. Here index $i$ runs over all the corrections in the $q$ $(p)$ quadrature that contributed analog information over both the left and right qubits of the specific pair. Then the total likelihood of each of the four error patterns consistent with the error syndrome is just the product of $P_j$'s for $j$'s corresponding to the erroneous qubits and $1-P_j$ for $j$'s which in the given pattern do not have an error. Then we pick the error pattern with the largest likelihood and flip the corresponding classical bit(s) in the 7-bit measurement outcome in that quadrature. After that we read off the logical measurement outcome by xor-ing all the 7 bits as for the [[7,1,3]]-code the logical $X$ ($Z$) operator has the form $XXXXXXX$ ($ZZZZZZZ$).

We note that, as described in Section~\ref{sec:innerleaves}, in the final CC-amplification we measure both the GKP and the [[7,1,3]] code stabilizers destructively and the correction is done on the classical data. This means that the only type of error that can happen after the correction is a logical error on the second level, which effectively corresponds to a flip of the measurement outcome. This in turn will correspond to an $X$ ($Z$) error on the qubit teleported through our entanglement swapping on the inner leaves. Hence in the simulation we can verify whether the BSM was performed reliably or resulted in a logical $X$ or $Z$ error.

\begin{table}
\begin{tabular}{ |p{2.25cm}||p{4.5cm}|p{0.5cm}|p{5.5cm}|p{1.75cm}| p{1cm}|p{1cm}| }
 \hline
 \multicolumn{7}{|c|}{Logical error channels per each physical GKP inner-leaf qubit} \\
 \hline
 Logical channel & description& $Z$ or $X$ err. &standard deviation $\sigma$ & no of times & post-select.& analog info for [[7,1,3]] corr.\\
 \hline
 $\mathcal{N}^s_{c}$&   single qubit error propagating from the fusion in step c)   & $Z$   &$\sigma^2_{Z} = 3\sigma_{GKP}^2 +\frac{1 - \eta_d}{\eta_d}$ & 1 &yes & no\\
 $\mathcal{N}^{corr}_{c}$&   correlated error between inner leaf qubits propagating from the fusion in step c)   & $Z$   &$\sigma^2_{Z} = 3\sigma_{GKP}^2 +\frac{1 - \eta_d}{\eta_d}$ & see TABLE~\ref{tab:correlated_Z_errors} &yes & no\\
 $\mathcal{N}_{d,d',e,f}$ &single qubit errors propagating from fusions in steps d), d'), e) and f) & $Z$&  $\sigma^2_{Z} = 3\sigma_{GKP}^2 +\frac{1 - \eta_d}{\eta_d}$ or $\;\;\;\;\;\,\,\,\;\;\;\;\;$ $\sigma^2_{Z} = 2\sigma_{GKP}^2 +\frac{1 - \eta_d}{\eta_d}$ & see TABLE~\ref{tab:uncorr_Z_errors} & yes & no \\
 $\mathcal{N}_{TEC}\circ \mathcal{N}^{\text{nat}}_{b}$ &   First GKP TEC correcting native errors after Steane EC in step b) and the first segment of the lossy storage channel. This channel only appears if $nL>1$.   & $Z$ and $X$   &$\sigma^2_{Z} = 3\sigma_{GKP}^2 + (1-\eta) + \frac{1 - \eta_d}{\eta_d}$, $\;\;\;\;\;\,\,\,\;\;\; \,\,\,\,\,\,\,\,\,\,\,\,\,\,\,\,\,\,\,\,\,\,\,\,$  $\sigma^2_{X} = 2\sigma_{GKP}^2 + (1-\eta) + \frac{1 - \eta_d}{\eta_d}$ where ${\eta=\exp\left(-1/(n_{\text{per-km}} \times L_0\right))}$  & 1 &no & yes\\
 $\mathcal{N}_{TEC}\circ \mathcal{N}^{\text{nat}}_{TEC}$ &   Intermediate GKP TEC correcting native errors from the previous GKP TEC and the subsequent lossy storage channel. This channel only appears if $n_{\text{per-km}}L>2$.   & $Z$ and $X$   &$\sigma^2 = 2\sigma_{GKP}^2 + (1-\eta) + \frac{1 - \eta_d}{\eta_d}$ where ${\eta=\exp\left(-1/(n_{\text{per-km}} \times L_0\right))}$  & $n_{\text{per-km}}L-2$ &no & yes\\
 
 $\mathcal{N}_{BSM}\circ \mathcal{N}^{\text{nat}}_{TEC}$ &   Final CC-amplification channel on two GKP qubits from two different cubes that are fused together correcting native errors from the previous GKP TEC and the subsequent lossy storage channel. The variance given here is the total joint channel simulating noise for both inner leaf qubits involved in the BSM. This channel only appears if $n_{\text{per-km}}L>1$.   & $Z$ and $X$   &$\sigma^2 = 2\sigma_{GKP}^2 + \frac{1 - \eta\eta_d}{\eta\eta_d}$, where ${\eta=\exp\left(-1/(n_{\text{per-km}} \times L_0\right))}$  & $1$ (per pair of GKP qubits from two different cubes) &no & yes\\
 
 $\mathcal{N}_{BSM}\circ \mathcal{N}^{\text{nat}}_{b}$ & CC-amplification channel on two GKP qubits from two different cubes that are fused together correcting native errors after Steane EC in step b) and the subsequent lossy storage channel. The variance given here is the total joint channel simulating noise for both inner leaf qubits involved in the BSM. This channel only appears if $n_{\text{per-km}}L=1$.   & $Z$ and $X$   &$\sigma^2_{Z} = 4\sigma_{GKP}^2 + \frac{1 - \eta\eta_d}{\eta\eta_d}$,  $\sigma^2_{X} = 2\sigma_{GKP}^2 + \frac{1 - \eta\eta_d}{\eta\eta_d}$, where ${\eta=\exp\left(-1/(n_{\text{per-km}} \times L_0\right))}$  & $1$ (per pair of GKP qubits from two different cubes) &no & yes\\
 \hline
\end{tabular}
\begin{tabular}{ |p{2.25cm}||p{15.0cm}| }
$H$ & Hadamard acting on qubits 1-4 in FIG.~\ref{fig:InnerLeaves} transfers all the $Z$ errors from channels $\mathcal{N}_{c}^s, \, \mathcal{N}_{c}^{corr}, \, \mathcal{N}_{d,d',e,f}$ onto $X$ errors on these GKP qubits. Moreover, the channels described above involving $\mathcal{N}_b^{\text{nat}}$ have their errors exchanged between $X$ and $Z$ on these qubits.  \\
\hline
\end{tabular}
\caption{Error processes affecting inner leaves.}
\label{tab:tableInnerLeaves}
\end{table}

\section{Inner leaf information in terms of typical sequences}
\label{sec:typicalsequences}

In this appendix we provide additional insights into how the inner leaf information enables for maintaining positive rate for much larger distances. A good way of explaining this phenomenon is using typical sequences. The specific sequences we are referring to here are the sequences of bits in $\vec{s}_X$ and $\vec{s}_Z$, since the inner leaf information allows us to extract the secret key (entanglement) separately from the bins with different values of $\abs{\vec{s}_{X}}$ and $\abs{\vec{s}_{Z}}$.

For any $\epsilon > 0$, the typical sequences $x_1, x_2, ..., x_n$ drawn from an i.i.d.~distribution $X$ are defined as those sequences whose probability $p(x_1, x_2, ..., x_n)$ satisfies:
\begin{equation}
    2^{-n(H(X) + \epsilon)} \le p(x_1, x_2, ..., x_n) \le 2^{-n(H(X) - \epsilon)} \, ,
\label{eq:TypSeq}
\end{equation}
where $H(X)$ is the Shannon entropy of the distribution $X$. The probability of a randomly drawn sequence belonging to this set approaches 1 as $n \rightarrow \infty$. In our case of sequences $\vec{s}_X$ and $\vec{s}_Z$ the underlying distribution is the Bernoulli distribution for each trial and then one can show that $2^{-nH(X)} = p_{av}$ where $p_{av} = t^{nt}(1-t)^{n(1-t)}$ is the probability of an average sequence with $nt$ 1s (corresponding in our case to $nt$ repeaters with an error syndrome and $n(1-t)$ repeaters with zero-error syndrome) for the Bernoulli distribution with parameter $t$. Hence the typical sequences are the sequences centered around the average sequences. For finite $n$ the probability of a randomly sampled sequence belonging to the set of typical sequences depends on $\epsilon$. For the case of Bernoulli distribution it might be more helpful to define the sequences included in the typical set in terms of the weights of the corresponding bit strings rather than $\epsilon$. The two are linked by a simple linear relationship. This can be seen as follows. For $\epsilon=0$ we obtain only the sequences with the average number of 0's and 1's. Let us assume $t<0.5$ which is the case for us. Then the sequence with probability
\begin{equation}
    P = 2^{-n(H(X) + \epsilon)}
\label{eq:ProbMore1s}
\end{equation}
for $\epsilon>0$ will include more 1's than the average sequence and occur with the smaller probability than the average sequence. The probability of a sequence with $nz$ more 1's than the average sequence is given by:
\begin{equation}
    P = t^{n(t+z)}(1-t)^{n(1-t-z)}
\label{eq:ProbnzMore1s}
\end{equation}
Hence equating Eq.~\eqref{eq:ProbMore1s} and Eq.~\eqref{eq:ProbnzMore1s} we find that
\begin{equation}
    \epsilon = z (\log_2(1-t) - \log_2(t)).
\label{eq:epsiloVsz}
\end{equation}
Similarly, we can see that the probability of a sequence with $nz$ less 1's than the average sequence is given by:
\begin{equation}
    P = 2^{-n(H(X) - \epsilon)} \, .
\end{equation}
This shows that for a Bernoulli distribution parameterised by $t$ there is a simple linear relationship between defining the typical set in terms of sequences with specific number of 1's, and defining them in terms of their probabilities as defined in Eq.~\eqref{eq:TypSeq}. Specifically, we can define the typical set to include all the sequences whose number of 1's is between $n(t-z)$ and $n(t+z)$. This corresponds to a typical set defined in Eq.~\eqref{eq:TypSeq} in terms of the probabilities of the corresponding typical sequences with $\epsilon$ given in Eq.~\ref{eq:epsiloVsz}. Clearly we require that $z<\min(t,1-t)$.

\begin{figure}
\includegraphics[width =\columnwidth]{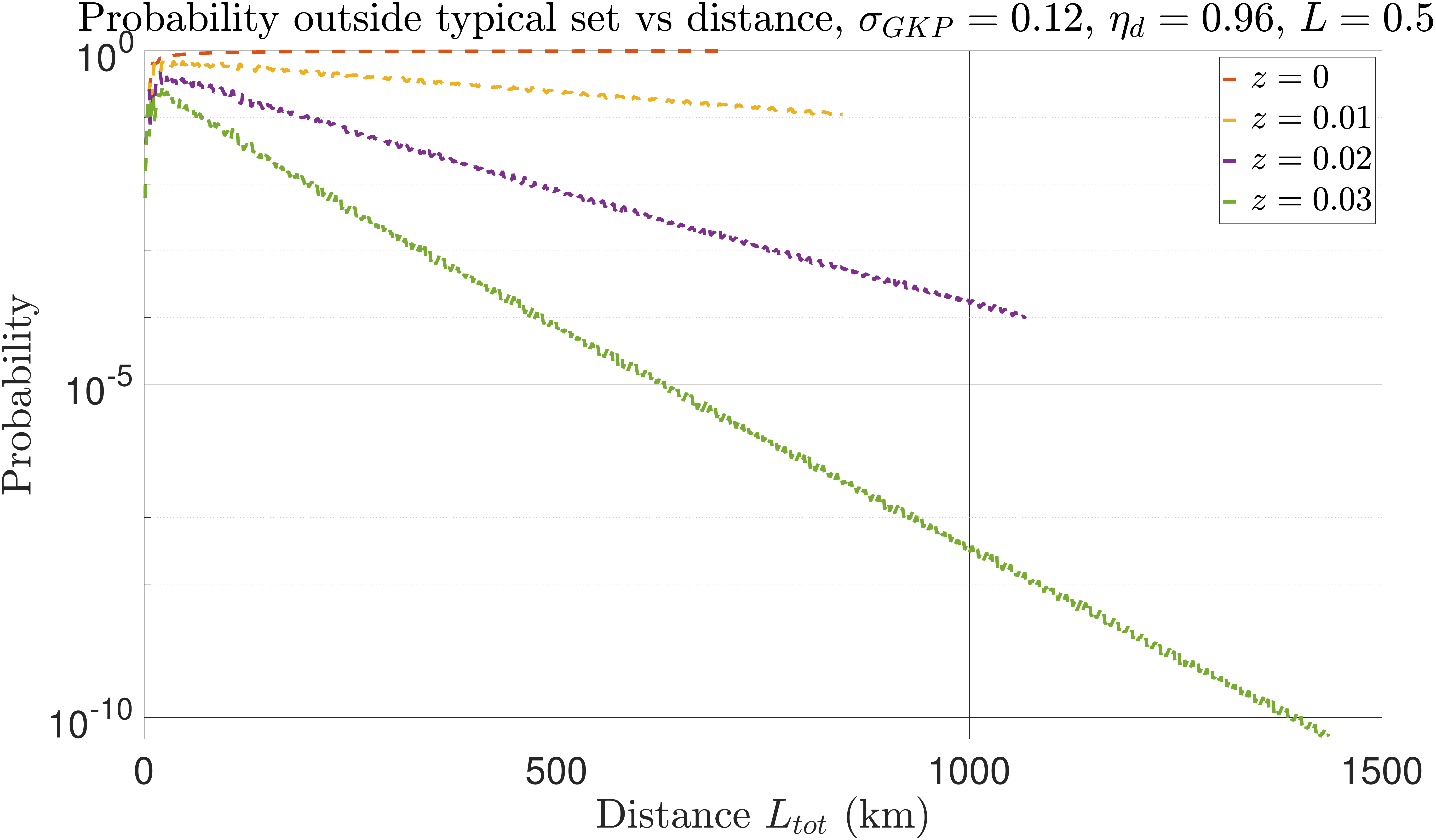}
\caption{Probabilities of at least one of $\vec{s}_X$ and $\vec{s}_Z$ sequences being outside of the typical set as a function of total distance $L_{\text{tot}}$ for different typical sequences parameterised by $z$. The curves stop at the value of $L_{\text{tot}}$ at which the corresponding rate $R$ drops to zero.}
\label{fig:ProbOutTypSet}
\end{figure}

In FIG.~\ref{fig:ProbOutTypSet} we plot the probability of being outside the typical set for at least one of the two bit-strings $\vec{s}_X$ and $\vec{s}_Z$ for different values of $z$ (the same value of $z$ is used for both bit-strings) for the parameters considered in FIG.~\ref{fig:InnerLeafInfo}. For these parameters $t_Z = 0.0537$ and $t_X = 0.0581$. This gives us an indication what values of $z$ would correspond to including the best all-zero sequence for $\vec{s}_X$ and $\vec{s}_Z$ strings.

\begin{figure}
\includegraphics[width =\columnwidth]{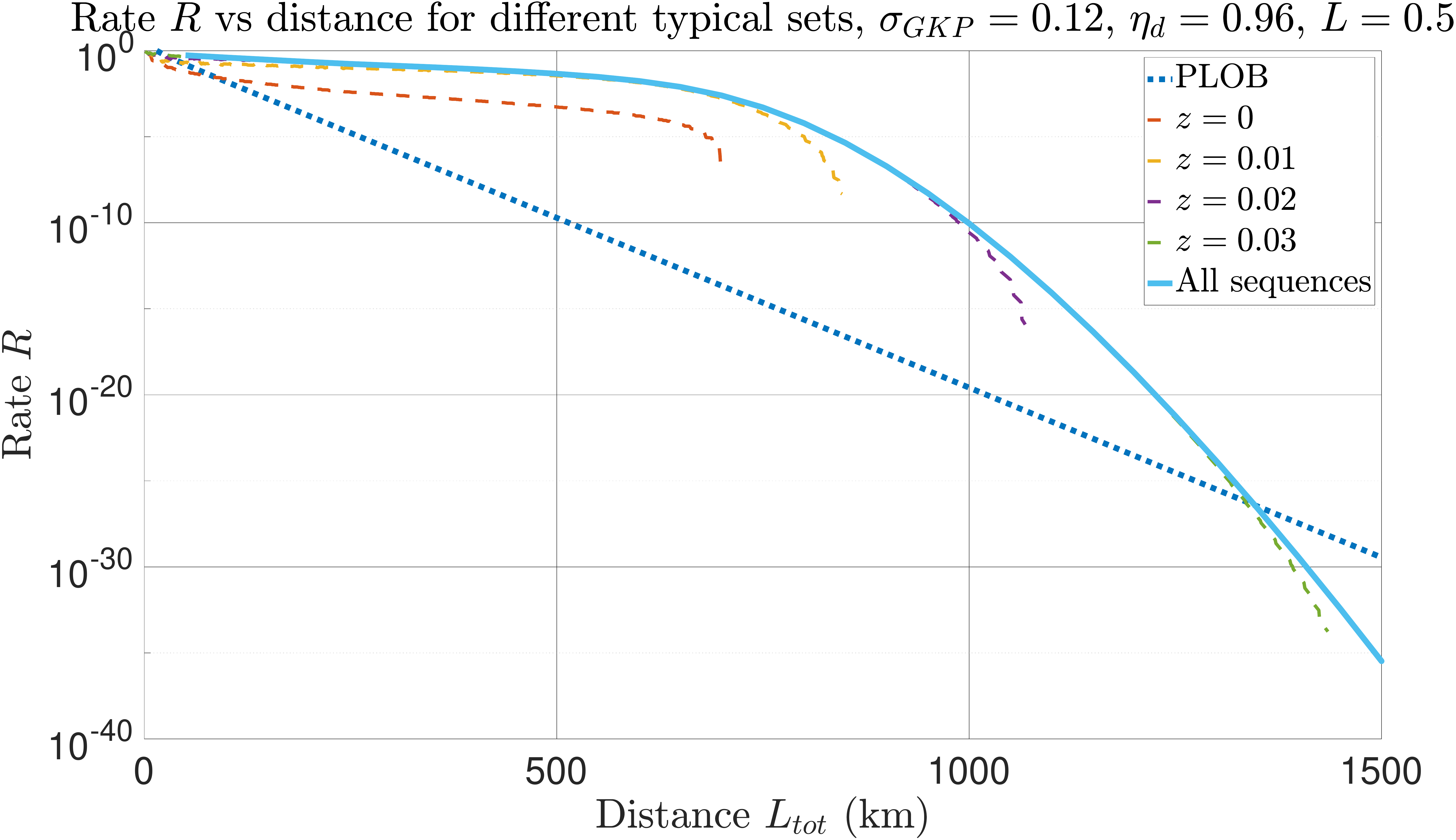}
\caption{Rate $R$ as a function of total distance $L_{\text{tot}}$ for different typical sequences parameterised by $z$. Additionally we also re-plot the rate that includes all the sequences and the PLOB bound. We note that the sudden break of the dashed curves is due to a rapid drop of the corresponding rate to zero which is so quick that it cannot be captured by the resolution of our data.}
\label{fig:RateTypicalSequences}
\end{figure}

In FIG.~\ref{fig:RateTypicalSequences} on the other hand we plot the rate achieved when post-selecting only the bins belonging to the corresponding typical set and then extracting entanglement (secret key) from each of those bins separately. Additionally we also re-plot the full rate that includes all the bins. On both plots we only plot the curves for distances for which the rate is non-zero.

We find that increasing $z$ and hence $\epsilon$ increases the achievable distance. This is precisely because for every typical set defined for a fixed $z$ there exists a distance for which the QBER is too large for all the sequences belonging to that typical set. By increasing $z$ we enlarge the typical set to include those before atypical sequences for which the QBER for this distance is still much smaller hence allowing for key/entanglement extraction. Hence at every distance at which for a given $z$ the performance starts rapidly dropping to zero, we can rescue it by including those atypical sequences with smaller error. We see that including all the sequences allows for achieving the largest distance.

It is interesting to note that the probabilities of those atypical sequences that can rescue the rate are very small. This can be seen from the fact that for most values of $z$ and for the distance at which the corresponding rate drops to zero (end of the curves in FIG.~\ref{fig:ProbOutTypSet}), the probability of being outside of the typical set for at least one of $\vec{s}_X$ and $\vec{s}_Z$ is generally small.
Yet the fact that for that same distance the rate can be made positive by increasing $z$ shows that those previously excluded sequences, for which the probability of obtaining any of them is very small, include bins which can generate non-zero rate at these distances. In fact we see that the ends of the curves in FIG.~\ref{fig:ProbOutTypSet} could be connected by a curve that would be close to a straight line. This also shows that the probability of being outside of the typical set for at least one of $\vec{s}_X$ and $\vec{s}_Z$ for the largest value of $z$ for which the typical set cannot generate any key/entanglement for some distance $L_{\text{tot}}$, decays exponentially with that distance as well.

Here we also explain in more detail why for large repeater separation of $L=5$ km with very good parameters the achievable distance with respect to the PLOB bound is almost the same as the one with respect to the threshold of $10^{-6}$. This is because of the following feature of the inner leaf information. Firstly we note that for the best, most highly ranked links, the error rate for the single elementary link corresponding to $s=0$ (Steane code zero-error syndrome) is in most cases at least one and often two or three orders of magnitude smaller than the corresponding error for the $s=1$ case. Therefore the smallest non-zero key that we can expect to obtain will correspond to the largest distance for which we can still extract key from the $\vec{s}=\bar{0}$ sequence for both $X$ and $Z$ (we note that here we cut off if that will lead to rate below $10^{-30}$). If the repeaters are placed more densely, these bit strings grow faster with distance, e.g.~in FIG.~\ref{fig:AchievableDistances} we see that for $\eta_d = 0.99$ and $\sigma_{\text{GKP}} = 0.15$ the optimal repeater separation is $L=0.5$ km. For that configuration $t_Z \approx 0.026$ and $t_X = 0.031$. This means that at the achievable distance with respect to PLOB of 7450 km the probability of having $\vec{s}=\bar{0}$ sequence for both $Z$ and $X$ is $(1-t_Z)^{2 \times 7450} \times (1-t_X)^{2 \times 7450}$. This is a completely negligible number and since at that distance our rate is still above $10^{-30}$, we see that other bit strings still contribute at this distance. This shows us that the distance at which only the $\vec{s}=\bar{0}$ bit string will contribute to the rate is much larger than 7450 km and the probability of that bit string there will be even smaller than at 7450 km. Hence, we see that the reason that for this configuration the rate decays so gradually leading to a significant distance range over which the rate is below $10^{-6}$ and above PLOB and $10^{-30}$ is that the probabilities of individual bins/sequences can be very small but from many of these bins one can still extract significant amount of key/entanglement. Hence we conjecture that the slow decay of the overall rate is dictated by the decay of the probabilities of being in the specific bins. 

For the case of large repeater separation $L=5$ and $\eta_d = 0.99$ and $\sigma_{\text{GKP}} = 0.12$ on the other hand, the rate drops from around $10^{-6}$ at the distance of 2300 km to $10^{-10}$ at 2350 km and then immediately to zero afterwards. This is because
the values of $t$ is slightly lower in this case, they are $t_Z \approx 0.023$ and $t_X \approx 0.024$ but more importantly because the repeaters are now ten times further away. Hence at the achievable distances of 2300-2350 km the probability of having $\vec{s}=\bar{0}$ sequence for both $Z$ and $X$ is $(1-t_Z)^{2300/5} \times (1-t_X)^{2300/5} \approx 10^{-10}$. Since the secret fraction decays fast, for a distance of 2400 km it is already zero for that bin. Hence the fact that for these parameters the probabilities of the individual sequences are much higher means that there is no finer graining possible and the rate decays very fast leaving little or no gap between the achievable distance defined with respect to the rate being above $10^{-6}$ and $10^{-30}$ with beating PLOB. 

\section{All achievable distances}
\label{sec:AllAchievableDistanceTABLEs}

\begin{figure}
    \centering
    \includegraphics[width=1\textwidth]{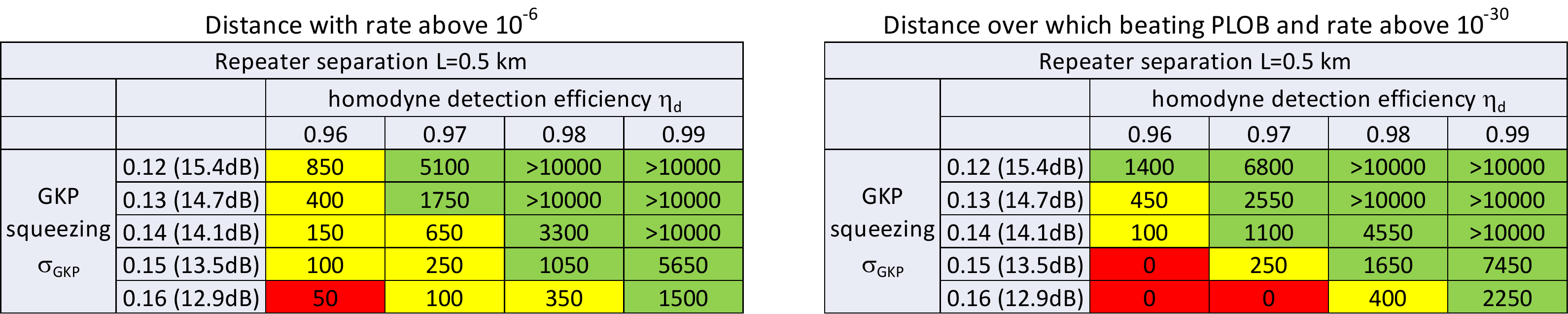}
    \caption{Achievable distances for $L=0.5$ km}
    \label{fig:AchievableDistancesL05}
\end{figure}

\begin{figure}
    \centering
    \includegraphics[width=1\textwidth]{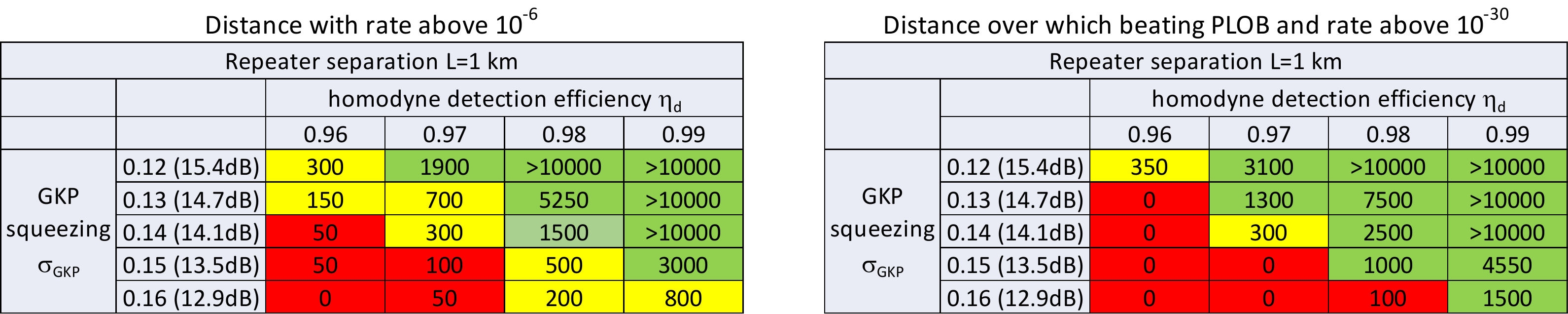}
    \caption{Achievable distances for $L=1$ km}
    \label{fig:AchievableDistancesL10}
\end{figure}

\begin{figure}
    \centering
    \includegraphics[width=1\textwidth]{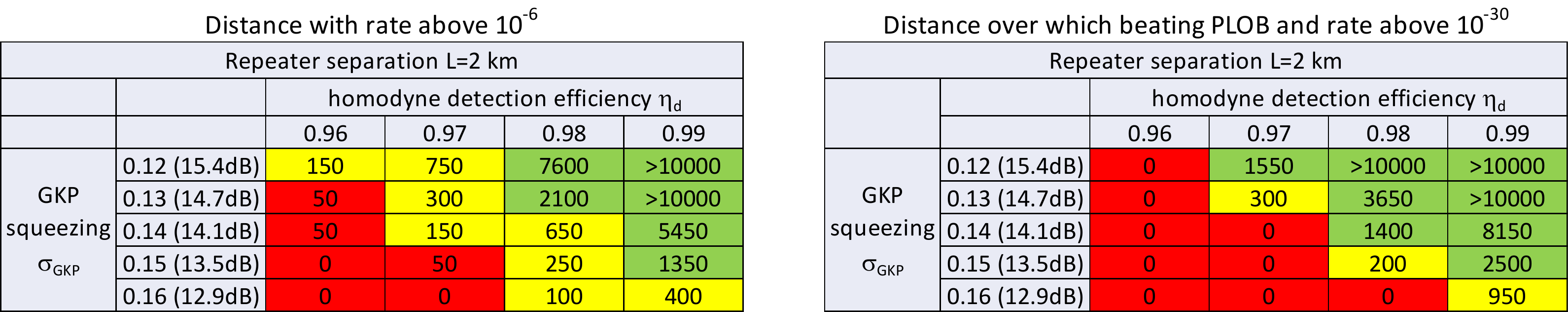}
    \caption{Achievable distances for $L=2$ km}
    \label{fig:AchievableDistancesL20}
\end{figure}

\begin{figure}
    \centering
    \includegraphics[width=1\textwidth]{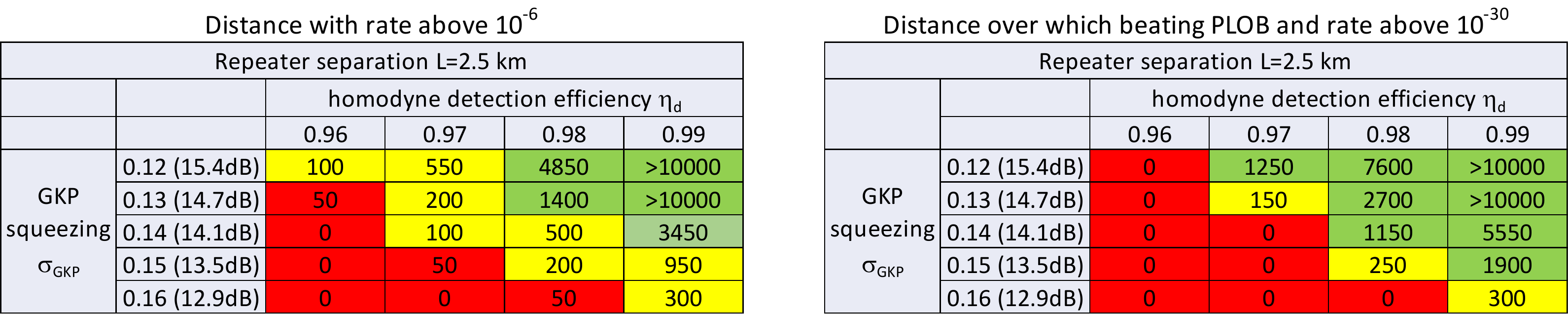}
    \caption{Achievable distances for $L=2.5$ km}
    \label{fig:AchievableDistancesL25}
\end{figure}

\begin{figure}
    \centering
    \includegraphics[width=1\textwidth]{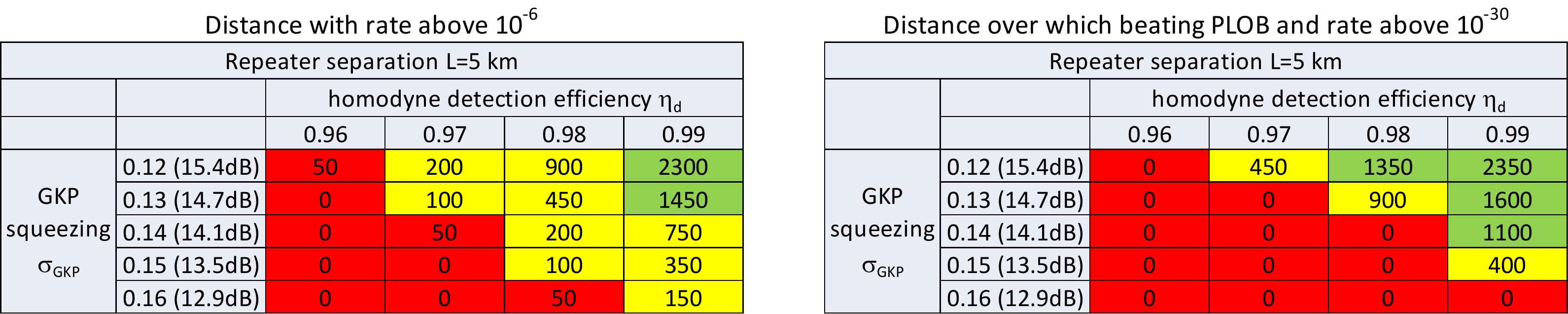}
    \caption{Achievable distances for $L=5$ km}
    \label{fig:AchievableDistancesL50}
\end{figure}

In this Appendix we provide the tables with achievable distances for all the considered configurations. The results are shown in FIG.~\ref{fig:AchievableDistancesL05}, FIG.~\ref{fig:AchievableDistancesL10}, FIG.~\ref{fig:AchievableDistancesL20}, FIG.~\ref{fig:AchievableDistancesL25} and  FIG.~\ref{fig:AchievableDistancesL50}.


\section{Simulation accuracy}
\label{sec:SimAcc}

In this appendix we describe how we quantify the accuracy of our Monte-Carlo simulations. We specify the desired accuracy and run the simulation until that accuracy is reached. Specifically, we start with 10 simulation runs. After that we estimate the standard error on the obtained quantities. If that error estimate satisfies the desired accuracy, we output the results. Otherwise we rerun the simulation with 10 times more runs and again estimate the standard error on the obtained data. We repeat this procedure until we have simulated so many runs that the desired level of accuracy has been reached.

We estimate the standard error on all $2k$ $Z$ and $X$-flip probabilities (after combining the inner-leaf $Q_{X/Z, \text{inner}}(s=0,1)$ and outer-leaf $Q_{X/Z, \text{outer}}(j=\{1,...,k\})$ flip probabilities using Eq.~\eqref{eq:elementaryLinkFlip}) over the single link ($k$ values for $k$ multiplexed ranked links and then for each ranked link we have two values corresponding to the case with $s=0$ and $s=1$ for the inner leaf information). We also estimate the standard error on the probability of the inner leaf error syndrome $t_X$ and $t_Z$. Clearly all the variables are Bernoulli variables so the standard error estimate is given by:
\begin{equation}
\begin{aligned}
    \Delta \tilde{Q}_{X/Z}(s,j) &= \sqrt{\frac{\tilde{Q}_{X/Z}(s,j)(1-\tilde{Q}_{X/Z}(s,j))}{N_{X/Z}(s)}} \, , \\
    \Delta \tilde{t}_{X/Z} &= \sqrt{\frac{\tilde{t}_{X/Z}(1-\tilde{t}_{X/Z}))}{N_{\text{tot}}}} \, .
\end{aligned}
\end{equation}
where tilde indicates that we are dealing with estimates based on currently available simulation output and $N$ is the number of simulation runs. We also note that the number of simulation runs for the error estimate when $s=0$ and when $s=1$ is not fixed but rather depends on the value of $\tilde{t}_{X/Z}$. Specifically, when setting a number of total simulation runs, it is only $N_{\text{tot}}$ that we set. The values of $N_{X/Z}(s)$ depend on $N_{\text{tot}}$ as follows:
\begin{equation}
\begin{aligned}
    N_{X/Z}(s=0) &= N_{\text{tot}}(1-\tilde{t}_{X/Z}) \, , \\
    N_{X/Z}(s=1) &= N_{\text{tot}}\tilde{t}_{X/Z} \, .
\end{aligned}
\end{equation}
Clearly, since we expect $\tilde{t}_{X/Z}$ to be small, we will have much less data points for the $s=1$ case than the $s=0$ case when fixing $N_{\text{tot}}$. However, the actual value of the error will be much larger for the $s=1$ than the $s=0$ case so much less simulation runs will be sufficient for the first than for the second case in order to achieve the same relative accuracy.

After calculating these estimates we calculate the relative error estimates $\Delta \tilde{Q}_{X/Z}(s,j)/\tilde{Q}_{X/Z}(s,j)$ and $\Delta \tilde{t}_{X/Z}/\tilde{t}_{X/Z}$. We can then check whether $\Delta \tilde{Q}_{X/Z}(s,j)/\tilde{Q}_{X/Z}(s,j) < b(s,j)_{X/Z}$ and $\Delta \tilde{t}_{X/Z}/\tilde{t}_{X/Z} < h_{X/Z} $ are satisfied for preset thresholds $b(s,j)_{X/Z}$ and $h_{X/Z}$. If at least one of the above $2k+1$ inequalities is not satisfied we increase $N_{\text{tot}}$ by a factor of 10 and rerun the simulation.

We note that for all but one of our simulations we set the accuracy thresholds as $b(s,j)_{X/Z} = 0.1$ for all $s$ and $j$ and $h_{X/Z} = 0.001$. The specific simulation with better accuracy was run for the data shown in FIG.~\ref{fig:RateVsMultiplexing}, where the data obtained for the ranking strategy has been simulated with accuracy given by $b(j)_{X/Z} = 0.05$ for all $j$ leading to the error bars shown on the plot.

\end{document}